\documentclass[11pt]{article}

\usepackage{amssymb}
\usepackage{amsthm}
\usepackage{amsmath}

\usepackage[mathscr]{eucal}

\usepackage{graphicx}
\usepackage{psfrag}
\usepackage{subfigure}

\usepackage{fancyheadings}

\renewcommand{\sectionmark}[1]%
        {\markboth%
                {}%
                {{\rm\thesection}\quad{\sc #1}}}

\newcommand{\proofend}{\hspace*{\fill}\rule{0.2cm}{0.2cm}}

\newcommand{\barM}{\bar{M}}
\newcommand{\barL}{\bar{\Lambda}}
\newcommand{\tildeM}{\tilde{M}}
\newcommand{\tildeC}{\tilde{C}}
\newcommand{\tilder}{\tilde{r}}

\newcommand{\threemetric}{{}^3\!g}

\renewcommand{\theenumi}{\roman{enumi}}

\newcounter{store}

\theoremstyle{plain}
\newtheorem{theorem}{Theorem}[section]
\newtheorem{corollary}[theorem]{Corollary}
\newtheorem{lemma}[theorem]{Lemma}
\newtheorem{proposition}[theorem]{Proposition}
\newtheorem{conjecture}[theorem]{Conjecture}
\newtheorem*{definition}{Definition}

\theoremstyle{remark}

\newtheorem*{remark}{Remark}

\begin{document}

\title{\huge \sc CMC-Slicings of Kottler-Schwarzschild-de Sitter Cosmologies}

\author{ \\
{\Large \sc Robert Beig}\thanks{Electronic address: {\tt Robert.Beig@univie.ac.at}} \\[0.5ex]
Institute for Theoretical Physics, University of Vienna, \\
Boltzmanngasse 5, A-1090 Vienna, Austria
\and \\
{\Large\sc J.\ Mark Heinzle}\thanks{Electronic address:  {\tt Mark.Heinzle@aei.mpg.de}} \\[0.5ex]
Max-Planck-Institute for Gravitational Physics, \\
Am M\"uhlenberg 1, D-14476 Golm, Germany \\[2ex] }

\date{}
\maketitle
\begin{abstract}

There is constructed, for each member of a
one-parameter family of cosmological
models, which is obtained from the Kottler-Schwarzschild-de Sitter
spacetime by identification under discrete isometries, a slicing
by spherically symmetric Cauchy hypersurfaces of constant mean
curvature. These slicings are unique up to the action of the
static Killing vector.
Analytical and numerical results are found as to when different leaves
of these slicings do not intersect, i.e. when the slicings form foliations.

\end{abstract}

\vspace{2cm}

\begin{center}
Keywords: \begin{minipage}[t]{10cm}
Constant mean curvature -- CMC-slicing -- CMC-foliation -- cosmological constant -- Schwarzschild-de Sitter
\end{minipage}
\end{center}

\vfill
\newpage
\section{Introduction}
\label{introduction}

Consider a globally hyperbolic spacetime which is spatially
compact.  One can ask the question whether there exists a slicing
by Cauchy surfaces of constant mean curvature (in short:
CMC-slicing) and what are its properties. Since CMC-slicings have
a wide variety of uses in general relativity, this question has
received a lot of attention, we refer to \cite{Rendall:1996} and
\cite{Andersson:2004} for recent overviews.

In the present paper we study this problem for a class of spatially compact spacetimes,
which are quotient spaces
of the Kottler-Schwarzschild-de Sitter family of spacetimes \cite{Kottler:1918} under
certain discrete subgroups of their isometry group.
These spacetimes can be viewed as models of cosmological spacetimes containing a black hole.
Studies of CMC-slicings which we are aware of
treat cases where the spacetime has --- in the future and/or past --- either an all-encompassing crushing
singularity or is geodesically
complete, whereas in Kottler-Schwarzschild-de Sitter both types of asymptotic behavior coexist.
Furthermore, the Kottler-Schwarzschild-de Sitter spacetimes, due to the presence of a positive
cosmological constant, violate the so-called timelike convergence condition (strong energy
condition) which most papers assume.

Before we state the main results of the present work,
let us define and describe the family of Kottler-Schwarzschild-de Sitter cosmologies we consider:

The Kottler-Schwarzschild-de Sitter metric reads
\begin{equation}\label{KSSdSmetric}
d s^2 = -V d t^2 + V^{-1} d r^2 + r^2 d\Omega^2 \qquad\text{where}\quad V= V(r) = 1-\frac{2 M}{r} -\frac{\Lambda r^2}{3}\:.
\end{equation}
The cosmological constant $\Lambda>0$ and the constant $M>0$ are required to satisfy
\begin{equation}
\label{range} 9 M^2 \Lambda <1\:;
\end{equation}
then the function $V(r)$ is positive in the interval $(r_{\mathrm{b}}, r_{\mathrm{c}})$,
and $V(r) = 0$ at the ``black hole horizon'' $r = r_{\mathrm{b}}$
and at the ``cosmological horizon'' $r= r_{\mathrm{c}}$.
The region $r_{\mathrm{b}} < r < r_{\mathrm{c}}$ is a static region of the spacetime
with Killing vector $\xi = \partial_t$.
(When $9 M^2 \Lambda >1$, there are no horizons, the spacetime is not static but homogeneous.)
It is straightforward to see that
\begin{equation}
2 M < r_{\mathrm{b}} < 3 M < \frac{1}{\sqrt{\Lambda}} < r_{\mathrm{c}} < \frac{3}{\sqrt{\Lambda}}\:.
\end{equation}
In the limit where $\Lambda$ goes to zero, the spacetime metric tends
to Schwarzschild and $r_{\mathrm{b}}$ tends to $2M$.
In the limit where $M$ goes to zero, the metric
becomes de Sitter and $r_{\mathrm{c}}$ tends to $3/\sqrt{\Lambda}$.

\begin{figure}[htp]
    \centering
    \subfigure[$0<r<r_{\mathrm{c}}$]{\label{tile0}\psfrag{A}[bc][cc][0.6][-45]{$r=r_{\mathrm{c}}$\,, $t=\infty$}
      \psfrag{B}[bc][cc][0.6][45]{$r=r_{\mathrm{c}}$\,, $t=-\infty$}
      \psfrag{C}[tc][cc][0.6][-45]{$r=r_{\mathrm{c}}$\,, $t=\infty$}
      \psfrag{D}[tc][cc][0.6][45]{$r=r_{\mathrm{c}}$\,, $t=-\infty$}
      \psfrag{J}[cc][cc][0.5][-45]{$r=r_{\mathrm{b}}$\,, $t=-\infty$}
      \psfrag{K}[cc][cc][0.5][45]{$r=r_{\mathrm{b}}$\,, $t=\infty$}
      \psfrag{L}[cc][tc][0.5][45]{$r=r_{\mathrm{b}}$\,, $t=\infty$}
      \psfrag{M}[cc][tc][0.5][-45]{$r=r_{\mathrm{b}}$\,, $t=-\infty$}
      \psfrag{E}[cc][tc][0.6][0]{$t>0$}
      \psfrag{F}[cc][cc][0.6][0]{$t=0$}
      \psfrag{G}[cc][cc][0.6][0]{$t<0$}
      \psfrag{H}[cc][tc][0.6][0]{$t<0$}
      \psfrag{I}[cc][tc][0.6][0]{$t>0$}
      \psfrag{N}[cc][cc][0.5][-70]{$t<0$}
      \psfrag{O}[cc][cc][0.5][90]{$t=0$}
      \psfrag{P}[cc][tc][0.5][70]{$t>0$}
      \psfrag{Q}[cc][cc][0.7][0]{$r=0$}
      \includegraphics[width=0.6\textwidth]{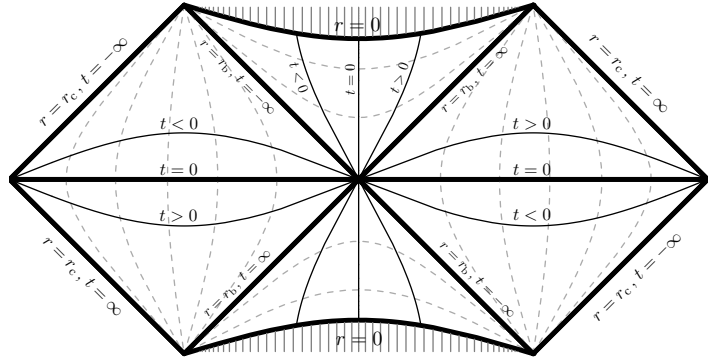}}
    \subfigure[$r_{\mathrm{b}}<r<\infty$]{\label{tile0p}\psfrag{A}[cc][cc][0.7][0]{$r=\infty$}
      \psfrag{B}[cc][cc][0.7][0]{$r=\infty$}
      \psfrag{C}[bc][cc][0.6][-45]{$r=r_{\mathrm{b}}$\,, $t=-\infty$}
      \psfrag{D}[tc][cc][0.6][45]{$r=r_{\mathrm{b}}$\,, $t=\infty$}
      \psfrag{E}[bc][cc][0.6][45]{$r=r_{\mathrm{b}}$\,, $t=\infty$}
      \psfrag{F}[tc][cc][0.6][-45]{$r=r_{\mathrm{b}}$\,, $t=-\infty$}
      \psfrag{G}[cc][cc][0.5][-45]{$r=r_{\mathrm{c}}$\,, $t=\infty$}
      \psfrag{H}[cc][cc][0.5][45]{$r=r_{\mathrm{c}}$\,, $t=-\infty$}
      \psfrag{I}[cc][cc][0.5][45]{$r=r_{\mathrm{c}}$\,, $t=-\infty$}
      \psfrag{J}[cc][cc][0.5][-45]{$r=r_{\mathrm{c}}$\,, $t=\infty$}
      \psfrag{K}[cc][tc][0.6][0]{$t<0$}
      \psfrag{L}[cc][cc][0.6][0]{$t=0$}
      \psfrag{M}[cc][cc][0.6][0]{$t>0$}
      \psfrag{N}[cc][tc][0.6][0]{$t>0$}
      \psfrag{O}[cc][cc][0.6][0]{$t=0$}
      \psfrag{P}[cc][cc][0.6][0]{$t<0$}
      \psfrag{Q}[cc][cc][0.5][-68]{$t>0$}
      \psfrag{R}[cc][cc][0.5][90]{$t=0$}
      \psfrag{S}[cc][tc][0.5][65]{$t<0$}
      \includegraphics[width=0.6\textwidth]{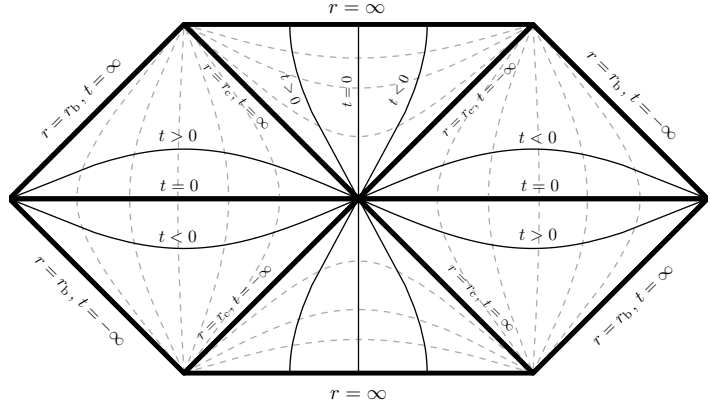}}
    \caption{{\small The figures show the compactified regions $0<r<r_{\mathrm{c}}$ and $r_{\mathrm{b}}<r<\infty$.
    Solid lines represent hypersurfaces $t=\mathrm{const}$,
        dashed lines are hypersurfaces $r=\mathrm{const}$.}}
        \label{tile00p}
\end{figure}

\begin{figure}[htp]
    \psfrag{r0}[cc][cc][0.8][0]{$r=0$}
    \psfrag{ri}[cc][cc][0.8][0]{$r=\infty$}
    \psfrag{a}[cc][cc][0.6][-45]{$r=r_{\mathrm{b}}$}
    \psfrag{b}[cc][cc][0.6][45]{$r=r_{\mathrm{b}}$}
    \psfrag{c}[cc][cc][0.6][45]{$r=r_{\mathrm{b}}$}
    \psfrag{d}[cc][cc][0.6][-45]{$r=r_{\mathrm{b}}$}
    \psfrag{e}[cc][cc][0.6][-45]{$r=r_{\mathrm{c}}$}
    \psfrag{f}[cc][cc][0.6][45]{$r=r_{\mathrm{c}}$}
    \psfrag{g}[cc][cc][0.6][45]{$r=r_{\mathrm{c}}$}
    \psfrag{h}[cc][cc][0.6][-45]{$r=r_{\mathrm{c}}$}
    \psfrag{t0}[cc][cc][0.6][90]{$t=0$}
    \psfrag{t1}[cc][cc][0.6][80]{$t=2\mathrm{T}$}
    \psfrag{id}[cc][cc][0.6][0]{identification}
    \centering
    \includegraphics[width=0.9\textwidth]{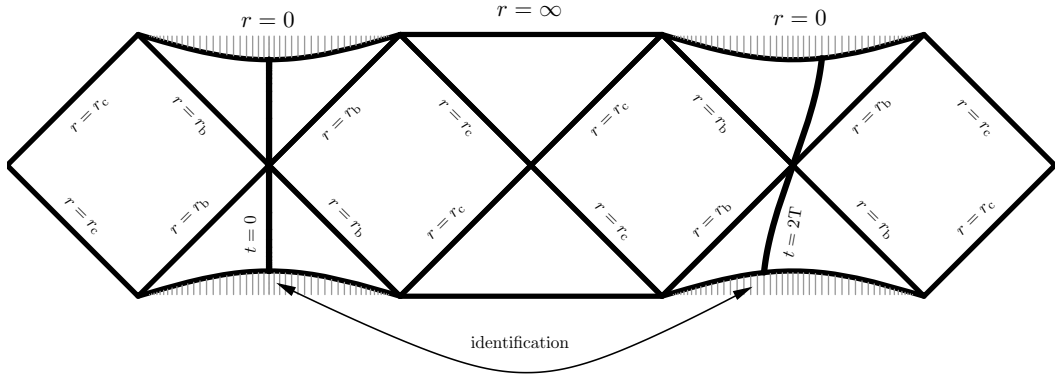}
    \caption{{\small The compactified Kottler-Schwarzschild-de Sitter spacetime KSSdS[T].}}
        \label{tile01}
\end{figure}

The static region of the spacetime has an analytic extension reminiscent of the Kruskal extension
of the Schwarzschild spacetime and the de Sitter spacetime. A common way of depicting
the arising spacetime uses two charts which cover the regions $0<r<r_{\mathrm{c}}$ and $r_{\mathrm{b}}<r<\infty$. The
conformal compactification of the region $0<r<r_{\mathrm{c}}$ is depicted in Fig.~\ref{tile0}; it contains the
curvature singularity $r=0$; the conformal compactification of $r_{\mathrm{b}}<r<\infty$ is shown in
Fig.~\ref{tile0p}. The overlap of the charts is $r_{\mathrm{b}} < r< r_{\mathrm{c}}$. As is well-known, the constructed
spacetime corresponding to the union of Figs.~\ref{tile0} and~\ref{tile0p} can be smoothly (in fact
analytically) extended in a ``periodic--in--r''-fashion. Thereby one obtains an inextendible,
globally hyperbolic spacetime of topology $\mathbb{R} \times \mathbb{R} \times S^2$ satisfying
$G_{\mu\nu} + \Lambda g_{\mu \nu} = 0$, see Fig.~\ref{tile01}. We call this spacetime the
Kottler-Schwarzschild-de Sitter spacetime KSSdS.

On KSSdS there exists an isometric action of $\mathbb{R} \times \mathrm{SO}(3)$. The dashed lines
in Figs.~\ref{tile0} and~\ref{tile0p} are orbits under the $\mathbb{R}$-factor in this action, i.e.
under the static Killing vector $\xi=\partial_t$. Henceforth this action will be called ``Killing
flow'' for brevity. Note that the
Killing vector $\xi$ is globally defined; it is null on the Killing horizons $r=r_{\mathrm{b}}$ and $r=r_{\mathrm{c}}$
which emanate from the bifurcation 2-spheres at which $\xi$ vanishes. Furthermore there exist
discrete isometries (``reflections'') leaving fixed the hypersurfaces $t= \mathrm{T} =
\mathrm{const}$, which are given via $\mathrm{T} + t \mapsto \mathrm{T} - t$.

While $r$ is globally defined on KSSdS, $t$ blows up on the Killing horizons. In the static region
or in the black hole, white hole, or cosmological regions, $(r,t)$ forms a coordinate system. The
solid lines in Figs.~\ref{tile0} and~\ref{tile0p} represent hypersurfaces $t=\mathrm{const}$, which
are totally geodesic as fixed point sets of the discrete isometries. There exist two kinds:
spacelike and timelike $t=\mathrm{const}$ hypersurfaces; the latter we call $t=\mathrm{const}$
cylinders.

By virtue of the periodicity in $r$, the $t=0$ cylinders in two adjoining
copies of the region $0<r<r_{\mathrm{c}}$ can be identified,
which results in a smooth spacetime of topology
$\mathbb{R} \times S^1 \times S^2$, which we call
the cosmological Kottler-Schwarzschild-de Sitter spacetime KSSdS[0].
The spacetime KSSdS is the universal covering of KSSdS[0].
More generally, let $\mathrm{T} \in \mathbb{R}$ and identify
points of equal radius $r$ on a $t=0$ cylinder and a tilted
$t=2 \mathrm{T}$ cylinder in an adjacent copy of the region $0<r < r_{\mathrm{c}}$
on the r.h.\ side, see Fig.~\ref{tile01}.
Thereby we obtain a whole family of inextendible, globally hyperbolic,
cosmological spacetimes, which we call KSSdS[T].

Note that KSSdS[T] is a smooth, in fact analytic, spacetime:
consider a neighborhood of the $t=0$ cylinder in KSSdS. Via the Killing flow
this neighborhood is isometric to a neighborhood of the $t=2 \mathrm{T}$ cylinder,
hence the hypersurfaces $t=0$ and $t=2 \mathrm{T}$ agree not only in
their induced first and second fundamental forms, but also in all higher derivatives
of the fundamental forms. The identification of the two hypersurfaces thus
results in a smooth manifold.
Note that, while the spacetime KSSdS[0] is time-symmetric, KSSdS[T] with $\mathrm{T}\neq 0$
is not; however,
KSSdS[T] and $\text{KSSdS}[-\mathrm{T}]$ differ by time orientation only.

The spacetimes KSSdS[T] are not the only cosmological spacetimes
which arise as quotient spaces from KSSdS. Firstly, we note that
it is not necessary to identify $t=0$ with a $t=\mathrm{const}$
cylinder in an adjoining copy of the region $0<r<r_{\mathrm{c}}$;
indeed, an arbitrary number of intervening copies between the
identified copies is possible. However, the arising spacetimes do
not exhibit different structures as far as the properties of
CMC-slicings are concerned. Secondly, we may identify, in the
black/white hole or in the future/past cosmological region, points
mapped to each other by a discrete subgroup of the action under
$\xi$. Thereby we obtain cosmological spacetimes that are
completely different from the class KSSdS[T] considered here.
These spacetimes will be treated in forthcoming work by one of us
(J.M.H.) \cite{NonCompPaper}.

In this paper, a \textit{slicing} denotes a smooth family of 
smooth (spacelike) hypersurfaces. A
parametrization of a slicing is a smooth map $\Psi: I\times \Sigma \rightarrow \text{spacetime}$,
where $\Sigma$ is a 3-manifold and $I \subseteq \mathbb{R}$, such that for all $\tau\in I$,
$\Psi(\tau,\cdot)$ is an embedding. We require that $\Psi(\tau, \Sigma)$ is a hypersurface of the
slicing for all $\tau\in I$, i.e.\ by the parametrization of the slicing the hypersurfaces are
represented as level sets $\tau =\mathrm{const}$.
Note that a slicing is a \textit{foliation} iff
the map $\Psi$ is a diffeomorphism onto its image.

The content of the paper is as follows. In Sec.~\ref{data} we
study spherically symmetric compact CMC-slices in terms of the
associated initial data sets. We find that these compact
CMC-initial data sets are parametrized by two constants $(K,C)$,
which lie in a bounded connected open subset $\mathscr{KC}_0$ of
$\mathbb{R}^2$. The constant $K$ is the mean curvature of the
slice, i.e.\ the trace of the second fundamental form. The
interpretation of the constant $C$ is less immediate (see the note
following~(\ref{KijinKC})). Suffice it to say that initial data
generated by $(K,C)$ is umbilical, i.e.\ the extrinsic curvature
is proportional to the three-metric, iff $C=0$. The parameter
space $\mathscr{KC}_0$ is investigated in some detail in
Appendix~\ref{KCspaceapp}. In Sec.~\ref{embeddings} we discuss the
embedding of the compact initial data sets into the cosmological
spacetimes KSSdS[T]: we prove that each compact CMC-initial data
set is embeddable as a Cauchy CMC-hypersurface into KSSdS[T] for a
particular value of $\mathrm{T}$ and that the embedding is unique
modulo the Killing flow. Secs.~\ref{slicings} and~\ref{foliations}
are concerned with the formulation and the proof of the
\textbf{main results} of the present paper.

One main theorem can be stated in an informal manner as follows:
Each spacetime KSSdS[T] contains a unique non-trivial slicing of
Cauchy CMC-hyper\-surfaces. (Here, a CMC-slicing is called
trivial, if both $K$ and $C$ are constant along the slicing, and
uniqueness is again understood modulo the Killing flow.) The first
step in proving this theorem is undertaken in Sec.~\ref{slicings},
where we show, using the implicit function theorem, that if the
spacetime KSSdS[T] contains a compact CMC-hypersurface, then it
evolves into a unique CMC-slicing. Crucial for this is the
analysis of a (linear) ordinary differential equation, whose
solution is interpreted as the lapse function of the slicing. Due
to the violation of the timelike convergence condition, this
analysis is quite involved and thus deferred to
Appendix~\ref{thelapseequation}. We find that the CMC-slicing in
KSSdS[T] can be represented by a curve in the parameter space
$\mathscr{KC}_0$, which in turn is given as the T-level set of a
function $\mathcal{T}(K,C)$ on $\mathscr{KC}_0$. Thus there exists
an interplay between a global but finite dimensional picture,
where CMC-slicings are viewed in terms of their representation in
$\mathscr{KC}_0$ on the one hand, and a local but essentially
infinite dimensional view of slicings in spacetime on the other
hand. Based on these ideas, in the second step in the proof of the
theorem, in Sec.~\ref{foliations}, we show that each spacetime
KSSdS[T], $\mathrm{T}$ arbitrary, contains a compact CMC-slicing
and that this slicing is unique.

In another theorem in Sec.~\ref{foliations} we describe the
asymptotic behavior of the CMC-slicings: Along each slicing $K$
tends to $\pm\sqrt{3 \Lambda}$ in the future resp.\ in the past,
$C$ tends to explicitly known values that depend on $\Lambda$ and
$M$. In spacetime the hypersurfaces of the slicing approach
$r=(1/\sqrt{\Lambda})\left[1-\sqrt{1-3\sqrt{\Lambda}M}\right]$ in
the black hole (resp.\ white hole), and $r=\infty$ in the future
(resp.\ past) cosmological region (see Fig. 8). An essential
ingredient in the proof of the theorem is an asymptotic analysis
of the function $\mathcal{T}$.

Finally, in Sec.~\ref{foliations} and in Appendix~\ref{foliapp}, we discuss whether the compact
CMC-slicings in the spacetimes KSSdS[T] are foliations. We prove that each slicing is a foliation
at least during some time of its evolution. Moreover, if $|\mathrm{T}|$ is sufficiently large, then
the slicing cannot be a foliation for all times. In addition, we provide solid numerical evidence
that there are values for $(\Lambda,M)$ such that, if $|\mathrm{T}|$ is small, the CMC-slicing is a
foliation everywhere.

In this work we can, and of course do, make use of the spherical symmetry of the spacetimes. The
quantities we are seeking are all "essentially explicit" in this sense: they are either given by
quadratures of algebraic functions, or solutions to algebraic equations, or level sets of functions
which are in turn given by quadratures, and combinations of the above.
Still only a small part of our task can be performed by a direct investigation of these
expressions. Rather we require a somewhat delicate interplay between the analysis of
the explicit quantities and the geometric analysis.

\section{CMC-data}
\label{data}

In this section we investigate spherically symmetric CMC-initial data sets.
We find that there exists a two-parameter
family of \textit{compact} CMC-data;
we analyze the parameter space, and we discuss the main geometric properties
of the CMC-data sets in dependence on the parameters.

Consider a three-dimensional Riemannian manifold $\Sigma \cong J \times S^2$
endowed with a spherically symmetric 3-metric
\begin{equation}\label{3-metric}
\threemetric_{i j}\, d x^i d x^j = d l^2 + r^2(l) (d\vartheta^2 + \sin^2\vartheta \,d\varphi^2)\:;
\end{equation}
the coordinates $\vartheta$, $\varphi$ are usual angular coordinates, $l$ is
a ``radial'' coordinate which takes values in $J$, which is
(an open interval of) $\mathbb{R}$, or $J \cong S^1$.
Let there be given a symmetric tensor $K_{i j}$, the second fundamental form,
such that the mean curvature
\begin{equation}
K= \threemetric_{i j} K^{i j} =\mathrm{const}\:.
\end{equation}
By (locally) solving, along the lines of~\cite{Beig/OMurchadha:1998}, the vacuum constraints with positive
cosmological constant $\Lambda$
we obtain that $(\Sigma, \threemetric_{i j}, K_{i j})$ is a
spherically symmetric CMC-initial data set, iff there exists a
constant $C$ such that
\begin{equation}\label{KijinKC}
K_{i j} \, d x^i d x^j = \left(\frac{K}{3} + \frac{2 C}{r(l)^3}\right) d l^2 +
\left(\frac{K}{3} - \frac{C}{r(l)^3}\right) \: r(l)^2 (d\vartheta^2 + \sin^2\vartheta \,d\varphi^2)\:.
\end{equation}
Note that $K_{i j}$ is of the form $K_{i j} = (K/3)\: \threemetric_{i j} + C \,L_{i j}^{\mathrm{TT}}$,
i.e.\ it is the sum of a constant trace 
plus $C$ times a tensor that coincides with the unique spherically symmetric 
transverse traceless tensor w.r.t.\ $\threemetric$.
The initial data set is umbilical, i.e.\ $K_{ij} \sim g_{ij}$, iff $C=0$.
The function $r(l)$ is required to satisfy
\begin{equation}\label{rprime2}
r^{\prime\:2} = 1 -\frac{2 M}{r} - \frac{\Lambda r^2}{3} + \left( \frac{K r}{3} - \frac{C}{r^2}\right)^2
=: D(r) \:,
\end{equation}
where the prime denotes the derivative w.r.t.\ $l$.
The constant $M$ in~(\ref{rprime2}) coincides with the mass $M$ appearing in~(\ref{KSSdSmetric}),
which is a consequence of the subsequent considerations.

An initial data set $(\Sigma, \threemetric_{i j}, K_{i j})$ is
said to admit the Killing initial data (or KID) $(\alpha, X^i)$,
if $(\alpha, X^i)$ lies in the kernel of the (overdetermined)
operator given by the adjoint of the linearized constraint
operator (see~\cite{Chrusciel/Delay:2003}, also~\cite{Beig/Chrusciel:1997}).
The KID-condition, in the presence
of $\Lambda$, is equivalent to
\begin{equation}
2 \alpha K_{i j} + 2 D_{(i} X_{j)} = 0 \:;
\end{equation}
together with
\begin{equation}
D_i D_j \alpha + {\mathcal L}_X K_{i j} - \alpha  ({}^3\!R_{i j} +
K K_{i j} - 2 K_i^{\:l} K_{j l} -\Lambda \threemetric_{i j}) = 0,
\end{equation}
which, provided that $\alpha \neq 0$, is in turn equivalent to the
statement that the "Killing development", i.e.\ the stationary
metric
\begin{equation}
g_{\mu\nu} d x^\mu d x^\nu = (- \alpha^2 + X^2) d\tau^2 + 2 X_i
d\tau d x^i + \threemetric_{i j} d x^i d x^j
\end{equation}
satisfies $G_{\mu\nu} + \Lambda g_{\mu\nu}=0$. The present
CMC-initial data set does in fact admit a KID, namely
$(\alpha_\xi,X_\xi^i)$ given by
\begin{equation}\label{KID}
\alpha_\xi = r^{\prime}(l) \quad,\qquad
X_\xi^i \partial_i = x_\xi(l) \frac{\partial}{\partial l} =
-\left( \frac{K r(l)}{3} - \frac{C}{r(l)^2}\right)\frac{\partial}{\partial l} \:\,.
\end{equation}
The development of
the initial data set thus results in a spacetime with Killing
vector $\xi^\mu$,
\begin{equation}\label{KVxi}
\xi^\mu = \alpha_\xi n^\mu + X_\xi^\mu \partial_\mu \quad,\qquad
\xi^\mu \xi_\mu = - \alpha_\xi^2 + X_\xi^2 = - \left(1 -\frac{2 M}{r} -\frac{\Lambda r^2}{3}\right)\:\,.
\end{equation}
Here, $n^\mu$ denotes the unit normal of $\Sigma$ in the spacetime,
$X_\xi^2 = X_\xi^i X^\xi_i$.
The KID is ``static'': from
\begin{equation}
D_{[i} \left( \frac{X^\xi_{j]}}{- \alpha_\xi^2 + X_\xi^2}\right) = 0
\end{equation}
it follows that $\xi^\mu$ is hypersurface orthogonal.
In the Killing development of the initial data, $\Sigma$ is given by $\tau=0$, and the metric reads
\begin{equation}
g_{\mu\nu} d x^\mu d x^\nu =
(- \alpha_\xi^2 + X_\xi^2) d\tau^2 + 2 X^\xi_i d\tau d x^i + \threemetric_{i j} d x^i d x^j\:.
\end{equation}
By using $r$ and introducing the coordinate $t$ through
\begin{equation}\label{coordemb}
d \tau = d t - \frac{X^\xi_i}{-\alpha_\xi^2 + X_\xi^2} \, d x^i =
d t - V^{-1} \left( \frac{K r(l)}{3} - \frac{C}{r(l)^2}\right) d l
\end{equation}
we recover the original Schwarzschild-de Sitter metric~(\ref{KSSdSmetric}),
\begin{equation}
g_{\mu\nu} d x^\mu d x^\nu =
- \left( 1-\frac{2 M}{r} -\frac{\Lambda r^2}{3} \right) d t^2 +
 \left( 1-\frac{2 M}{r} -\frac{\Lambda r^2}{3} \right)^{-1} d r^2 + r^2 d\Omega^2\:.
\end{equation}

We record for later use the identity
\begin{equation} \label{dt}
\nabla_\mu t = - V^{-1} \xi_\mu,
\end{equation}
which follows from the previous discussion.
As another particular consequence we conclude
that the constant $M$ introduced in~(\ref{rprime2}) is to be interpreted as the mass
$M$ appearing in the Schwarzschild-de Sitter metric.

In the following we consider the cosmological constant $\Lambda$ and the mass $M$ as given constants.
A CMC-initial data set $(\Sigma, \threemetric_{i j}, K_{i j})$
arises from the choice of a pair $(K,C)$, by which $D(r)$ and thus $r(l)$ is
defined (modulo an irrelevant translational freedom in $l$), cf.~(\ref{rprime2}).

\begin{figure}[htp]
    \psfrag{D}[cc][cc][0.8][0]{$D(r)$}
    \psfrag{r}[cc][cc][0.8][0]{$r$}
    \psfrag{a}[cc][cc][0.7][0]{$r_{-}$ }
    \psfrag{b}[cc][cc][0.7][0]{$r_{\min}$ }
    \psfrag{c}[cc][cc][0.7][0]{$r_{+}$}
    \psfrag{d}[cc][cc][0.7][0]{$r_{\max}$}
    \psfrag{q}[cc][cc][0.8][0]{$r^\prime$}
    \psfrag{u}[cc][cc][0.7][0]{$r_{\rm{one}}$}
    \centering
    \subfigure[]{\label{niceDelta}
      \includegraphics[width=0.3\textwidth]{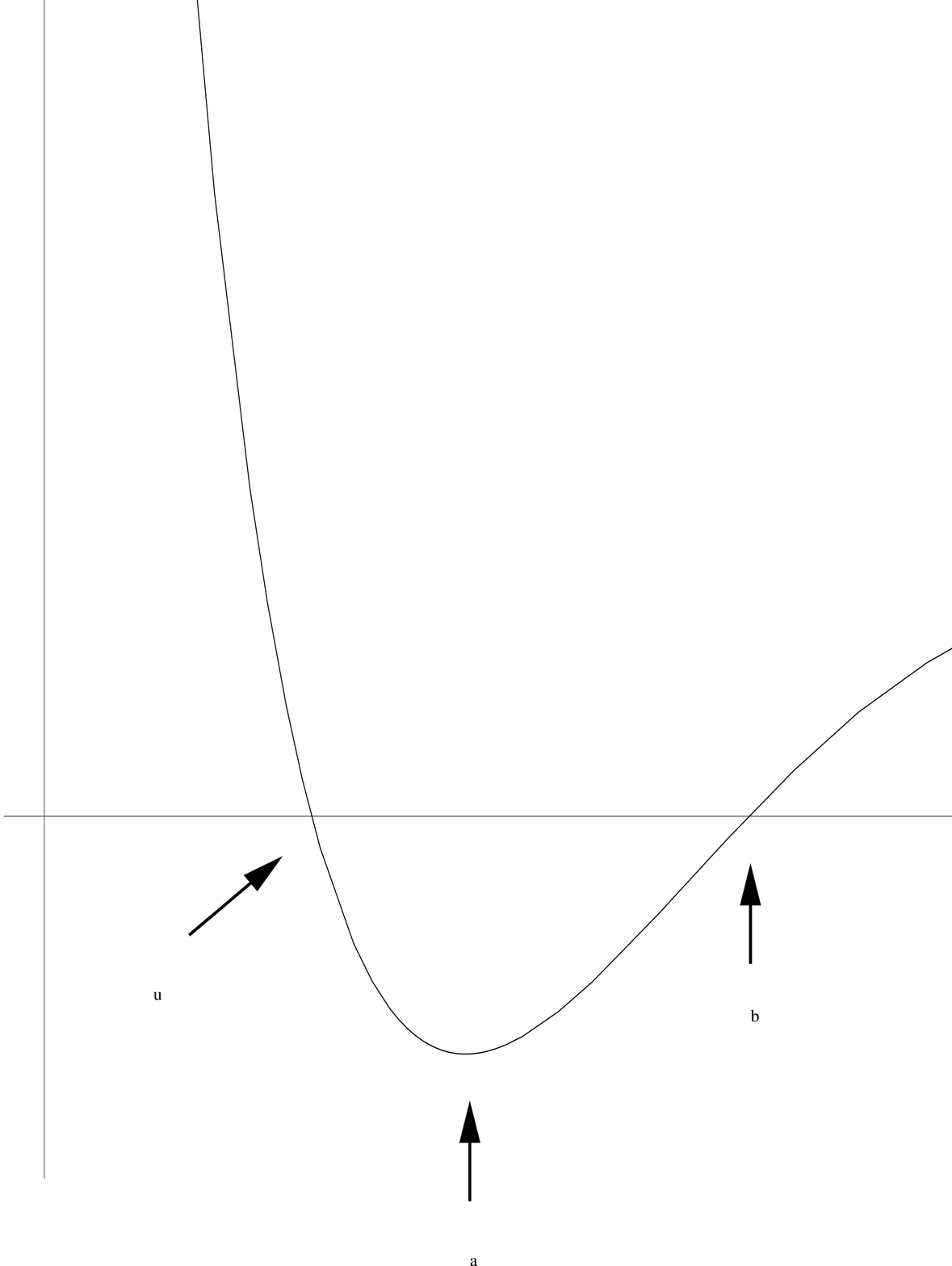}}\qquad\qquad
    \subfigure[]{\label{eggfig}
      \includegraphics[width=0.3\textwidth]{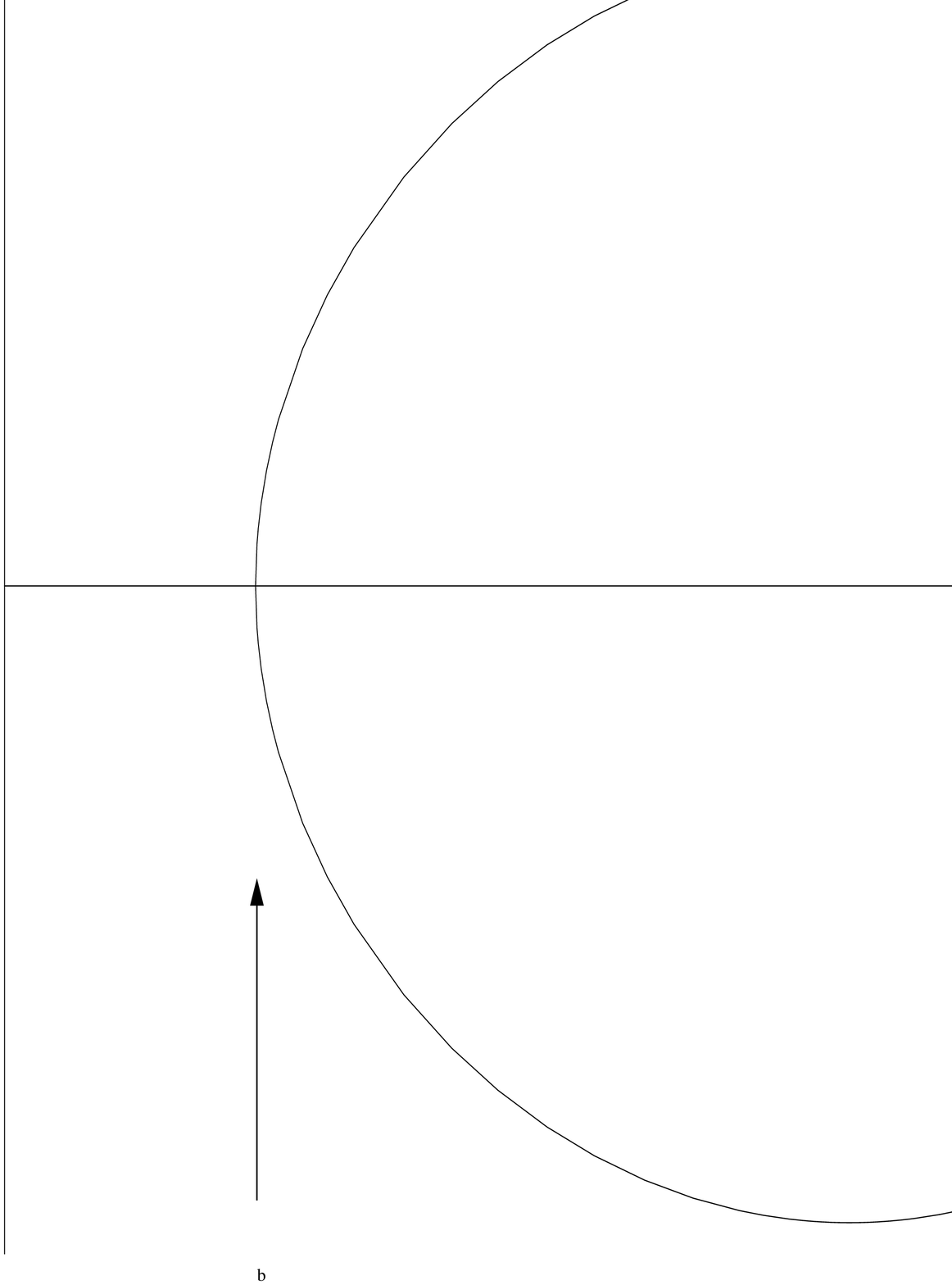}}
    \caption{{\small For $(K, C)\in \mathscr{KC}$,
        $D(r)$ is positive between two positive zeros, $r_{\min}$/$r_{\max}$, and
        $dD/d r$ does not vanish at $r_{\min}$/$r_{\max}$.
        Then $D(r)$ gives rise to a (smooth) closed curve in the $(r,r^\prime)$-plane,
        and $r(l)$ becomes a periodic function.}}
        \label{closedrrprime}
\end{figure}

\begin{definition}
We define the parameter space of \textit{compact CMC-initial data} by
\begin{equation}
\mathscr{KC} :=\left(\{ (K,C)\:|\: (K,C) \:\,\text{generates compact CMC-data} \:\,
(\Sigma, \threemetric_{i j}, K_{i j})\:\, \text{with}\:\, \Sigma \cong S^1\times S^2 \}\right)^{\circ}\:\,.
\end{equation}
\end{definition}

\begin{proposition}
The parameter space
$\mathscr{KC}$ is the disjoint union of three open connected domains,
\begin{equation}
\mathscr{KC} = \mathscr{KC}_0 \cup \mathscr{KC}_1 \cup \mathscr{KC}_{-1}\:.
\end{equation}
$\mathscr{KC}_0$ is the connection component of $(K,C) = (0,0)$,
it is invariant under the inversion.
\end{proposition}

\begin{figure}[htp]
    \psfrag{A}[cc][cc][0.8][10]{$\leftarrow\, C_t(K)\, \rightarrow$}
    \psfrag{B}[cc][cc][0.8][10]{$\leftarrow\, C_b(K)\,\rightarrow$}
    \psfrag{E}[cc][cc][0.7][-90]{$K=\sqrt{3 \Lambda}$}
    \psfrag{F}[cc][cc][0.7][90]{$K=-\sqrt{3\Lambda}$}
    \psfrag{1}[cc][cc][0.5][0]{$\sqrt{\Lambda}$}
    \psfrag{-1}[cc][cc][0.5][0]{$\sqrt{\Lambda}$}
    \psfrag{K}[cc][cc][0.7][0]{$K$}
    \psfrag{C}[cc][cc][0.7][0]{$C$}
    \psfrag{0.15}[rc][rc][0.5][0]{}
    \psfrag{-0.15}[rc][rc][0.5][0]{}
    \centering
    \includegraphics[width=0.5\textwidth,height=0.2\textheight]{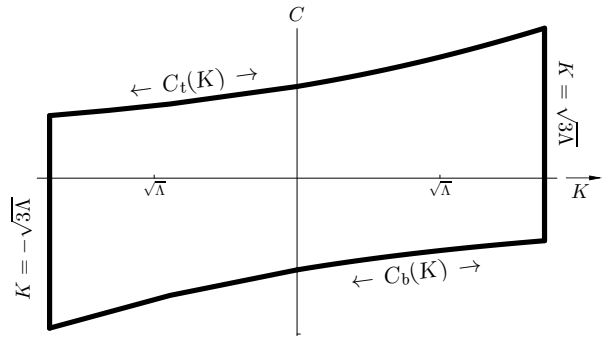}
    \caption{{\small $\mathscr{KC}_0$ for $\Lambda=1$, $M=1/4$; it is enclosed by $C_{b,t}(K)$
        and $K=\pm\sqrt{3\Lambda}$.}}
        \label{CKBer}
\end{figure}

For proofs we refer to App.~\ref{KCspaceapp}.
In this paper, we focus on compact CMC-initial data
generated by $(K,C)\in \mathscr{KC}_0$.
The space $\mathscr{KC}_0$ is depicted in Fig.~\ref{CKBer}; it
is enclosed
by the curves $C_t$ and $C_b$, and
the vertical straight lines $K=\pm \sqrt{3 \Lambda}$.
(In the following we refrain from making a distinction between the curves $C_{b,t}$
and the functions $C_t(K)$, $C_b(K) = -C_t(-K)$, that parametrize the curves.)
Some properties of the functions $C_{b,t}(K)$ are discussed in App.~\ref{KCspaceapp}.

If and only if
$(K,C)\in \mathscr{KC}$, the function $D(r)$
exhibits the form depicted in Fig.~\ref{niceDelta}:
in particular, $D(r)$ possesses two positive (simple) zeros,
$r_{\min}$ and $r_{\max}$, such that $D(r) > 0$ in the
interval $(r_{\min}, r_{\max})$.
Accordingly, when viewed over $r\in [r_{\min},r_{\max}]$,
$r^\prime = \pm \sqrt{D(r)}$
describes a closed curve, cf.~Fig.~\ref{eggfig}, so that
$r(l)$ becomes a periodic function
that oscillates between $r_{\min}$ and $r_{\max}$;
we denote the period by $2 L$,
\begin{equation}
L = \int\limits_{r_{\min}}^{r_{\max}} D^{-1/2}(r)\: d r\:.
\end{equation}
Without loss of generality we assume that
\begin{equation}\label{r0isrmin}
r(0) = r_{\min} \qquad\text{so that} \quad r(\pm L) = r_{\max} \:;
\end{equation}
it follows that $r(l)$ is an \textit{even periodic} function, which
is implicitly given through
\begin{equation}
l(r) = \pm \int_{r_{\min}}^r D^{-1/2}(\hat{r})\: d \hat{r}\:.
\end{equation}
By the natural identification of $l=-L$ and $l=L$, the domain of the
function $r(l)$ becomes $S^1$, so that the
CMC-initial data $(\Sigma, \threemetric_{i j}, K_{i j})$ is \textit{compact} with
$\Sigma \cong S^1 \times S^2$.

\begin{remark}
From Fig.~\ref{niceDelta} we see that a pair $(K,C)$ that generates compact CMC-data,
in general also gives rise to non-compact CMC-data, where $r(l)$ ranges in $(0,r_{\mathrm{one}}]$.
A CMC-data set of this type is embeddable in KSSdS as a hypersurface that runs into the
singularity. A detailed classification of all possible types of CMC-data sets 
and their embeddings in KSSdS[T] and other cosmological KSSdS-spacetimes
will be presented in~\cite{NonCompPaper} 
\end{remark}

\begin{figure}[htp]
    \psfrag{D}[cc][cc][0.6][0]{$D(r)$}
    \psfrag{r}[cc][cc][0.6][0]{$r$}
    \centering
    \subfigure[$(K,C)\in C_{b,t}$]{\label{Deltaclass4}
      \includegraphics[width=0.25\textwidth]{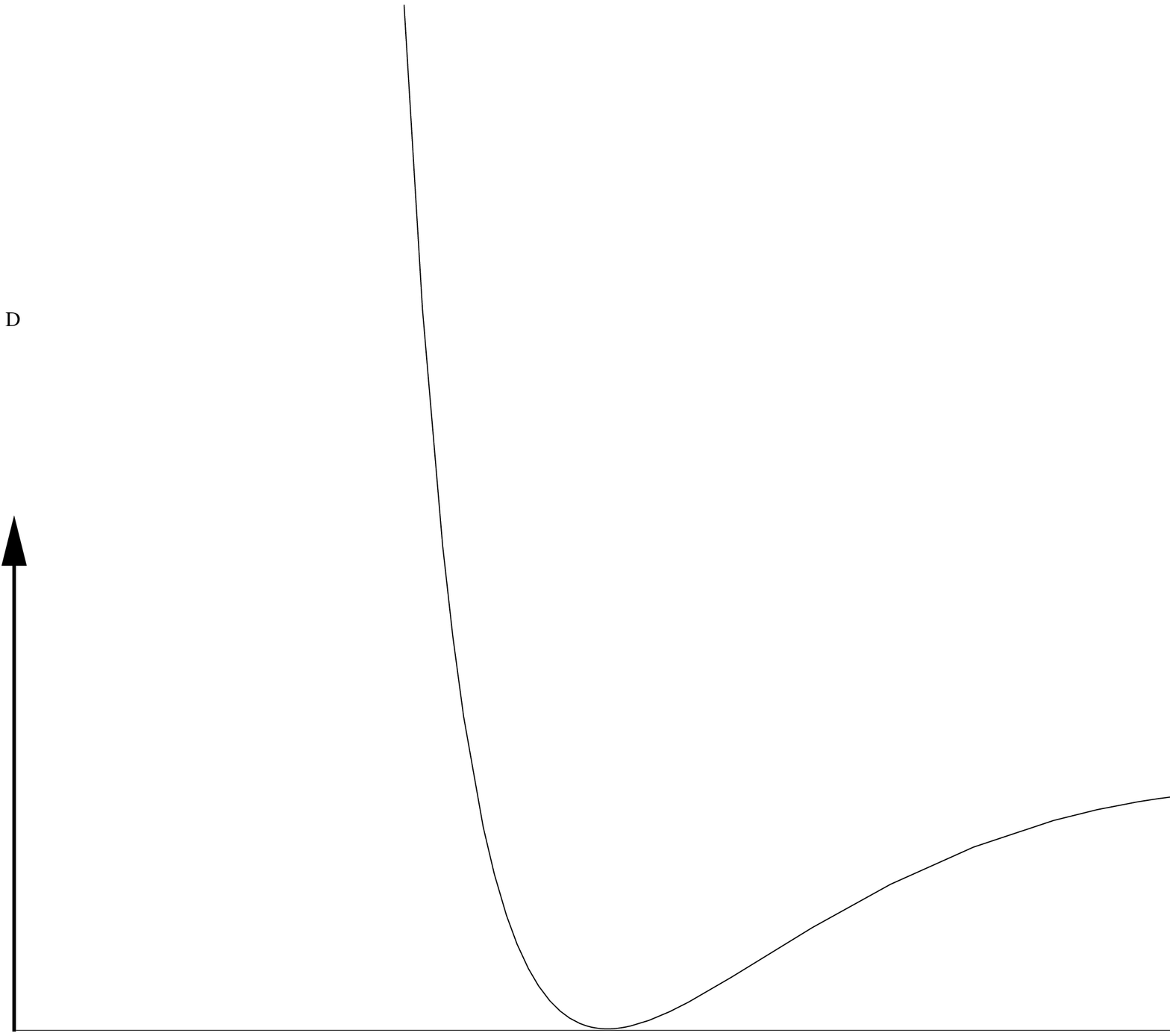}}\qquad
    \subfigure[$K=\pm \sqrt{3\Lambda}$]{\label{Deltaclass6}
      \includegraphics[width=0.25\textwidth]{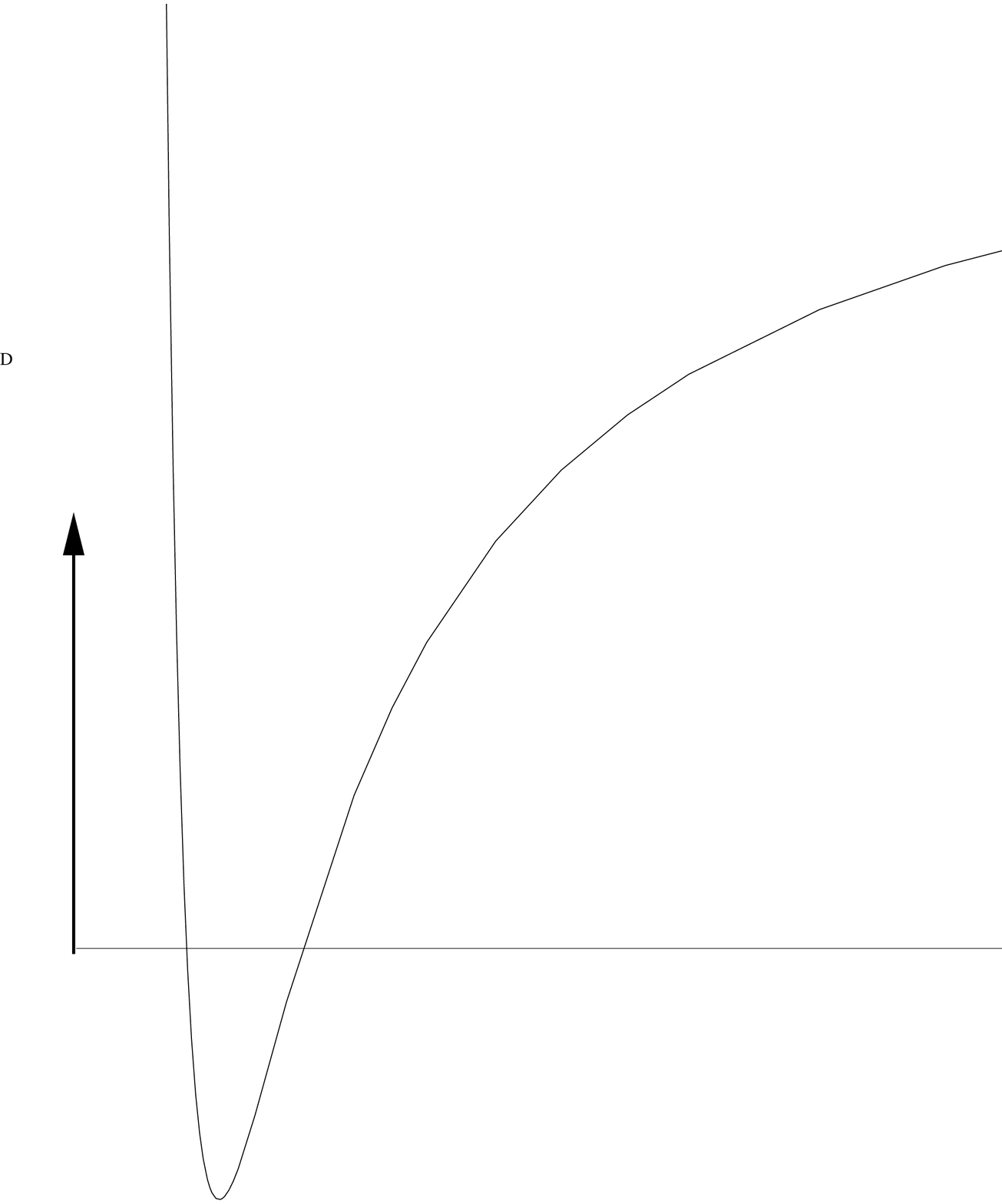}}
    \caption{{\small The profiles of $D(r)$ on $\partial(\mathscr{KC}_0)$.}}
        \label{Deltaborderline}
\end{figure}

When $(K,C)\in \partial(\mathscr{KC})$, the
profile of the function $D(r)$ is a borderline case
of the profile~\ref{niceDelta}.
On the boundaries $C_{b,t}$, the function has the form~\ref{Deltaclass4}, i.e.\
the zero $r_{\min}$ is a double zero.
The solution of $r^\prime = \pm \sqrt{D(r)}$
that is relevant for our purposes is
$r(l)\equiv \mathrm{const} = r_{\min}$: by identifying
$l=l_0$ with $l=l_1$ for any $l_0$, $l_1$, we obtain
a compact CMC-initial data set $(\Sigma,\threemetric_{i j}, K_{i j})$,
$\Sigma \cong S^1 \times S^2$.
On the boundaries $K=\pm \sqrt{3 \Lambda}$, $D(r)$ possesses
a simple zero $r_{\min}$, but ``$r_{\max} = \infty$'', i.e.\
$D \rightarrow 1$ ($r\rightarrow \infty$), see Fig.~\ref{Deltaclass6}.
In this case, no compact CMC-data is generated.
The features of $D(r)$ described above occur simultaneously on
the intersections of the boundaries.

For every compact CMC-initial data set
$(S^1\times S^2, \threemetric_{i j}, K_{i j})$ associated with $(K,C)\in\mathscr{KC}_0$,
the radius $r$ of the spheres of symmetry varies
between values $r_{\min}$ and $r_{\max}$.
We conclude the section by discussing $r_{\min}(K,C)$ and $r_{\max}(K,C)$
in dependence on $(K,C)\in\overline{\mathscr{KC}_0}$.

Most importantly, we note that
$r_{\min}\leq r_{\mathrm{b}}$ and that $r_{\min} = r_{\mathrm{b}}$ iff $(K,C)$ is such
that $C= (r_{\mathrm{b}}^3/3) K$.
This ``line of maximal $r_{\min}$'' divides
$\overline{\mathscr{KC}_0}$ in two regions,
an upper (left) half, characterized by $[ K r_{\min}/3 - C/r_{\min}^2 ] < 0$,
and a lower (right) half with $[ K r_{\min}/3 - C/r_{\min}^2 ] > 0$,
see Fig.~\ref{divrmin}.
The function $r_{\min}$ assumes its global minimum on $\overline{\mathscr{KC}_0}$
at the point $(K,C) = (\sqrt{3 \Lambda}, C_b(\sqrt{3 \Lambda}))$
and at the reflected point; the minimal value is given by~(\ref{rminextremal}).
$r_{\min}$ decreases/increases along the boundaries
of $\mathscr{KC}_0$ as depicted in Fig.~\ref{divrmin}.
$r_{\max}\geq r_{\mathrm{c}}$ and $r_{\max} = r_{\mathrm{c}}$ iff
$(K,C)$ is such
that $C= (r_{\mathrm{c}}^3/3) K$.
This ``line of minimal $r_{\max}$'' divides
$\overline{\mathscr{KC}_0}$ in two regions, an (upper) left half, where $[ K r_{\max}/3 - C/r_{\max}^2 ] < 0$
and a (lower) right half, where $[ K r_{\max}/3 - C/r_{\max}^2 ] > 0$, see Fig.~\ref{divrmax}.
$r_{\max}$ is unbounded on $\mathscr{KC}_0$, it diverges like $\sqrt{3}/\sqrt{\barL}$
as $K \rightarrow\pm\sqrt{3\Lambda}$; $\barL = \Lambda- K^2/3$.
Both $r_{\min}$ and $r_{\max}$
are constant along straight lines in $\mathscr{KC}_0$, see Fig.~\ref{linesofrminmax}.

$r_{\min}$ can be given explicitly,
\begin{equation}\label{rminextremal}
r_{\min} =  \frac{1}{\sqrt{\Lambda}} \left(\pm 1 \mp  \sqrt{1 \mp 3 \sqrt{\Lambda} M}\right)\:,
\qquad
\text{for $K= \sqrt{3 \Lambda}$, $C= C_{b,t}(\sqrt{3\Lambda})$}.
\end{equation}
Here, the upper sign applies to $C_{t}$, the lower to $C_b$.

\section{Embeddings}
\label{embeddings}

In this section we investigate embeddings of compact CMC-initial data
$(\Sigma, \threemetric_{i j}, K_{i j})$ as CMC-hypersurfaces in
Schwarzschild-de Sitter spacetime.

Let $(K,C) \in \mathscr{KC}_0$; the pair generates compact CMC-data
$(\Sigma \cong S^1\times S^2, \threemetric_{i j}, K_{i j})$.
We consider the universal covering
$(\tilde{\Sigma} \cong \mathbb{R}\times S^2,\threemetric_{i j}, K_{i j})$ of
the data by regarding $r(l)$ as a periodic even function on $\mathbb{R}$.
The initial data set $(\tilde{\Sigma},\threemetric_{i j}, K_{i j})$ is
embedded as a CMC-hypersurface $\tilde{\mathcal{S}}$ in KSSdS via
\begin{equation}\label{tofl}
r = r(l) \qquad\qquad t = t(l) := \int\limits_0^l V^{-1}(r(\hat{l})) \:
\left(\frac{K r(\hat{l})}{3} -\frac{C}{r(\hat{l})^2}\right)  d \hat{l} \:,
\end{equation}
which follows from~(\ref{coordemb}).
The integral is understood in the principal value sense, so that $t(l)$ is well-defined
for all $l$ with $V(r(l))\neq 0$; this suffices to uniquely define
the embedded hypersurface in KSSdS.
(Alternatively, the embedding can be given in Kruskal type coordinates.)
If $(K,C)$ is such that $r_{\min} = r_{\mathrm{b}}$, then $V^{-1}$ is singular at $l=0$, but also
$(K r/3-C/r^2) = 0$ when $l\rightarrow 0$, cf.~Fig.~\ref{divrmin}.
Using de l'Hospital's rule we see
$\lim_{l\rightarrow 0} V^{-1}(r(l)) [K r(l)/3 -C/r^2(l)] = -V^{\prime\:-1}(r_{\mathrm{b}}) (K/3)$,
i.e.\ the integrand is bounded as $l\rightarrow 0$.
Similarly, the integrand is always bounded as $l\rightarrow L$.

The CMC-hypersurface $\tilde{\mathcal{S}}$ defined by~(\ref{tofl}) is not the unique
embedding of $\tilde{\Sigma}$: integrating~(\ref{coordemb})
leaves the freedom of choosing an integration constant so that
$t$ is replaced by $t - \mathrm{const}$ in~(\ref{tofl}).
The one-parameter freedom is associated with the Killing flow,
hence, modulo the Killing isometries
the embedding is indeed unique.

\begin{figure}[htp]
        \psfrag{r0}[cc][cc][0.8][0]{$r=0$}
    \psfrag{ri}[cc][cc][0.8][0]{$r=\infty$}
    \psfrag{a}[cc][cc][0.5][37]{$r=r_{\min}$}
    \psfrag{t0}[cc][cc][0.6][-90]{$t=0$}
    \psfrag{t1}[cc][cc][0.6][-78]{$t=2 \mathcal{T}$}
    \psfrag{id}[cc][cc][0.7][0]{identification}
    \centering
    \includegraphics[width=0.95\textwidth]{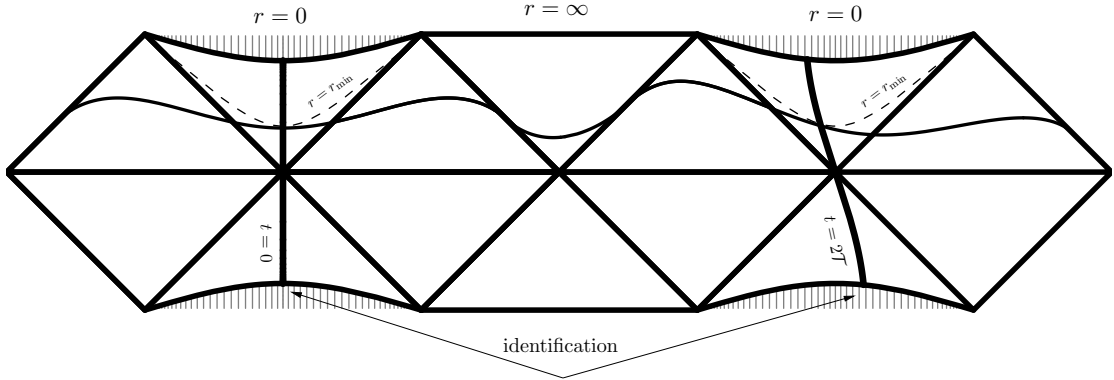}
    \caption{{\small The embedding of the compact CMC-initial data with $K=1$ and $C=0.1$
        in KSSdS[T] (with $\Lambda=1$, $M=1/4$).
        The hypersurface is \textit{not} null at the horizon $r_{\mathrm{c}}$: for the figure we have used
        the coordinate system of the region $0<r < r_{\mathrm{c}}$ and extended it to $0<r<\infty$ to obtain global
        coordinates; however, these coordinates break down at $r=r_{\mathrm{c}}$.}}
        \label{CMCCauchy}
\end{figure}

Let $(K,C) \in \mathscr{KC}_0$. Then $r(l)$ oscillates between $r_{\min} \leq r_{\mathrm{b}}$
and $r_{\max} \geq r_{\mathrm{c}}$, therefore $\tilde{\mathcal{S}}$ is a \textit{Cauchy hypersurface}
in KSSdS, see~Fig.~\ref{CMCCauchy}.

The (future pointing) unit normal of $\tilde{\mathcal{S}}$ is given by
\begin{equation}\label{unitnor}
n^\mu \partial_\mu = r^\prime V^{-1}\, \frac{\partial}{\partial t} +
\,\left(\frac{K r}{3} -\frac{C}{r^2}\right)\:\frac{\partial}{\partial r} \:;
\end{equation}
it is straightforward to check $\nabla_\mu n^\mu = K$. The
CMC-hypersurface $\tilde{\mathcal{S}}$ passes through the white 
hole region [black hole region] if $(K r_{\min}/3 -C/r_{\min}^2) >0$ [$(K r_{\min}/3 -C/r_{\min}^2) <0$]; when $(K
r_{\min}/3 -C/r_{\min}^2) = 0$, $r_{\min} =r_{\mathrm{b}}$ and the
hypersurface passes through the bifurcation sphere; compare with
Fig.~\ref{divrmin}. Similarly, the hypersurface passes through the
future [past] cosmological region if $(K r_{\max}/3 -C/r_{\max}^2)
>0$ [$K r_{\max}/3 -C/r_{\max}^2 <0$], cf.~Fig.~\ref{divrmax}.
These relations guarantee that $\tilde{\mathcal{S}}$ cannot
oscillate between the black hole and the white hole (or the future
and the past cosmological regions): either $\tilde{\mathcal{S}}$
passes through the black hole regions or through the white hole
regions for all $l = 2 n L$, $n\in\mathbb{N}$, where $r(l) =
r_{\min}$. However, it will occur that $\tilde{\mathcal{S}}$ runs
through the black hole region and through the past cosmological
horizon (or vice versa). This can be seen from the combination of
Figs.~\ref{divrmin} and~\ref{divrmax}.

The embedding of a compact CMC-initial data set $(\Sigma, \threemetric_{i j}, K_{i j})$,
$\Sigma\cong S^1\times S^2$, as a CMC-hypersurface $\mathcal{S}$ in KSSdS[T] is more delicate.

Consider~(\ref{tofl}) and make the following
\begin{definition}
\begin{equation}\label{Tdef}
\mathcal{T} :=  t(L) =
\int\limits_0^L V^{-1}(r(l)) \:\left(\frac{K r(l)}{3} -\frac{C}{r(l)^2}\right)  d l =
\int\limits_{r_{\min}}^{r_{\max}} V^{-1} D^{-1/2}(r)
\:\left(\frac{K r}{3} -\frac{C}{r^2}\right)  d r\:.
\end{equation}
\end{definition}
Like $r_{\min}$ and $r_{\max}$, the function $\mathcal{T}$ depends on the pair $(K,C)\in\mathscr{KC}_0$
characterizing the CMC-data.
For $(K,C)\in C_{b,t}$, $\mathcal{T}$ is not canonically defined;
each choice of $l_0$, $l_1$, which leads to compact data, cf.~Sec.~\ref{data},
is associated with a different value of $\mathcal{T}$.

\begin{proposition}\label{embedprop}
Let $(K,C) \in \mathscr{KC}_0$. Then the associated compact CMC-initial
data set $(\Sigma\cong S^1\times S^2, \threemetric_{i j}, K_{i j})$
is embeddable as a CMC-hypersurface $\mathcal{S}$ in KSSdS[T]
if and only if $T = \mathcal{T}$.
The hypersurface is a Cauchy hypersurface, and the
embedding is unique modulo the Killing flow.
\end{proposition}

\proof
Consider the universal covering $\tilde{\Sigma}$ of the data and the embedded
hypersurface $\tilde{\mathcal{S}}$ as given by~(\ref{tofl}).
The hypersurface is invariant under
the two discrete isometries $t \mapsto -t$ and
$2 \mathcal{T} + t \mapsto 2 \mathcal{T} -t$ of the spacetime; this is because
$r(l)$ is even about $l=0$ and even about $l=2 L$
according to its periodicity.
Since the unit normal vectors of $\tilde{\mathcal{S}}$ at $l=0$ and $l=2 L$ must
be invariant under the respective isometries as well, they must be tangential
to the fixed point sets $t=0$ and $t=2 \mathcal{T}$ of the isometries: we conclude
that $n^\mu \partial_\mu \propto \partial_t$ at $l=0$ and $l=2 L$.
The identification of $t=0$ with $t = 2 \mathcal{T}$ in KSSdS
thus results in a CMC-hypersurface $\mathcal{S}$ in KSSdS[$\mathcal{T}$]
which is isometric to $\Sigma$ by construction.
The proof of the remaining claims is trivial.
\proofend

The argument used in the proof
will re-appear in the proof of Theorem~\ref{localslicing}; alternatively,
we could have used~(\ref{unitnor}).

\begin{remark}
Consider (the universal covering of) the CMC-initial data generated
by $(K,C) \in C_{b,t} \subset \partial(\mathscr{KC})$.
Since the data is characterized by
a function $r(l)\equiv \mathrm{const}$,
the embedding is a $r=\mathrm{const}$ hypersurface, which is
contained in the black/white hole region of KSSdS[T] (for arbitrary $\mathrm{T}$).
The compact interpretation of the data set, which is obtained by an identification
of $l=l_0$ with $l=l_1$ for any $l_0$, $l_1$,
is not naturally embeddable in KSSdS[T].
\end{remark}

\section{Slicings}
\label{slicings}

Proposition~\ref{embedprop} states that there exist spacetimes KSSdS[T] containing a compact
CMC-hypersurface. The aim of this section is to
investigate whether these spacetimes contain
\textit{CMC-slicings}, i.e.\ smooth families of CMC-hypersurfaces.

Any CMC-hypersurface can be
evolved into a CMC-slicing by the Killing flow; however, this CMC-slicing is
trivial in the sense that $K$ and $C$ are constant along the slicing.
The existence of non-trivial slicings is shown in the following theorem:
any compact CMC-hypersurface of KSSdS[T] evolves into a compact CMC-slicing,
along which the mean curvature $K$ is monotonic;
furthermore, this
slicing is necessarily unique modulo the Killing flow.

\begin{theorem}\label{localslicing}
Consider a spacetime KSSdS[T] that contains a compact CMC-hyper\-surface $\mathcal{S}$.
Then there exists a unique local slicing of KSSdS[T] by hypersurfaces $\mathcal{S}_\tau$,
$\tau \in (-\bar{\tau}, \bar{\tau})$, such that
\begin{enumerate}
\item\label{S0} $\mathcal{S}_0 =\mathcal{S}$,
\item\label{Staucomp} $\mathcal{S}_\tau$ is a compact CMC-hypersurface for all $\tau \in (-\bar{\tau}, \bar{\tau})$,
\item\label{Srefl}$\mathcal{S}_\tau$ is reflection symmetric for all $\tau$.
\setcounter{store}{\value{enumi}}
\end{enumerate}
Along the slicing $\mathcal{S}_\tau$ the mean curvature is a strictly
monotonic function, i.e.\
\begin{enumerate}
\setcounter{enumi}{\value{store}}
\item\label{Kmono} $K(\tau)$ is strictly monotonic.
\end{enumerate}
Furthermore, every slicing
$\mathcal{S}_\tau^\prime$ that satisfies~(\ref{S0}) and~(\ref{Staucomp}) arises from
$\mathcal{S}_\tau$ by combining the flow of $\mathcal{S}_\tau$ with an appropriate admixture of the
Killing flow.
\end{theorem}

\begin{remark}
The requirement~(\ref{Srefl}) is a convenient way of fixing a representative
within the equivalence class of slicings that satisfy~(\ref{S0}) and~(\ref{Staucomp}).
The compact CMC-hypersurface
$\mathcal{S}$ contains a totally geodesic 2-sphere at $r=r_{\min}$,
which evolves into a cylinder represented by $\tau\mapsto r_{\min}(\tau)$.
The slicing $\mathcal{S}_\tau$ is reflection symmetric if
the cylinder is totally geodesic, i.e.\ coincides with $t=\mathrm{const}$;
without loss of generality we always assume reflection symmetry w.r.t.\ $t=0$.
\end{remark}

\proof
The CMC-hypersurface $\mathcal{S}$ is characterized by
$(K_{\mathcal{S}}, C_{\mathcal{S}}) \in \mathscr{KC}_0$;
the induced metric is $d l^2 + r_{\mathcal{S}}^{\,2}(l) \:d\Omega^2$.
$\mathcal{S}$ can be represented by~(\ref{tofl}), it is invariant under the reflection
$t \mapsto -t$.
We introduce Gaus\nolinebreak sian coordinates in a neighborhood of $\mathcal{S}$:
the metric reads
\begin{equation}\label{metricGauss}
g_{\mu\nu} d x^\mu d x^\nu = -d\sigma^2 +s^2(\sigma,\lambda) d \lambda^2 + r^2(\sigma,\lambda) d\Omega^2\:;
\end{equation}
$\sigma \in (-\bar{\sigma}, \bar{\sigma})$.
The hypersurface $\mathcal{S}$ is represented by
$\sigma = 0$; $\lambda =l$ on $\mathcal{S}$. We obtain by using the identity
$\partial_\sigma g_{ij} = 2 K_{ij}$ and~(\ref{KijinKC}),
\begin{equation}\label{lamini}
s(0,\lambda) =1, \quad  r(0,\lambda) = r_{\mathcal{S}}(\lambda),\quad (\partial_\sigma{s})(0,\lambda) =
\frac{K_{\mathcal{S}}}{3}+\frac{2 C_{\mathcal{S}}}{r^3_{\mathcal{S}}(\lambda)},\quad
(\partial_\sigma{r})(0,\lambda) = \frac{K_{\mathcal{S}} r_{\mathcal{S}}(\lambda)}{3}-
\frac{C_{\mathcal{S}}}{r_{\mathcal{S}}^{\,2}(\lambda)}\:.
\end{equation}
We choose $\bar{\sigma}$ so that $\inf_{|\sigma|<\bar{\sigma}} s(\sigma,\lambda)\geq\mathrm{const}$ and
$\inf_{|\sigma|<\bar{\sigma}} r(\sigma,\lambda)\geq \mathrm{const}$ $\forall \lambda$.
The normal vector of $\mathcal{S}$ at
$\lambda=0$ is tangential to the cylinder $t=0$, since $\mathcal{S}$ is invariant under the
discrete isometry $t \mapsto -t$. The cylinder $t=0$, being the fixed point set of a discrete
isometry, is totally geodesic, therefore the normal
geodesics passing through $\lambda=0$ lie on $t=0$, hence the hypersurfaces $\lambda=0$ and $t = 0$
coincide; analogously, $\lambda=L_{\mathcal{S}}$ corresponds to $t = \mathcal{T}_{\mathcal{S}}\: (=
\mathrm{T})$. From this fact that the Gaussian coordinates are adapted to the discrete symmetry, it
thus follows that $s(\sigma, \lambda)$ and $r(\sigma,\lambda)$ are even functions in $\lambda$ for
all $\sigma$; analogously, the functions are even about $\lambda=L_{\mathcal{S}}$.

In a neighborhood of $\mathcal{S}$ in KSSdS[T], a compact hypersurface can be described by the equations
\begin{equation}\label{sighy}
\sigma = \varphi(l)\:, \qquad\lambda = l\:, \qquad\qquad \text{where} \quad \varphi: S^1 \rightarrow
(-\bar{\sigma}, \bar{\sigma})\:.
\end{equation}
We define the mean curvature operator $\mathcal{K}$,
\begin{equation}\label{meancurvoper}
\mathcal{K}[\varphi]:= \frac{1}{\sqrt{s^2 - \varphi^{\prime\:2}}} \, \left( \partial_\sigma{s} +
\frac{2 s \partial_\sigma{r}}{r} \right) - \frac{\varphi^\prime}{(s^2 -
\varphi^{\prime\:2})^{3/2}}+ \frac{2 r^\prime}{r s} \frac{\varphi^\prime}{\sqrt{s^2 -
\varphi^{\prime\:2}}} + \frac{1}{s} \, \left(\frac{\varphi^\prime}{\sqrt{s^2 -
\varphi^{\prime\:2}}}\right)' \:;
\end{equation}
for $\varphi$, the prime denotes a derivative w.r.t.~$l$,
for $s$ and $r$, a derivative w.r.t.~the second argument;
the suppressed arguments of the functions $s$, $r$, and its derivatives, are
$(\varphi(l),l)$. For a given $\varphi$, $\mathcal{K}[\varphi]$ is a function $S^1\rightarrow
\mathbb{R}$ describing the mean curvature of the hypersurface $\sigma - \varphi(\lambda) = 0$.
The hypersurface $\mathcal{S}$ is represented by $\varphi=0$; from~(\ref{lamini}) we obtain
$\mathcal{K}[0] = K_{\mathcal{S}}$. In order to show the claims of the theorem we solve the
prescribed mean curvature equation $\mathcal{K}[\varphi] = K \:(\equiv \mathrm{const})$ in a
neighborhood of $\mathcal{S}$.

Consider the Sobolev space $H^2(S^1) = W^{2,2}(S^1)$.
The Sobolev inequalities ensure that
$\psi \in H^2(S^1)$ implies $\psi \in \mathcal{C}^{1,1/2}(S^1)$, where
$\mathcal{C}^{1,1/2}$ denotes the relevant H\"older space;
we have the estimate
$\|\psi\|_{\mathcal{C}^{1,1/2}} \leq \mathrm{const}\, \|\psi\|_{H^2}$ for all $\psi$.
Since $\|\psi\|_{L^{\infty}} + \|\psi^\prime\|_{L^\infty} \leq \|\psi\|_{\mathcal{C}^{1,1/2}}$,
the set
${}^\delta\!H^2(S^1) :=
\{\psi \in H^2(S^1)\:|\: \|\psi\|_{L^{\infty}} + \|\psi^\prime\|_{L^\infty} < \delta = \mathrm{const}\}$
is a neighborhood of the origin in $H^2(S^1)$.
Let $H^k_{\mathrm{even}}(S^1)$ denote the space of even functions $\psi\in H^k(S^1)$;
as a closed subspace of $H^k(S^1)$ it is a Banach space. Finally define
${}^\delta\!H^2_{\mathrm{even}}(S^1) := H^2_{\mathrm{even}}(S^1) \cap {}^\delta\!H^2(S^1)$.
We choose $\delta$ sufficiently small,
so that $\|\psi\|_{L^{\infty}} < \bar{\sigma}$ and
$\|s^2 - \psi^{\prime\:2}\|_{L^{\infty}} \geq \mathrm{const} > 0$.
The quasilinear operator $\mathcal{K}$, viewed as a map
\begin{equation}
\mathcal{K}: {}^\delta\!H^2_{\mathrm{even}}(S^1) \rightarrow H^0_{\mathrm{even}}(S^1)\:.
\end{equation}
is $C^1$-differentiable.
The argument is standard and can be inferred, e.g.\ from Thm.~4.1 in~\cite{Valent:1987}.
The (Fr\'echet) derivative of $\mathcal{K}$
at a point $\varphi$ is a linear operator,
\begin{equation}
\mathcal{K}^\prime[\varphi]: H^2_{\mathrm{even}}(S^1) \rightarrow H^0_{\mathrm{even}}(S^1)\:,
\end{equation}
which, for $\varphi$ regular enough, is given by
\begin{equation}\label{mathcalKprime}
\mathcal{K}^\prime[\varphi] (\dot{\varphi}) = \Big(\Delta_{(\threemetric_{\varphi})}
+ \Lambda - K_{\varphi\,i j}\:K_\varphi^{i j}\Big)
\left(\frac{s}{\sqrt{s^2-\varphi^{\prime\:2}}} \:\dot{\varphi}\right)\:\,
\end{equation}
with $\dot{\varphi} \in  H^2_{\mathrm{even}}(S^1)$.
The expressions $\threemetric_{\varphi}$ and $K_\varphi$ are the
metric and extrinsic curvature of the surface $\sigma - \varphi(\lambda)=0$; the Laplacian is the one associated
with $\threemetric_{\varphi}$. It is convenient to obtain Eq.~(\ref{mathcalKprime}) through
geometric arguments, cf.~the remark following Corollary~\ref{lapsecorr}.

Since $\varphi\equiv 0$ represents the CMC-hypersurface $\mathcal{S}$
we obtain
\begin{equation}\label{mathcalKprime0}
\mathcal{K}^\prime[0] = \Delta + a \qquad \text{where}\quad
a(l) = \Lambda - \frac{K_{\mathcal{S}}^2}{3} - \frac{6 C_{\mathcal{S}}^2}{r^6(l)}\:.
\end{equation}
(Note that there do not exist pairs $(K_{\mathcal{S}},C_{\mathcal{S}})\in \mathscr{KC}_0$ such
that $a(l) \leq 0$ $\forall l$.)
The operator
$\mathcal{K}^\prime[0]$ is elliptic, the Fredholm alternative holds. In Lemma~\ref{isomlemma} we
show that $\mathrm{ker}\, \mathcal{K}^\prime[0]$ is trivial, i.e.\ the homogeneous equation $\Delta
\alpha + a \alpha =0$ admits only the trivial solution $\alpha=0$. It follows that
$\mathcal{K}^\prime[0]$ is an isomorphism, and we are able to apply the inverse function theorem,
see, e.g.~\cite{Renardy/Rogers:2004}: there exists an open neighborhood $V$ of
$\mathcal{K}[0]$ in
$H^0_{\mathrm{even}}(S^1)$ and a unique continuously differentiable mapping $\mathcal{K}^{-1}:
V\rightarrow {}^\delta\!H^2_{\mathrm{even}}(S^1)$ with the property that
$\mathcal{K}[\mathcal{K}^{-1}(\kappa)] = \kappa$ for all $\kappa\in V$.

Let $K$ be a smooth real function of $\tau\in\mathbb{R}$ which
is strictly monotonic and satisfies
$K(0) = K_{\mathcal{S}}$; let
$(-\bar{\tau},\bar{\tau})$ be an interval such that
$K(\tau)\in V$ for all $\tau \in (-\bar{\tau},\bar{\tau})$;
note that $K(\tau)$ is interpreted as a constant function
for each $\tau$. Then
\begin{equation}\label{thevarphitaus}
\varphi_\tau:=\mathcal{K}^{-1}(K(\tau))
\end{equation}
uniquely defines a family
of ${}^\delta\!H^2_{\mathrm{even}}(S^1)$-functions $\{\varphi_\tau\:|\:\tau \in (-\bar{\tau},\bar{\tau})\}$, where
$\varphi_{\tau_1} \neq \varphi_{\tau_2}$ for $\tau_1\neq \tau_2$.
For each $\tau$,
$\varphi_\tau$ is smooth, by elliptic regularity.
Moreover, the mapping $\tau\rightarrow\varphi_\tau$
is continuously differentiable by construction.
Therefore $\{\varphi_\tau\}$
defines a unique local slicing of KSSdS[T]
by compact CMC-hypersurfaces
\begin{equation}\label{theSs}
\mathcal{S}_\tau:= \{ (\sigma, \lambda, \Omega) \in \mathrm{KSSdS[T]} \:|\: \sigma =
\varphi_\tau(l), \lambda = l \}\:,
\end{equation}
$\tau \in (-\bar{\tau}, \bar{\tau})$;
$\mathcal{S}_{\tau_1} \neq \mathcal{S}_{\tau_2}$ for $\tau_1\neq \tau_2$;
since $\varphi_0 \equiv 0$,
$\mathcal{S}_0 =\mathcal{S}$.
By construction, since $\varphi_\tau$ is even for each $\tau$,
and since $\lambda=0$ coincides with $t=0$,
$\mathcal{S}_\tau$ is invariant under the discrete isometry $t\mapsto -t$
for all $\tau$.
Hence, the properties~(\ref{S0})--(\ref{Srefl}) are proved;
also~(\ref{Kmono}) is a direct consequence of the construction.
The remaining claim follows immediately from the considerations
of Sec.~\ref{data} and Prop.~\ref{embedprop}.
\proofend

\begin{remark}
In the proof, $K(\tau)$ is strictly monotonic, which does not
imply that $\dot{K} \neq 0$ a priori.
However, by an appropriate redefinition of the parameter $\tau$,
$\dot{K}\neq 0$ can always be achieved.
Henceforth  we will always adopt the convention that $\dot{K}>0$.
\end{remark}

\begin{lemma}\label{isomlemma}
Consider the equation
\begin{equation}\label{homlapseeq}
\Delta \alpha + a \alpha = 0 \:,
\end{equation}
where $a(l) = \Lambda - K^2/3 - 6 C^2/r^6(l)$,
and let $\alpha\in H^2_{\mathrm{even}}(S^1)$ be a solution.
Then $\alpha \equiv 0$.
\end{lemma}

For the proof we refer to Appendix~\ref{thelapseequation}.

\begin{corollary}\label{lapsecorr}
The equation
\begin{equation}\label{lapseeq}
\Delta \alpha + a \alpha = \dot{K} \qquad (\dot{K} = \mathrm{const} > 0)
\end{equation}
possesses a unique even solution $\alpha:S^1\rightarrow \mathbb{R}$
for all $(K,C)\in\mathscr{KC}_0$.
\end{corollary}

\proof
The equation is elliptic, the homogeneous equation possesses only the
trivial solution. The Fredholm alternative holds,
hence existence and uniqueness of an even periodic solution $\alpha$ follows.
\proofend

Equation~(\ref{lapseeq}) is obtained by differentiating the mean curvature
$\threemetric^{i j} K_{i j}$ and using the evolution equations along the slicing.
The solution of~(\ref{lapseeq}) is the lapse function of the slicing $\mathcal{S}_\tau$
described in Theorem~\ref{localslicing}:
in fact, when we set
\begin{equation}\label{Phiparam}
\Phi: (\tau,l,\Omega) \mapsto (\sigma, \lambda,\Omega) = \Phi(\tau,l,\Omega) =\big(\varphi(\tau,l), l,\Omega\big)\:,
\end{equation}
where $\varphi(\tau,\cdot) = \varphi_\tau(\cdot)$,
and note that
the future unit (co)normal of the slicing $\mathcal{S}_\tau$ is
given by
\begin{equation}
n_\mu dx^\mu = \frac{s}{\sqrt{s^2-\varphi^{\prime\:2}}} \: \left(- d\sigma + \varphi^\prime\, d
\lambda\right)\:,
\end{equation}
we obtain
\begin{equation}\label{alphaandvarphidot}
\alpha = \frac{s}{\sqrt{s^2-\varphi^{\prime\:2}}}\: \dot{\varphi}
\end{equation}
from the decomposition $\dot{\Phi}^\mu = \alpha n^\mu + \Phi^\mu_{\:,i} X^i$
(where the dot denotes $\partial/\partial\tau$).
Hence~(\ref{mathcalKprime}) and~(\ref{lapseeq}) coincide.

The shift vector of the slicing $\mathcal{S}_\tau$ depends on the ``spatial gauge'' we are imposing.
Let us briefly elaborate on this issue.
Consider the parametrization of the slicing
\begin{equation}\label{Phiparam2}
\Phi: (\tau,l,\Omega) \mapsto (\sigma, \lambda,\Omega) = \Phi(\tau,l,\Omega) =\big(\varphi(\tau,\beta_\tau(l)), \beta_\tau(l),\Omega\big)\:,
\end{equation}
where $\beta_\tau(\cdot): [-L_{\mathcal{S}},L_{\mathcal{S}}]_{\sim}\cong S^1\rightarrow S^1$ is
one-to-one for each $\tau$; when the parametrization is adapted to the reflection symmetry of the
slicing, $\beta_\tau(0) = 0$ must hold; by virtue of the $S^1$-periodicity,
$\beta_\tau(\pm L_{\mathcal{S}}) = \pm L_{\mathcal{S}}$ for all $\tau$. On each hypersurface $\mathcal{S}_\tau$,
$l$ is a coordinate,
however, the 3-metric on $\mathcal{S}_\tau$ is not of the
form $d l^2 + r^2 d\Omega^2$ for general $\beta_\tau$.
Now consider
$\dot{\Phi}^\mu \partial_\mu = (\dot{\varphi} + \dot{\beta} \varphi^\prime)\partial_\sigma + \dot{\beta}\partial_\lambda$
and
$\Phi^\mu_{\:,l}\, \partial_\mu = \varphi^\prime \beta^\prime \partial_\sigma + \beta^\prime \partial_\lambda$,
where $\beta(\tau,\cdot) = \beta_\tau(\cdot)$.
While the lapse $\alpha$ is given by~(\ref{alphaandvarphidot}),
the shift vector field $X^i \partial_i = x \partial_l$, $x= x(\tau,l)$,
depends on the gauge $\beta_\tau$. It is given explicitly by
\begin{equation}
x = -\frac{1}{\beta^\prime} \frac{\varphi^\prime\dot{\varphi}}{s^2-\varphi^{\prime\:2}} +
\frac{\dot{\beta}}{\beta^\prime} = - \frac{\varphi^\prime}{\beta^\prime} \,
\frac{\alpha}{s \sqrt{s^2-\varphi^{\prime\:2}}} +
\frac{\dot{\beta}}{\beta^\prime}\:,
\end{equation}
where we again suppress the arguments $\big(\tau,\beta(\tau,l)\big)$ on the r.h.\ side. Note that
$x(\tau,0) = 0$ and $x(\tau,\pm L_{\mathcal{S}}) = 0$, since $\varphi^\prime(\tau,\pm
L_{\mathcal{S}}) =0$ and $\beta(\tau,\pm L_{\mathcal{S}}) = 0$.
We mention three gauges: when $\beta_\tau(l) = l$ we recover the gauge used in
Theorem~\ref{localslicing}, cf.~(\ref{sighy}), and $x = -\alpha s^{-1} (s^2 -
\varphi^{\prime\:2})^{-1/2} \varphi^{\prime}$. Another interesting parametrization is the one given
by  $\dot{\beta} = [s^2 - \varphi^{\prime\:2}]^{-1} \varphi^{\prime} \dot{\varphi}$;
this gives vanishing shift, i.e.\ an ``Eulerian'' gauge.
As a third possibility, the gauge given by
$\beta^\prime = [s^2 - \varphi^{\prime\:2}]^{-1/2}$ suggests itself. Here the spatial metric reads
$d l^2 + r^2 d\Omega$ for all $\tau$, as is easily seen by considering the pull-back
of~(\ref{metricGauss}); we call this the ``isotropic gauge''. The shift, which reads $x = -\alpha
\varphi^\prime +\sqrt{s^2-\varphi^{\prime\:2}} \dot{\beta}$, satisfies $x^\prime = -\alpha (K/3 -2
C/r^3)$, as can be shown by using~(\ref{lamini}) or by invoking the identity $\partial_\tau
(\Phi^\mu{},_i \Phi^\nu{},_j g_{\mu \nu}) = 2 \alpha K_{ij} + 2 D_{(i} X_{j)}$. But beware: the
isotropic gauge is not consistent with the $S^1$-periodicity requirement, and thus $x(\tau,\pm
L_{\mathcal{S}})\neq 0$.

The following relations are recorded for later use.
Consider $r=r(\Phi(\tau,l))$ and $t=t(\Phi(\tau,l))$ in any gauge.
Restricted to $\mathcal{S}$ (i.e.\ for $\tau=0$ or $\sigma=0$), we have
\begin{equation}\label{rdotandalpha}
\frac{\partial r}{\partial \tau} =
\left( \frac{K_{\mathcal{S}} r}{3}-\frac{C_{\mathcal{S}}}{r^2}\right)\, \alpha + x  \,r^\prime
\: ,\qquad
\frac{\partial t}{\partial \tau}  =
V^{-1}
\left[ r^\prime \alpha +  \left( \frac{K_{\mathcal{S}} r}{3}-\frac{C_{\mathcal{S}}}{r^2}\right) x \right]\:.
\end{equation}
To prove the first relation we compute
\begin{equation}
\dot{r}\big|_{\tau=0} = (\partial_\sigma r)\big|_{\sigma=0} \, (\dot{\varphi} + \varphi^\prime \dot{\beta}) \big|_{\tau =0}
+ (\partial_\lambda r)\big|_{\sigma=0}\: \dot{\beta}\big|_{\tau =0} = \left[\,
\left( \frac{K_{\mathcal{S}} r}{3} - \frac{C_{\mathcal{S}}}{r^2} \right) \alpha + r^\prime x\:\right]_{\tau=0}\:,
\end{equation}
where we have used~(\ref{lamini}) and the fact that $\varphi(0,l)\equiv 0$ and $\beta(0,l)\equiv
l$. To show the second relation we note that the Killing vector is given by $\xi = r^\prime
\partial_\sigma - [K_{\mathcal{S}} r/3 - C_{\mathcal{S}}/r^2]\partial_\lambda$ for $\sigma =0$,
cf.~(\ref{KVxi}). Hence for $\xi_\nu d x^\nu$ we obtain $\xi_\sigma = -r^\prime$ and $\xi_\lambda =
- [K_{\mathcal{S}} r/3 - C_{\mathcal{S}}/r^2]$, both evaluated at $\sigma =0$,
cf.~(\ref{metricGauss}). Now, $\partial_\nu t = -V^{-1} \xi_\nu$, cf.~(\ref{dt}), therefore
\begin{equation}
\dot{t}\big|_{\tau=0} = (\partial_\sigma t)\big|_{\sigma=0} \, \alpha\big|_{\tau=0}
+ (\partial_\lambda t)\big|_{\sigma=0}\: x\big|_{\tau=0} = V^{-1}
\left[ r^\prime \alpha +  \left( \frac{K_{\mathcal{S}} r}{3}-\frac{C_{\mathcal{S}}}{r^2}\right) x \right]_{\tau=0}\:.
\end{equation}

\begin{proposition}\label{slicingcurve}
Consider a spacetime KSSdS[T] that contains a CMC-slicing
$\mathcal{S}_\tau$, $\tau\in (-\bar{\tau},\bar{\tau})$, satisfying
the properties~(\ref{S0})--(\ref{Srefl})
of Theorem~\ref{localslicing}. This slicing
uniquely corresponds to a smooth curve in $\mathscr{KC}_0$,
\begin{equation}
(-\bar{\tau},\bar{\tau}) \ni \tau \mapsto (K,C)(\tau) \in \mathscr{KC}_0\:,
\end{equation}
such that $(K,C)(0) = (K_{\mathcal{S}}, C_{\mathcal{S}})$. The tangent vector
$(\dot{K},\dot{C})(\tau)$ of the curve is given through
\begin{equation}\label{Cdoteqmin}
-\frac{\dot{K}(\tau) r_{\min}(\tau)}{3} +\frac{\dot{C}(\tau)}{r_{\min}^2(\tau)} =
r^{\prime\prime}_{\min}(\tau)\, \alpha_{\min}(\tau)\:,
\end{equation}
where $r_{\min}(\tau) = r(\tau;0)$, 
$r^{\prime\prime}_{\min}(\tau) = r^{\prime\prime}(\tau;0) =
(1/2) (dD(\tau;r)/dr)|_{r=r_{\min}(\tau)}$,
$\alpha_{\min}(\tau) = \alpha(\tau;0)$; $\alpha(\tau;l)$
is the lapse function of the slicing.
\end{proposition}

\proof
$\mathcal{S}_\tau$ is a compact CMC-hypersurface in KSSdS[T] for all $\tau$, hence the
CMC-data is characterized by $(K(\tau)), C(\tau)) \in \mathscr{KC}_0$ for all $\tau$. Accordingly,
$\mathcal{S}_\tau$ generates a mapping $(-\bar{\tau},\bar{\tau}) \ni \tau \mapsto
(K,C)(\tau)=(K(\tau),C(\tau)) \in \mathscr{KC}_0$; clearly,
$(K,C)(0)=(K(0),C(0))=(K_\mathcal{S},C_{\mathcal{S}})$. By virtue of the uniqueness property of
$\mathcal{S}_\tau$ established in Theorem~\ref{localslicing}, the connection between the slicing
$\mathcal{S}_\tau$ and the map $(K,C)(\tau)$ is one-to-one. To prove that $(K,C)(\tau)$ is a smooth
curve in $\mathscr{KC}_0$, consider
\begin{equation}
K_{i j}(\tau;r) K^{i j}(\tau;r) = \frac{K^2(\tau)}{3} +\frac{6 C^2(\tau)}{r^6}\:,
\end{equation}
cf.~(\ref{KijinKC}). Along the (smooth) slicing $\mathcal{S}_\tau$, $K_{i j} K^{i j}$ and $K$ are
smooth functions, therefore $C(\tau)$ and thus $(K,C)(\tau)$ is smooth.

Let $\alpha(\tau;l)$ be the lapse function of the slicing $\mathcal{S}_\tau$ and $X^i \partial_i =
x(\tau;l) \partial_l$ the shift vector in an arbitrary gauge.
Consider $\partial_\tau [K_{i j}(\tau;l) K^{i j}(\tau;l)]$. We have on one hand that
\begin{equation}\label{KijKijdoteq1}
(K_{i j} K^{i j})\dot{} = \frac{2}{3} \, K \dot{K} + \frac{12 C}{r^6}\, \dot{C} -\frac{36 C^2}{r^7}\, \dot{r}\:,
\end{equation}
where $\dot{r}$ is given by~(\ref{rdotandalpha}). On the other hand,
from the evolution equations and by employing~(\ref{lapseeq}),
\begin{equation}\label{KijKijdoteq2}
(K_{i j} K^{i j})\dot{} =
-\frac{12 \,C \,\alpha}{r^6} \,(-r D(r) + r - 3 M)  -\frac{12\, C r^\prime \alpha^\prime}{r^4} +
2 \left(\frac{K}{3} + \frac{2 C}{r^3}\right) \dot{K} - \frac{36 C^2}{r^7} r^\prime x \:.
\end{equation}
Equating~(\ref{KijKijdoteq1}) and~(\ref{KijKijdoteq2}), the terms involving the shift cancel, and
we obtain
\begin{equation}\label{Cdoteq}
r^\prime(\tau;l) \alpha^\prime(\tau;l) - r^{\prime\prime}(\tau;l) \alpha(\tau;l) =
\frac{\dot{K}(\tau) r(\tau;l)}{3} -\frac{\dot{C}(\tau)}{r^2(\tau;l)}\:,
\end{equation}
from which Eq.~(\ref{Cdoteqmin}) ensues by evaluation at $l=0$. \proofend

\begin{definition}
Equation~(\ref{Cdoteqmin}), i.e.\
\begin{equation}\label{orienteddirection}
-\frac{\dot{K} r_{\min}}{3} +\frac{\dot{C}}{r_{\min}^2} =
r^{\prime\prime}_{\min}\, \alpha_{\min}\:,
\end{equation}
together with $\dot{K}>0$, defines a unique oriented direction field on $\mathscr{KC}_0$. This is
because $r_{\min}$, $r_{\min}^{\prime\prime}$, and $\alpha_{\min}$ can be regarded as functions of
$(K,C)\in\mathscr{KC}_0$. (Note in particular that the lapse function is determined by
equation~(\ref{lapseeq}), which only relies on the initial data sets.)
\end{definition}

Proposition~\ref{slicingcurve} states that a slicing $\mathcal{S}_\tau$ is uniquely represented by
a (local) integral curve of the direction field. The following proposition turns this into a global
statement:

\begin{proposition}
Consider a spacetime KSSdS[T] that contains a CMC-slicing $\mathcal{S}_\tau$, $\tau\in
(-\bar{\tau},\bar{\tau})$, satisfying the properties~(\ref{S0})--(\ref{Srefl}) of
Theorem~\ref{localslicing}. This slicing can be maximally extended to a slicing $\mathcal{S}_\tau$,
$\tau\in (\tau_-,\tau_+)\supseteq(-\bar{\tau},\bar{\tau})$, satisfying~(\ref{S0})--(\ref{Srefl}),
such that
\begin{equation}
(K,C)(\tau) \rightarrow \partial(\mathscr{KC}_0) \quad (\tau\rightarrow \tau_\pm)\:.
\end{equation}
Thus, the maximal extension of a slicing uniquely corresponds
to a maximal integral curve of the oriented direction field in $\mathscr{KC}_0$.
\end{proposition}

\proof
Let $(\tau_-,\tau_+)$ be the maximal interval of existence of the slicing
$\mathcal{S}_\tau$ provided by Theorem~\ref{localslicing},
and consider the associated curve $(K,C)(\tau)$ in $\mathscr{KC}_0$,
which is an open piece of a maximal integral curve $(K,C)_{\mathrm{mic}}(\tau)$
of the direction field
defined on $\mathscr{KC}_0$.
Assume that $(K,C)(\tau)\rightarrow (K_+,C_+)\in\mathscr{KC}_0$
as $\tau \rightarrow \tau_+$. Since $\mathcal{S}_\tau$
is a hypersurface in KSSdS[T], $\mathcal{T}(\tau) = \mathrm{T}$
for all $\tau \in (\tau_-,\tau_+)$.
As a function of $K$ and $C$ on
$\mathscr{KC}_0$, $\mathcal{T}$ is continuous, cf.~(\ref{Tdef});
it follows that
$\mathcal{T}(K_+,C_+) = \lim_{\tau\rightarrow\tau_+} \mathcal{T}(K(\tau),C(\tau)) = \mathrm{T}$.
Thus, by Theorem~\ref{localslicing}, there exists a
unique local slicing $\hat{\mathcal{S}}_\pi$, $\pi\in(-\bar{\pi},\bar{\pi})$,
in KSSdS[T], such that
$\hat{\mathcal{S}}_0$ is the CMC-hypersurface characterized by $(K_+, C_+)$.
The slicing $\hat{\mathcal{S}}_\pi$ is represented by an open
piece of the maximal integral curve $(K,C)_{\mathrm{mic}}$
passing through $(K_+,C_+)$;
hence $\mathcal{S}_\tau \cup \hat{\mathcal{S}}_\pi$ is a slicing of KSSdS[T]
that extends $\mathcal{S}_\tau$. This is a contradiction to the assumption.
The argument for $\tau_-$ is identical.
\proofend

\section{Properties of the slicings, and foliations}
\label{foliations}

In this section we prove that each spacetime KSSdS[T] contains a unique (maximally extended)
slicing by compact CMC-hypersurfaces, and we show that this slicing is a foliation for a certain range of the
time parameter. However, we begin with a discussion of the asymptotic behavior of slicings.

\begin{lemma}\label{Tlemma}
Consider the function
\begin{equation}
\mathcal{T}(K,C) = \int\limits_{r_{\min}(K,C)}^{r_{\max}(K,C)} V^{-1}(r)\: D^{-\frac{1}{2}}(K,C;r)\,
\:\left(\frac{K r}{3} -\frac{C}{r^2}\right)  d r\:,
\end{equation}
cf.~(\ref{Tdef}). Let $[0,1)\ni\nu \mapsto (K,C)(\nu) \in \mathscr{KC}_0$ be a curve
such that $(K,C)(\nu) \rightarrow (K_\partial, C_\partial)\in \partial(\mathscr{KC}_0)$
as $\nu\rightarrow 1$.
Then
\begin{subequations}
\begin{align}
\mathcal{T} \rightarrow +\infty  & \qquad\text{if} \quad
(K_\partial,C_\partial) \in \Big( C_t \cup (K=-\sqrt{3\Lambda})\Big) \\
\mathcal{T} \rightarrow -\infty  & \qquad\text{if} \quad
(K_\partial,C_\partial) \in \Big( C_b \cup (K=+\sqrt{3\Lambda})\Big)\:,
\end{align}
\end{subequations}
where, however, the points $(\sqrt{3\Lambda},C_t(\sqrt{3 \Lambda}))$ and
$(-\sqrt{3\Lambda},C_b(-\sqrt{3 \Lambda}))$ are excluded.
\end{lemma}

\begin{figure}[htp]
    \psfrag{Cb}[cc][cc][0.7][12]{$\longleftarrow C_b \longrightarrow$}
    \psfrag{Ct}[cc][cc][0.7][9]{$\longleftarrow C_t \longrightarrow$}
        \psfrag{A}[cc][cc][0.8][6]{$\mathcal{T} = \infty$}
        \psfrag{B}[cc][cc][0.8][13]{$\mathcal{T}= \infty$}
        \psfrag{C}[cc][cc][0.8][6]{$\mathcal{T} = -\infty$}
        \psfrag{D}[cc][cc][0.8][16]{$\mathcal{T} = -\infty$}
        \psfrag{E}[cc][cc][0.8][90]{$\mathcal{T} = \infty$}
        \psfrag{F}[cc][cc][0.8][-90]{$\mathcal{T} = -\infty$}
    \psfrag{c}[cc][cc][0.6][0]{$C$}
        \psfrag{K}[cc][cc][0.6][0]{$K$}
    \psfrag{Kd}[cc][cc][0.6][0]{$\dot{K}>0$}
    \centering
    \includegraphics[width=0.7\textwidth]{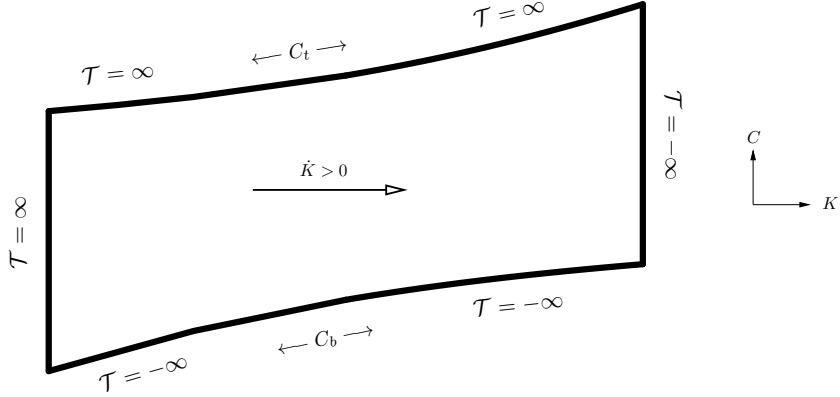}
    \caption{{\small On $\mathscr{KC}_0$
        the direction field $(\dot{K},\dot{C})$ satisfies $\dot{K} > 0$.
        On the boundaries $\mathcal{T} = \pm \infty$.}}
        \label{direconKC}
\end{figure}

\proof
Let $(K_\partial,C_\partial) \in C_t$ with $K_\partial \in (-\sqrt{3\Lambda},+\sqrt{3\Lambda})$.
Choose $\epsilon > 0$ small, and write
\begin{equation}
\mathcal{T}(\nu)
= \int\limits_{r_{\min}(\nu)}^{r_{\min}(\nu)+\epsilon}
\frac{V^{-1}}{\sqrt{D(\nu;r)}}\:\left(\frac{K(\nu) r}{3} -\frac{C(\nu)}{r^2}\right) d r
+ \int\limits_{r_{\min}(\nu)+\epsilon}^{r_{\max}(\nu)}\frac{V^{-1}}{\sqrt{D(\nu;r)}} \:
\left(\frac{K(\nu) r}{3} -\frac{C(\nu)}{r^2}\right) d r \:.
\end{equation}
The second integral converges to a constant as $\nu\rightarrow 1$, since
$r_{\min}(\nu)$ and $r_{\max}(\nu)$ converge to the values of
$r_{\min}$ and $r_{\max}$ at $(K_\partial, C_\partial)$,
respectively, and the integrand converges uniformly.
In $r\in(r_{\min}(\nu) , r_{\min}(\nu)+\epsilon)$ we make the expansion
\begin{equation}
D(\nu;r) = D^\prime(\nu;r_{\min}) (r-r_{\min}) +
\frac{D^{\prime\prime}(\nu;r_{\min})}{2} (r-r_{\min})^2 +
O((r-r_{\min})^3)\:,
\end{equation}
where $r_{\min}=r_{\min}(\nu)$ and
the prime denotes differentiation w.r.t.\ $r$ in the present context.
Then the first integral becomes
\begin{equation}\label{Tdiv1}
\left[V^{-1}(r_{\min}) \left(\frac{K r_{\min}}{3} -\frac{C}{r_{\min}^2}\right)\right]
\Bigg|_{(K_\partial, C_\partial)}\:
\times
\left[ -\frac{\sqrt{2}\:\log[D^\prime(\nu;r_{\min})]}{\sqrt{D^{\prime\prime}(\nu;r_{\min})}} \,
 \right]
+\, D_\epsilon + O(D^\prime(\nu;r_{\min}))\:.
\end{equation}
The first factor is a positive number, since $K_\partial r_{\min}/3 -C_\partial/r_{\min}^2$ is
negative, cf.~Fig.~\ref{divrmin}; $D_\epsilon$ is a constant. 
With $\nu\rightarrow 1$, $D^\prime(\nu; r_{\min}(\nu)) \rightarrow 0$,
cf.~Fig.~\ref{Deltaclass4}; the equation $D^\prime(r_{\min}) = 0$ is in fact the defining
equation for $C_{b,t}$, see~Appendix~\ref{KCspaceapp}. Therefore, $\mathcal{T}(\nu)$ diverges
like $-\log[D^\prime(\nu;r_{\min}(\nu))]$, i.e.\ $\mathcal{T}\rightarrow \infty$ as
$\nu\rightarrow 1$. Analogously, we are able to prove that $\mathcal{T} = -\infty$ on $C_b$ when
$K_\partial\neq \pm\sqrt{3\Lambda}$; the different sign results from the different sign of 
the quantity $(K r_{\min}/3 -C/r_{\min}^2)$; see~Fig.~\ref{direconKC}.

Now let $K_\partial = \pm \sqrt{3\Lambda}$,
$C_\partial\in (C_b(\pm\sqrt{3\Lambda}),C_t(\pm\sqrt{3\Lambda}))$.
Along the curve $(K,C)(\nu)$, $r_{\max}(\nu)$ diverges like
$\sqrt{3}/\sqrt{\Lambda-K(\nu)^2/3}$, cf.~Sec.~\ref{data}.
Choose $r_0 > r_{\mathrm{c}}$ such that $D(\nu;r) < 4$, $r - 2 M >0$, and
$|K(\nu)| r^3 >  6 |C(\nu)|$ on $r\in[r_0,r_{\max}(\nu)/2]$ for all $\nu$ sufficiently close to $1$.
Then,
\begin{align}\nonumber
\mathcal{T} - \int\limits_{r_{\min}}^{r_0}\frac{V^{-1}}{\sqrt{D}}\:\left(\frac{K r}{3} -\frac{C}{r^2}\right) d r
& \lessgtr \int\limits_{r_0}^{r_{\max}/2}
\frac{V^{-1}}{\sqrt{D}}\:\left(\frac{K r}{3} -\frac{C}{r^2}\right) d r \\ \label{Tdiv2}
& \lessgtr  \frac{1}{4} \int\limits_{r_0}^{r_{\max}/2}
\left(-\frac{\Lambda r^2}{3}\right)^{-1} \left(\frac{K r}{3}\right) d r
= -\frac{1}{4} \frac{K}{\Lambda} \log r \Big|_{r_0}^{r_{\max}/2}\:,
\end{align}
where we have suppressed the dependence on $\nu$.
We infer that $\mathcal{T}(\nu)$ diverges at least like
$-(\mathrm{sign} K) \log(\Lambda-K(\nu)^2/3)$ as $\nu \rightarrow 1$; hence
$\mathcal{T} \rightarrow -\infty$ when $K_\partial = \sqrt{3\Lambda}$ and
$\mathcal{T} \rightarrow \infty$ when $K_\partial = -\sqrt{3\Lambda}$.
A combination of the arguments used for~(\ref{Tdiv1}) and~(\ref{Tdiv2}) also yields
that $\mathcal{T} \rightarrow -\infty$ when
$(K_\partial,C_\partial) = (\sqrt{3\Lambda}, C_b(\sqrt{3\Lambda}))$
and $\mathcal{T} \rightarrow \infty$ when
$(K_\partial,C_\partial) \rightarrow (-\sqrt{3\Lambda}, C_t(-\sqrt{3\Lambda}))$.
\proofend

\begin{figure}[htp]
    \psfrag{r0}[cc][cc][0.8][0]{$r=0$}
    \psfrag{r1}[cc][cc][0.8][0]{$r=\infty$}
    \psfrag{a}[cc][cc][0.60][-45]{$r=r_{\mathrm{b}}$}
    \psfrag{b}[cc][cc][0.60][45]{$r=r_{\mathrm{b}}$}
    \psfrag{c}[cc][cc][0.60][45]{$r=r_{\mathrm{b}}$}
    \psfrag{d}[cc][cc][0.60][-45]{$r=r_{\mathrm{b}}$}
    \psfrag{e}[cc][cc][0.60][-45]{$r=r_{\mathrm{c}}$}
    \psfrag{f}[cc][cc][0.60][45]{$r=r_{\mathrm{c}}$}
    \psfrag{g}[cc][cc][0.60][45]{$r=r_{\mathrm{c}}$}
    \psfrag{h}[cc][cc][0.60][-45]{$r=r_{\mathrm{c}}$}
    \centering
    \includegraphics[width=0.9\textwidth]{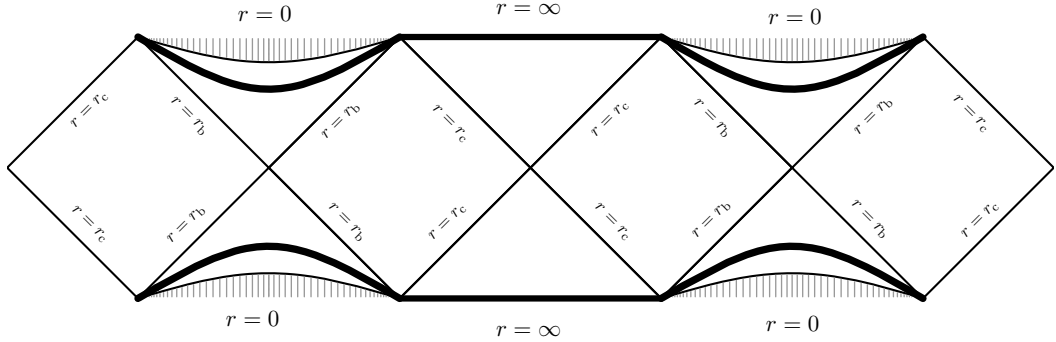}
    \caption{{\small The asymptotic hypersurfaces $\mathcal{S}_{\pm}$ of the slicing $\mathcal{S}_\tau$.}}
        \label{asyhyp}
\end{figure}

\begin{proposition}\label{slicingends}
Consider a spacetime KSSdS[T] that contains a CMC-slicing
satisfying the properties~(\ref{S0})--(\ref{Srefl})
of Theorem~\ref{localslicing}. Let $\mathcal{S}_\tau$, $\tau\in (\tau_-,\tau_+)$,
be the maximal extension.
Then
\begin{subequations}\label{asypropKC}
\begin{alignat}{3}
&(K,C)(\tau) \rightarrow (\sqrt{3\Lambda},C_t(\sqrt{3\Lambda})) & \qquad(\tau\rightarrow \tau_+)\:, \\
&(K,C)(\tau) \rightarrow (-\sqrt{3\Lambda},C_b(-\sqrt{3\Lambda})) & \qquad(\tau\rightarrow \tau_-)\:.
\end{alignat}
\end{subequations}
Hereby, the hypersurfaces $\mathcal{S}_\tau$ converge
to the asymptotic hypersurfaces $\mathcal{S}_{\pm}$,
$\mathcal{S}_+$ in the future, $\mathcal{S}_-$ in the past,
\begin{equation}\label{Spm}
\mathcal{S}_\pm =
\left(r = \frac{1}{\sqrt{\Lambda}} \left[1 - \sqrt{1 - 3 \sqrt{\Lambda} M}\right]\right)
\:\cup\:
\left( r = \infty \right) \:\: \subseteq \:\overline{\mathrm{KSSdS[T]}}
\end{equation}
as $\tau \rightarrow \tau_{\pm}$, see Fig.~\ref{asyhyp}.
\end{proposition}

\begin{remark}
Proposition~\ref{slicingends} is equivalent to the statement that all maximal
integral curves of the oriented direction field on $\mathscr{KC}_0$
originate from the point $(-\sqrt{3\Lambda}, C_b(-\sqrt{3\Lambda}))$ and
end in the point $(\sqrt{3\Lambda}, C_t(\sqrt{3\Lambda}))$.
This is straightforward to prove:
\end{remark}

\proof
Each maximal integral curve $(K,C)(\tau)$ of the direction field on $\mathscr{KC}_0$
is characterized by $\mathcal{T}(\tau) \equiv \mathrm{T} = \mathrm{const}$,
since the slicing is embedded in KSSdS[T].
Therefore, the limit set of the curve $(K,C)(\tau)$ cannot contain a point on
$\partial(\mathscr{KC}_0)$ where $\mathcal{T}=\pm\infty$,
which leaves only the points $(-\sqrt{3\Lambda}, C_b(-\sqrt{3\Lambda}))$ and
$(\sqrt{3\Lambda}, C_t(\sqrt{3\Lambda}))$ by Lemma~\ref{Tlemma}.
Together with $\dot{K}>0$ this entails~(\ref{asypropKC}).

To show the second part of the assertion,
we recall that $\mathcal{S}_\tau$ can be represented by
$t(\tau;r)$ in KSSdS[T], cf.~(\ref{tofl}).
In analogy to the considerations in the proof of Lemma~\ref{Tlemma} we obtain
\begin{equation}
t(\tau;r_{\min}(\tau)+\epsilon) =
\int\limits_{r_{\min}(\tau)}^{r_{\min}(\tau)+\epsilon}
\frac{V^{-1}}{\sqrt{D(\tau;r)}}\:\left(\frac{K(\tau) r}{3} -\frac{C(\tau)}{r^2}\right) d r
\:\rightarrow\: \infty \qquad (\tau\rightarrow \tau_+)
\end{equation}
for all $\epsilon>0$;
moreover, $r_{\min}(\tau)\rightarrow r_{\min}(\tau_+)$ for $\tau\rightarrow\tau_+$.
Therefore, in the black hole region, $\mathcal{S}_\tau$ converges
to the hypersurface $r = r_{\min}(\tau_+)$ as $\tau\rightarrow \tau_+$,
where the convergence is uniform on each ``cone'' $\{t \:|\: t\in [-E,E]\:, E >0\}$.
From~(\ref{rminextremal}) we see that $r_{\min}(\tau_+) = (1 - \sqrt{1 - 3 \sqrt{\Lambda} M})/\sqrt{\Lambda}$.
Similar considerations apply for the cosmological region; in particular, $r_{\max}(\tau)\rightarrow \infty$
as $\tau\rightarrow\tau_+$. Hence the claim
is established.
\proofend

\begin{figure}[htp]
    \psfrag{A}[cc][cc][1][10]{$\leftarrow\, C_t(K)\, \rightarrow$}
    \psfrag{B}[cc][cc][1][10]{$\leftarrow\, C_b(K)\,\rightarrow$}
    \psfrag{E}[cc][cc][1][-90]{$K=\sqrt{3 \Lambda}$}
    \psfrag{F}[cc][cc][1][90]{$K=-\sqrt{3\Lambda}$}
    \centering
    \includegraphics[width=0.9\textwidth]{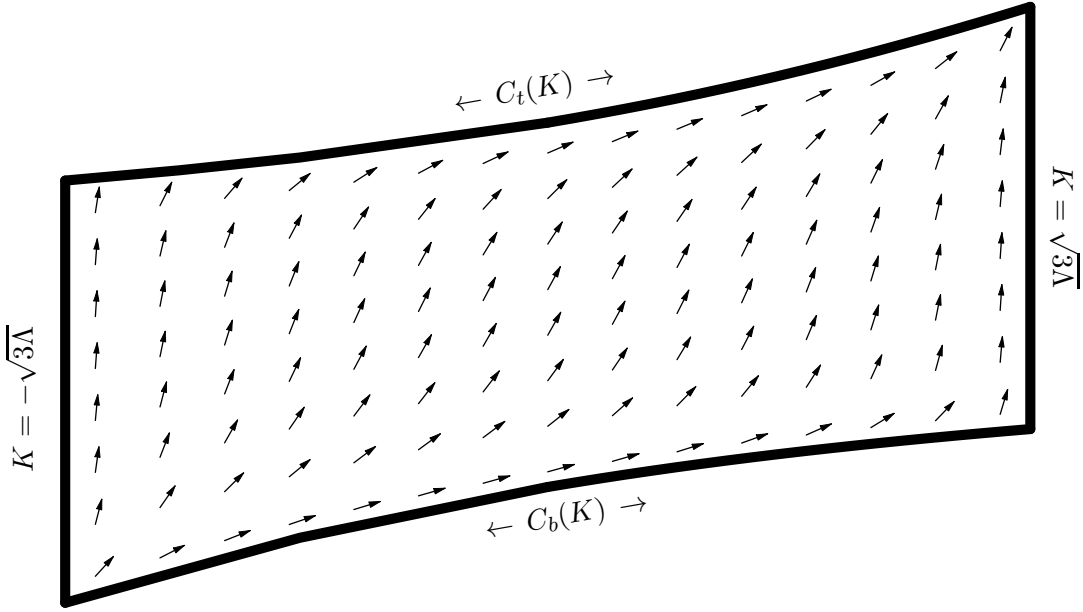}
    \caption{{\small The direction field $(\dot{K},\dot{C})$ on $\mathscr{KC}_0$ for $\Lambda=1$, $M=1/4$.}}
        \label{VFKC}
\end{figure}

\renewcommand{\theenumi}{\alph{enumi}}

\begin{theorem}\label{uniqueslicing}
Each spacetime KSSdS[T] contains a unique (maximally extended)
slicing $\mathcal{S}_\tau$, $\tau\in(\tau_-,\tau_+)$, such that
\begin{enumerate}
\item \label{abed} $\mathcal{S}_\tau$ is a compact CMC-hypersurface for all $\tau \in (\tau_-, \tau_+)$,
\item \label{bbed} $\mathcal{S}_\tau$ is reflection symmetric for all $\tau$.
\setcounter{store}{\value{enumi}}
\end{enumerate}
Along the slicing,
\begin{enumerate}
\setcounter{enumi}{\value{store}}
\item \label{cbed} $K(\tau)$ is strictly monotonically increasing.
\end{enumerate}
Every slicing $\mathcal{S}_\tau^\prime$ that satisfies~(\ref{abed})
arises from $\mathcal{S}_\tau$ by combining the flow of $\mathcal{S}_\tau$ with
an appropriate admixture of the Killing flow.
\end{theorem}

\begin{remark}
By definition, $\mathcal{T}$ is constant along the maximal
integral curves of the oriented direction field on $\mathscr{KC}_0$.
Below we prove that, for each $\text{T}$, the equation $\mathcal{T} = \text{T}$
defines a unique integral curve, which corresponds to a unique
slicing $\mathcal{S}_\tau$ in KSSdS[T], $\tau\in(\tau_-,\tau_+)$,
satisfying~(\ref{abed})--(\ref{cbed}).
The function $\mathcal{T}$ can thus be viewed as a ``Hamilton function'' for
the oriented direction field on $\mathscr{KC}_0$.
\end{remark}

\proof
Each maximal integral curve of the oriented direction field on $\mathscr{KC}_0$ has a unique
point of intersection with the line $K=\mathrm{const}$, since $\dot{K}>0$ everywhere. Hence the
pairs $(0,C)$ with $C\in(C_b(0), C_t(0))$ parametrize the family of integral curves in
$\mathscr{KC}_0$. In the following we prove that $\partial\mathcal{T}(K,C)/\partial C$ is positive
for all $(K,C)\in\mathscr{KC}_0$. By virtue of the asymptotic properties of $\mathcal{T}(K,C)$
established in Lemma~\ref{Tlemma}, this implies that $C\mapsto \mathcal{T}(0,C)$ is a bijection
between $(C_b(0), C_t(0))$ and $\mathbb{R}$. Hence, there exists a unique pair
$(0,C)\in\mathscr{KC}_0$, such that $\mathcal{T}(0,C) = \mathrm{T}$, and thus a unique maximal
integral curve in $\mathscr{KC}_0$, such that $\mathcal{T} =\text{T}$ along the curve. Since the
integral curve uniquely corresponds to a slicing $\mathcal{S}_\tau$
satisfying~(\ref{abed})--(\ref{cbed}), the claim of the theorem is established. We now show that
$\partial\mathcal{T}(K,C)/\partial C > 0$ for all $(K,C)\in\mathscr{KC}_0$.

Consider the initial data generated by $(K_{\mathcal{S}},C_{\mathcal{S}})\in \mathscr{KC}_0$
and let again $L_{\mathcal{S}} = L(K_{\mathcal{S}},C_{\mathcal{S}})$,
$\mathrm{T} =\mathcal{T}(K_{\mathcal{S}},C_{\mathcal{S}})$.
The universal covering of the data is embedded
as a Cauchy CMC-hypersurface $\mathcal{S}$ in the covering space KSSdS.
In a
neighborhood of $\mathcal{S}$ we introduce Gaussian coordinates $(\sigma,\lambda)$, cf.~the proof
of Theorem~\ref{localslicing}. Now consider the (auxiliary) spacetime given by 
$(-\bar{\sigma},\bar{\sigma})\times
(-L_\mathcal{S}-\varepsilon, L_\mathcal{S}+\varepsilon) \times S^2$, which we denote by
$\mathrm{KSSdS[T]}^\prime$; it is an open subset of
$\mathrm{KSSdS}$ and spatially incomplete. In $\mathrm{KSSdS[T]}^\prime$ the hypersurface
$\mathcal{S}$ does not generate a unique slicing, but evolves into a one-parameter family of
slicings; we are able to show this in a straightforward way by using the methods of the proof of
Theorem~\ref{localslicing}. We define the mean curvature operator $\mathcal{K}$ as
in~(\ref{meancurvoper}) and an operator $\mathcal{C}$,
\begin{equation}
\mathcal{C}^2[\varphi]:= \frac{r^6(\varphi(\lambda),\lambda)}{6}
\left(\mathcal{K}_{i j}[\varphi] \mathcal{K}^{i j}[\varphi] - \frac{\mathcal{K}[\varphi]^2}{3}\right)\:,
\end{equation}
which assigns to every hypersurface $\sigma=\varphi(\lambda)$ a function
$(-L_{\mathcal{S}}-\varepsilon,L_{\mathcal{S}}+\varepsilon) \ni \lambda \mapsto C[\varphi](\lambda)$;
for a CMC-hypersurface characterized by $(K,C)$, $\mathcal{C}[\varphi]\equiv C$.
By construction, $\mathcal{K}[0]=K_{\mathcal{S}}$ and $\mathcal{C}[0] = C_{\mathcal{S}}$.
$\mathcal{K}$ and $\mathcal{C}$ are $\mathcal{C}^1$ operators
on $H^2_{\mathrm{even}}\big((-L_{\mathcal{S}}-\varepsilon, L_{\mathcal{S}}+\varepsilon)\big)$,
$\mathcal{K}^\prime[0] = \Delta + a$, cf.~(\ref{mathcalKprime0}), and
\begin{equation}
\mathcal{C}^\prime[0] =  \frac{r^3}{3}\, \Delta - \nabla^i\left(\frac{r^3}{3}\right)\: \nabla_i + b  \:,
\end{equation}
where $b(\lambda) = M + K C/3 -4 C^2/r^3$ and $\nabla^i(r^3/3) \,\nabla_i = r^2 r^\prime
\partial_l$. The set $\mathrm{ker}\,\mathcal{K}^\prime[0]$ is not trivial, and neither is
$\mathrm{ker}\,\mathcal{C}^\prime[0]$. However, the joint map
$(\mathcal{K}^\prime,\mathcal{C}^\prime)[0]$ is an isomorphism; see
Appendix~\ref{thelapseequation}. We may apply the inverse function theorem: there exists a unique
continuously differentiable mapping $(\mathcal{K},\mathcal{C})^{-1}$ such that
$(\mathcal{K},\mathcal{C})[(\mathcal{K},\mathcal{C})^{-1}((\kappa,\gamma))]=(\kappa,\gamma)$ for
all $(\kappa,\gamma)$ of a neighborhood of $(K_{\mathcal{S}},C_{\mathcal{S}})$ in
$H^0_{\mathrm{even}}\times H^0_{\mathrm{even}}$. Hence, given a smooth function
$(-\bar{\tau},\bar{\tau})\ni\tau \mapsto (K(\tau),C(\tau))$ such that $(K(0),C(0)) =
(K_{\mathcal{S}},C_{\mathcal{S}})$, a unique slicing $\mathcal{S}_\tau$ in $\text{KSSdS[T]}^\prime$
is defined by using $\varphi_{\tau}:=(\mathcal{K},\mathcal{C})^{-1}(K(\tau),C(\tau))$.
By construction, the slicing $\mathcal{S}_\tau$ in $\text{KSSdS[T]}^\prime$ uniquely corresponds
to a curve $(K,C)(\tau)$ in a neighborhood of $(K_{\mathcal{S}}, C_{\mathcal{S}})\in\mathscr{KC}_0$.

Consider the hypersurface $\mathcal{S}$ and the slicing $\mathcal{S}_\tau$
in $\text{KSSdS[T]}^\prime$
such that $\mathcal{S}_0 = \mathcal{S}$ and $K(\tau) = K_{\mathcal{S}}$ for all $\tau$;
the slicing is represented by a curve $(K,C)(\tau) = (K_{\mathcal{S}}, C(\tau))$,
$C(0) = C_{\mathcal{S}}$, in $\mathscr{KC}_0$.
Let $\alpha(\tau,l)$ be the lapse function of the slicing
and $x(\tau,l)$ the shift vector in an arbitrary gauge.
Making use of~(\ref{rdotandalpha}) we find that
\begin{equation}
\frac{\partial r^\prime}{\partial \tau}(0,L_{\mathcal{S}}) =
\frac{\partial^2 r}{\partial l\, \partial \tau}(0,L_{\mathcal{S}}) =
\left(\frac{K_{\mathcal{S}} r_{\max}}{3} - \frac{C_{\mathcal{S}}}{r_{\max}^2}\right)\,\alpha^\prime(0,L_{\mathcal{S}})
+ x(0, L_{\mathcal{S}}) r^{\prime\prime}_{\max} \:,
\end{equation}
where $r_{\max}=r(0,L_{\mathcal{S}})$ and $r_{\max}^{\prime\prime} = r^{\prime\prime}(0,L_{\mathcal{S}})$.
Differentiation of the equation $r^\prime(\tau,L(\tau)) =0$ results in
\begin{equation}
\dot{L}(0) = - \left(\frac{K_{\mathcal{S}} r_{\max}}{3} - \frac{C_{\mathcal{S}}}{r_{\max}^2}\right)\,
\frac{\alpha^\prime(0,L_{\mathcal{S}})}{r_{\max}^{\prime\prime}} -x(0, L_{\mathcal{S}}) \:.
\end{equation}
We are now prepared to investigate the derivative of $\mathcal{T}(\tau)$ at $\tau=0$. The
definition $\mathcal{T}(\tau) = t(\tau,L(\tau))$ leads to
\begin{equation}
\dot{\mathcal{T}}(0) = \frac{\partial t}{\partial \tau}(0,L_{\mathcal{S}}) +
\frac{\partial t}{\partial l}(0,L_{\mathcal{S}})\: \dot{L}(0) =
\frac{\alpha^\prime(0,L_{\mathcal{S}})}{r_{\max}^{\prime\prime}} \,
\end{equation}
where we have employed~(\ref{tofl}) and~(\ref{rdotandalpha}). By virtue of
Eq.~(\ref{orienteddirection}) (which holds also for slicings of $\text{KSSdS[T]}^\prime$),
we have $\dot{C}(0) = r_{\min}^{\prime\prime} r_{\min}^2 \alpha(0,0)$, and thus
\begin{equation}
\frac{\partial \mathcal{T}(K,C)}{\partial C} \Big|_{(K_{\mathcal{S}}, C_{\mathcal{S}})} =
\frac{1}{r_{\min}^2 r_{\min}^{\prime\prime\:2}} \,
\frac{r_{\min}^{\prime\prime}}{r_{\max}^{\prime\prime}}
\frac{\alpha^\prime(0, L_{\mathcal{S}})}{\alpha(0,0)} =
\frac{r_{\min}^{\prime\prime}}{r_{\max}^{\prime\prime}}
\frac{\alpha_0^\prime(L_{\mathcal{S}})}{r_{\min}^2 r_{\min}^{\prime\prime\:2}} =
\frac{1}{2} \frac{\alpha_0^\prime(2 L_{\mathcal{S}})}{r_{\min}^2 r_{\min}^{\prime\prime\:2}}\:,
\end{equation}
where we have also used~(\ref{PiL0}) from App.~\ref{thelapseequation};
$\alpha_0$ denotes the solution of the homogeneous lapse
equation~(\ref{homlapseeq}), associated with $(K_{\mathcal{S}}, C_{\mathcal{S}})$, with initial
data $\alpha_0|_{l=0} =1$. By virtue of Lemma~\ref{isomlemma}, $\alpha_0$ is not periodic, so that
$\alpha_0^\prime(2 L_{\mathcal{S}}) \neq 0$; moreover, since $\alpha_0^\prime(2 L_{\mathcal{S}})
\neq 0$ holds irrespective of the choice of $(K,C)=(K_{\mathcal{S}},C_{\mathcal{S}})$, it has a
definite sign for all $(K,C) \in \mathscr{KC}_0$. Consequently, also $\partial
\mathcal{T}(K,C)/\partial C \gtrless 0$ for all $(K,C)$. Indeed $\partial\mathcal{T}/\partial C> 0$
by virtue of the asymptotic behavior of $\mathcal{T}(K,C)$ described in Lemma~\ref{Tlemma}. This
establishes the claim of the theorem. \proofend

\begin{remark}
The slicing $\mathcal{S}_\tau$ in KSSdS[0], which constitutes the only time-sym\-metric spacetime
in the family KSSdS[T], 
is represented by a distinguished integral curve in $\mathscr{KC}_0$ that is
invariant under the inversion. The curve passes through the origin
$(0,0)$ which is associated with a CMC-hypersurface of time-symmetry,
the hypersurface $t=0$, which connects the bifurcation 2-spheres.
\end{remark}

\begin{remark}
The approach we have taken to establish the main results displays a pronounced interplay 
between the geometric analysis and the analysis of the function $\mathcal{T}$.
We have chosen this approach because the direct investigation of this function is non-trivial; e.g.\ it is
not straightforward to show directly that it has no critical points, since methods
like the ones applied in~\cite{Beig/OMurchadha:1998} fail.
\end{remark}

\begin{theorem}\label{foliationthm}
In KSSdS[T] consider the unique maximally extended slicing
$\mathcal{S}_\tau$, $\tau\in (\tau_-,\tau_+)$,
satisfying the properties~(\ref{abed})--(\ref{cbed})
of Theorem~\ref{uniqueslicing}.
There exists an interval $(\bar{\tau}_-,\bar{\tau}_+)\subseteq (\tau_-,\tau_+)$,
such that $\mathcal{S}_\tau$, $\tau\in (\bar{\tau}_-,\bar{\tau}_+)$, is a foliation.
\end{theorem}

\proof Consider the lapse equation~(\ref{lapseeq}) for an umbilical pair $(K,C)\in\mathscr{KC}_0$,
i.e.\ $C=0$. In this case $a = \Lambda-K^2/3 = \mathrm{const}$, and
\begin{equation}\label{alphaCeq0}
\alpha\,\big|_{(K,C=0)} = \dot{K} \left(\Lambda-\frac{K^2}{3}\right)^{-1} = \mathrm{const} > 0
\end{equation}
is the unique even solution on the domain $S^1$ provided by Corollary~\ref{lapsecorr}.
Let $\alpha(\tau,l)$ be the lapse function of the slicing $\mathcal{S}_\tau$ in KSSdS[T]
which is represented by $(K,C)(\tau)$ in $\mathscr{KC}_0$.
The asymptotic properties~(\ref{asypropKC}) imply that there exists
$\tau_0$ such that $C(\tau_0) = 0$. At $\tau=\tau_0$,
$\alpha$ is given by~(\ref{alphaCeq0}), i.e.\
$\alpha(\tau_0,l) = \mathrm{const} > 0$.
Since $\alpha(\tau,l)$ continuously depends on $\tau$, cf.~(\ref{alphaandvarphidot}),
there exists an interval $(\bar{\tau}_-,\bar{\tau}_+) \ni \tau_0$, such that
$\alpha(\tau_0,l)  > 0$ for all $\tau \in (\bar{\tau}_-,\bar{\tau}_+)$.
Hence $\mathcal{S}_\tau$ is a foliation for $\tau \in (\bar{\tau}_-,\bar{\tau}_+)$.
\proofend

\begin{remark}
Since the lapse function is explicitly known for $C=0$, the direction field $(\dot{K},\dot{C})$ at
$C=0$ can be computed explicitly as well. From~(\ref{alphaCeq0}) we obtain
\begin{equation}
\dot{C}\big|_{(K,C=0)} =  \dot{K} M \left(\Lambda-\frac{K^2}{3}\right)^{-1} = \mathrm{const} > 0\:.
\end{equation}
We infer that every integral curve intersects the line $C=0$ exactly once,
hence the value of $\tau_0$ introduced in the proof of Theorem~\ref{foliationthm} is unique.
\end{remark}

In general, $(\bar{\tau}_-,\bar{\tau}_+) \neq (\tau_-,\tau_+)$ in Theorem~\ref{foliationthm};
this is proved in Appendix~\ref{foliapp}.
In particular, Corollary~\ref{folicor} shows that
$(\bar{\tau}_-,\bar{\tau}_+) \neq (\tau_-,\tau_+)$ for all spacetimes KSSdS[T] with $|\text{T}|$
large enough.

\begin{conjecture}\label{foliconj}
There exist numbers  $0<F_c< F_0<1$ such that the following statement holds:
if KSSdS[T] is a cosmological Kottler-Schwarzschild-de Sitter spacetime
with $9 M^2 \Lambda \in [F_c, F_0)$ and sufficiently small $|T|$, then
the maximally extended slicing
$\mathcal{S}_\tau$, $\tau\in(\tau_-,\tau_+)$, of Theorem~(\ref{uniqueslicing}) is a foliation.
\end{conjecture}

This conjecture is based on extensive numerical evidence, which we present in
Appendix~\ref{foliapp}.

\subsection*{Acknowledgements}
We would like to thank Christina Stanciulescu whose work~\cite{Stanciulescu:1998} and handwritten
notes were a very useful starting point for the present paper.
J.M.H.\ would also like to thank Alan Rendall for helpful discussions.

\begin{appendix}

\section{The parameter space $\mathscr{KC}_0$ of compact CMC-data}
\label{KCspaceapp}

\subsection{Properties of $\mathscr{KC}_0$}

By definition, $(K,C)\in\mathscr{KC}$ iff
there exists a neighborhood of $(K,C) \in \mathbb{R}^2$ such that
all pairs of that neighborhood generate compact CMC-initial data.
The function $r(l)$ that determines the initial data is given by the equation
\begin{equation}\label{rprimewieder}
r^{\prime} = \pm \sqrt{D(r)} \qquad\text{with}\qquad
D(r) = D(K,C;r)= 1 -\frac{2 \bar{M}}{r} - \frac{\bar{\Lambda} r^2}{3} +\frac{C^2}{r^4} \:,
\end{equation}
where $\barM = M + C K/3$ and $\barL= \Lambda -K^2/3$;
the initial data is compact iff $r(l)$ is periodic and thus interpretable
as a function $S^1\rightarrow \mathbb{R}$.
Therefore, $(K,C)\in\mathscr{KC}$ iff
the function $D(r)$
possesses two positive (simple)
zeros $r_{\min}$ and $r_{\max}$, such that $D(r)>0$ in $(r_{\min}, r_{\max})$.
(Accordingly, when viewed over $r\in [r_{\min},r_{\max}]$,
$r^\prime = \pm \sqrt{D(r)}$
describes a closed curve, cf.~Fig.~\ref{eggfig}, and
$r(l)$ becomes a periodic function
which oscillates between $r_{\min}$ and $r_{\max}$.)

Assume $C\neq 0$; then $D(r) \rightarrow \infty$ as $r\rightarrow 0$,
thus two positive extrema are necessary to obtain the desired profile of $D(r)$,
cf.~Fig.~\ref{niceDelta}.
The critical points of $D(r)$ are
\begin{equation}\label{rcupcap}
(r_{-})^3 = \frac{3}{2 \barL} \left(\barM - \sqrt{\barM^2-\frac{8}{3} \barL C^2}\right) \qquad
(r_{+})^3 = \frac{3}{2 \barL} \left(\barM + \sqrt{\barM^2-\frac{8}{3} \barL C^2}\right) \:,
\end{equation}
which are positive, iff
\begin{equation}\label{aprioricondis}
\barL > 0 \quad,\quad \barM > 0 \quad,\quad C^2 < \frac{3}{8} \frac{\barM^2}{\barL}\:.
\end{equation}
For future reference we note that
\begin{equation}\label{Kbounds}
\barL > 0 \quad\Leftrightarrow\quad K \in (-\sqrt{3\Lambda}, + \sqrt{3 \Lambda})\:.
\end{equation}
When the conditions~(\ref{aprioricondis}) are satisfied, $r_{-}$/$r_{+}$ is automatically a minimum/maximum of $D(r)$;
however, the conditions are not sufficient to ensure the profile~\ref{niceDelta} of $D(r)$.

Note that the case $C=0$ is simpler:
$D(r)$ possesses the desired profile iff $\barL>0$.
Proof: $D(r)\rightarrow-\infty$
($r\rightarrow 0$, $r\rightarrow\infty$); $r_-$ does not exist; $r_+ = \sqrt[3]{3 M/\barL}$
is a maximum iff $\barL>0$. Moreover,
$D(r_{+}) > 0$, since $9  M^2 \barL \leq 9 M^2 \Lambda < 1$.

We can effectively reduce the problem by one parameter by writing
\begin{equation}\label{getildet}
\tilde{D}(\tilde{r}) = 1 - \frac{2 \tildeM}{\tilder} - \frac{\tilder^2}{3} + \frac{\tildeC^2}{\tilder^4}
\qquad\text{with}\quad
\tilde{r} = \sqrt{\barL} r \:,\quad \tildeM = \barM \sqrt{\barL} \:,\quad \tilde{C} = C \barL \:.
\end{equation}
Investigating~(\ref{getildet}) we find that the
function $\tilde{D}(\tilde{r})$ has the desired profile if and only if
$(\tildeM, \tildeC)$ lies in a certain connected open set whose
boundaries are convex/concave functions.
Using these properties, via the variable transformation relating $(\tildeM, \tildeC)$
and $(K,C)$ for given $(M,\Lambda)$ we can prove
\begin{equation}
\mathscr{KC} = \mathscr{KC}_0 \cup \mathscr{KC}_1 \cup \mathscr{KC}_{-1}\:,
\end{equation}
where the $\mathscr{KC}_i$ are pairwise disjoint connected open sets;
$\mathscr{KC}_{-1}$ arises from $\mathscr{KC}_1$ by inversion at the origin.
$\mathscr{KC}_0$ is the connection component of $(K,C) = (0,0)$,
it is invariant under the inversion, cf.~Fig.~\ref{CKBer}.
$\mathscr{KC}_0$ is enclosed
by the functions $C_t(K)$ and $C_b(K) = -C_t(-K)$, and
the vertical straight lines $K=\pm \sqrt{3 \Lambda}$.

The functions
$C_{b,t}(K)$ are only know implicitly;
the defining equation for $C_{b,t}$ is $D(r_{-}) =0$, or, equivalently,
$\barL r_{-}^3 + 3 \barM = 2 r_-$, where $\barL$, $\barM$, and $r_-$ depend on $\big(K,C_{t,b}(K)\big)$.
In the limit of small $\barL$
we obtain the approximate solutions
\begin{equation}\nonumber
C_{b,t}(K) = \frac{\sqrt{3}}{K^2} \left(\pm 1 \mp  \sqrt{1\mp\sqrt{3} K M}\right)^2
\,\left(\pm 1 \pm \frac{3}{2}\,\frac{\barL}{K^2}
\frac{\left(\pm 1 \mp \sqrt{1\mp \sqrt{3} K M}\right)^2}{\sqrt{1 \mp\sqrt{3} K M}}\right) + O(\barL^2) \:,
\end{equation}
where the upper sign applies to $C_t$, the lower to $C_b$;
note that $|C_b(\sqrt{3 \Lambda})| < |C_t(\sqrt{3 \Lambda})|$.

The functions $C_{b,t}$ are strictly monotonically
increasing. To show this we differentiate the defining equation
$D(r_-) = 0$; we obtain
\begin{equation}\label{Ctmon}
\frac{\partial C_{b,t}}{\partial K} = \frac{1}{3}\, r_{-}^3\big|_{C_{b,t}} \:\,.
\end{equation}
For the sake of completeness we note that
$C_{t}(K)$ is a convex function.

\subsection{$r_{\min}$ and $r_{\max}$ on $\mathscr{KC}_0$}

We investigate
$r_{\min}$ and $r_{\max}$ as functions of $(K,C)\in\overline{\mathscr{KC}_0}$.
In the special case $C=0$, the function $D(r)$ reduces to
$D(r)= 1 -2 M/r -\barL r^2/3$,
thus
\begin{equation}\label{rminmaxxi}
r_{\min} = \frac{2}{\sqrt{\barL}}\: \cos \left(\frac{\xi + \pi}{3}\right)
\,,\quad
r_{\max} = \frac{2}{\sqrt{\barL}}\: \cos \left(\frac{\xi - \pi}{3}\right)
\qquad\text{where}\quad
\cos \xi := 3 M \sqrt{\barL}\:.
\end{equation}
Two subcases deserve special attention: when $K=0$ we have $\cos\xi = 3 M \sqrt{\Lambda}$
and $r_{\min}$, $r_{\max}$ coincide with the horizons of Schwarzschild-de Sitter,
$r_{\min} = r_{\mathrm{b}}$, $r_{\max} = r_{\mathrm{c}}$.
In the limit $K\rightarrow \pm \sqrt{3\Lambda}$,
(\ref{rminmaxxi}) becomes
\begin{equation}\label{rminmaxC=0}
r_{\min} = 2 M + \frac{8}{3} M^3 \barL + O(\barL^2) \qquad,\qquad
r_{\max} = \frac{\sqrt{3}}{\sqrt{\barL}} + O(\sqrt{\barL}) \:.
\end{equation}
Eq.~(\ref{rminmaxxi}) shows that $r_{\min} \downarrow$ and $r_{\max} \uparrow$ when $|K|
\uparrow$;
hence, for $C=0$, for all $K$, $r_{\min} \leq
r_{\mathrm{b}}$ and $r_{\max} \geq r_{\mathrm{c}}$.

By recalling that $D(r) = V(r) + [ K r/3 -C/r^2]^2$, we conclude
from $V(r) > 0 \:\, \forall r\in(r_{\mathrm{b}},r_{\mathrm{c}})$ that
$r_{\min}, r_{\max} \in(r_{\mathrm{b}},r_{\mathrm{c}})$ is excluded.
Since $r_{\min} \leq r_{\mathrm{b}}$ and $r_{\max} \geq r_{\mathrm{c}}$ for $C=0$,
and since $r_{\min}$ and $r_{\max}$ are
continuous functions of $(K,C)$ on the connected domain $\overline{\mathscr{KC}_0}$,
$r_{\min} \leq r_{\mathrm{b}}$ and $r_{\max} \geq r_{\mathrm{c}}$ must hold everywhere on $\overline{\mathscr{KC}_0}$.

\begin{figure}[htp]
    \centering
    \subfigure{%
      \psfrag{K}[cc][cc][0.5][0]{$K$}
      \psfrag{C}[cc][cc][0.5][0]{$C$}
      \psfrag{a}[cc][cc][0.7][0]{\fbox{$\frac{K r_{\min}}{3}- \frac{C}{r_{\min}^2} < 0$}}
      \psfrag{b}[cc][cc][0.7][0]{\fbox{$\frac{K r_{\min}}{3}- \frac{C}{r_{\min}^2} > 0$}}
      \label{divrmin}
      \includegraphics[width=0.45\textwidth]{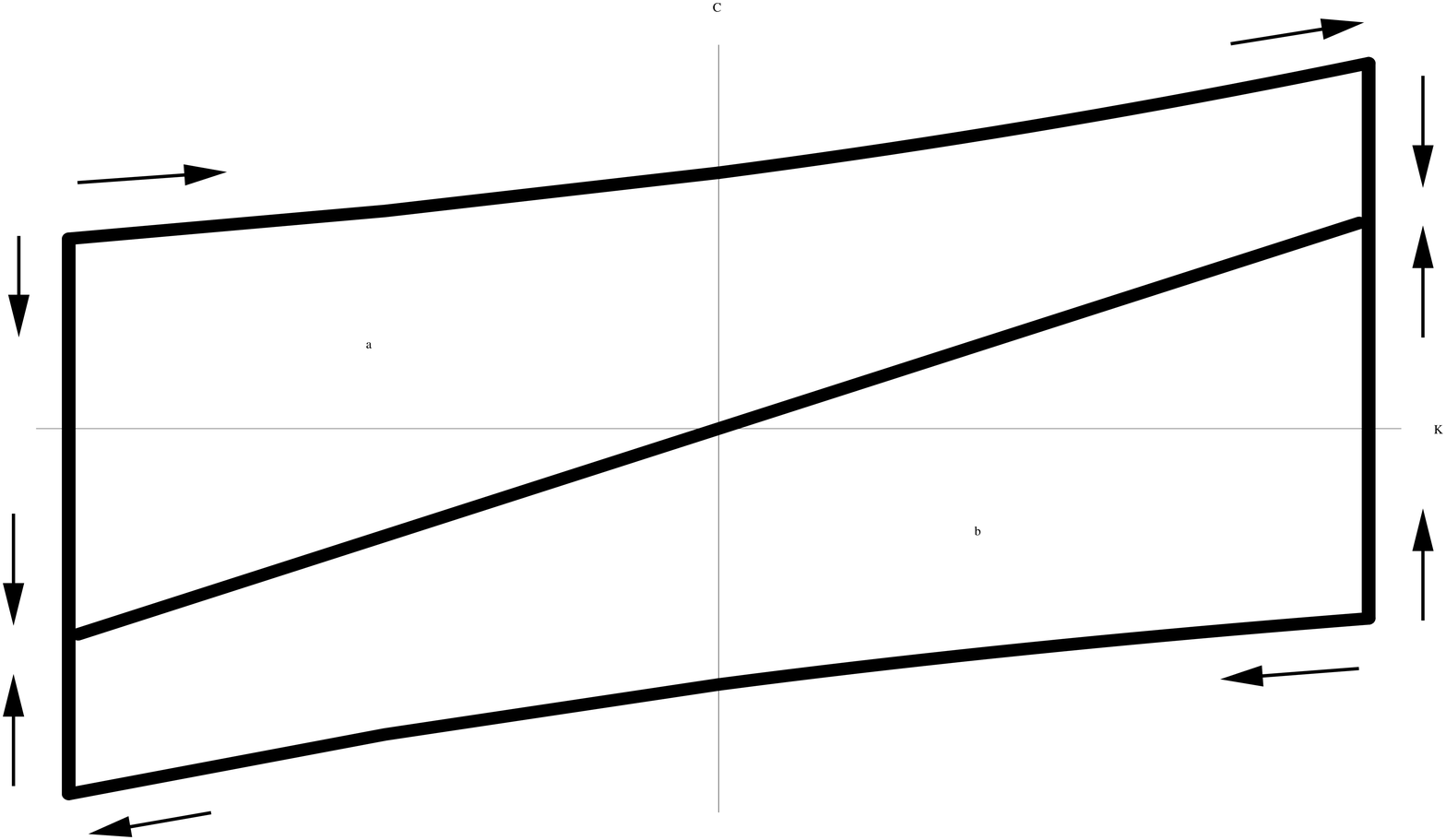}}\qquad
    \subfigure{%
      \psfrag{K}[cc][cc][0.5][0]{$K$}
      \psfrag{C}[cc][cc][0.5][0]{$C$}
      \psfrag{a}[cc][cc][0.7][0]{\fbox{$\frac{K r_{\max}}{3}- \frac{C}{r_{\max}^2} < 0$}}
      \psfrag{b}[cc][cc][0.7][0]{\fbox{$\frac{K r_{\max}}{3}- \frac{C}{r_{\max}^2} > 0$}}
      \label{divrmax}
      \includegraphics[width=0.45\textwidth]{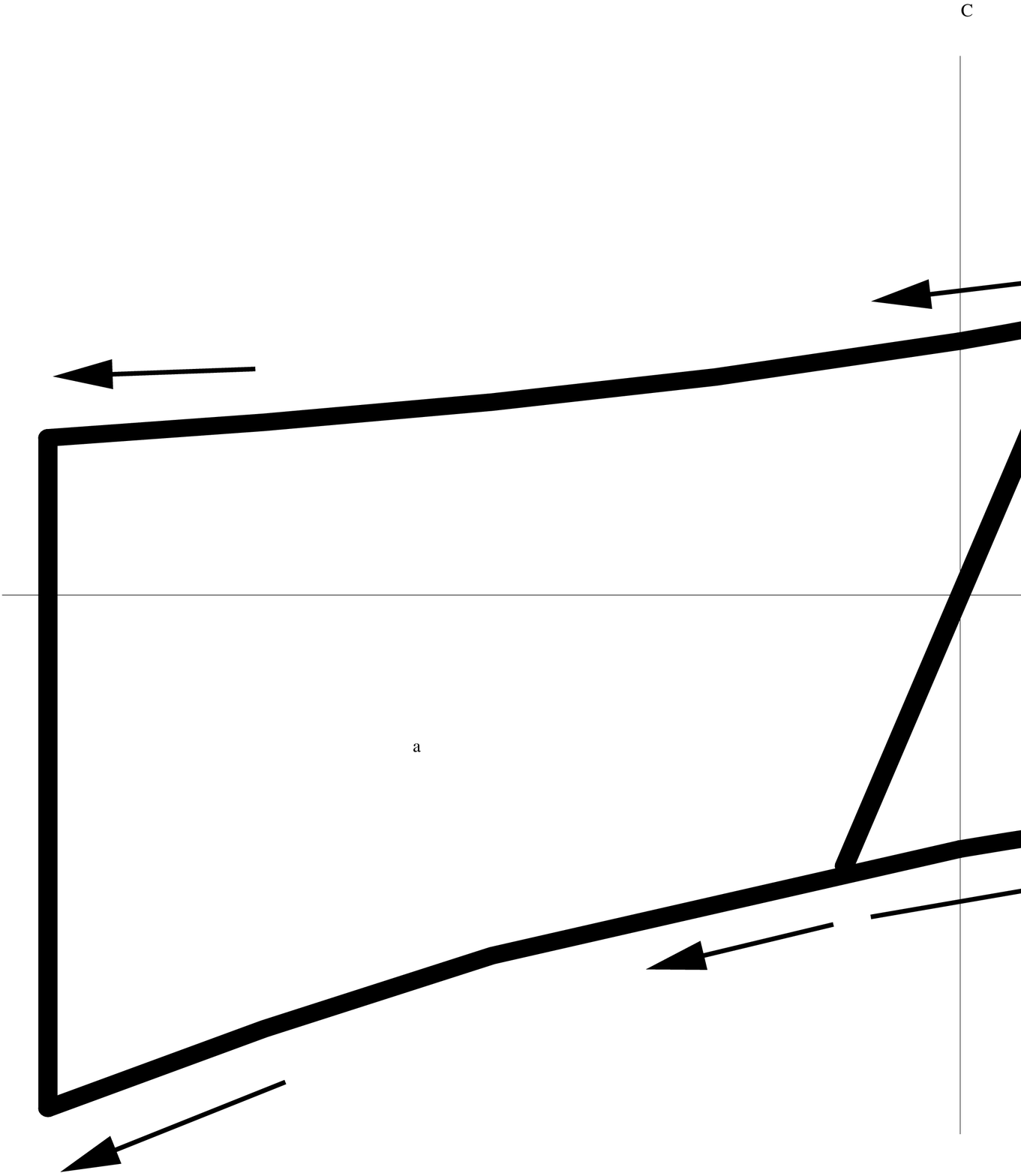}}
    \caption{{\small $r_{\min}$ is maximal, $r_{\min} = r_{\mathrm{b}}$, along a straight line. It
      divides $\overline{\mathscr{KC}_0}$ in two halves, which are
      characterized by $K r_{\min}/3 -C/r_{\min}^2 \gtrless 0$. On the boundaries,
      $r_{\min}$ increases as indicated by the arrows.
      For the first subfigure $\Lambda=1$, $M=1/5$.
      $r_{\max}$ is minimal, $r_{\min} = r_{\mathrm{c}}$, along another straight line, which
      defines the regions $K r_{\max}/3 -C/r_{\max}^2 \gtrless 0$. On the boundaries,
      $r_{\max}$ increases as indicated by the arrows; $r_{\max}\rightarrow \infty$
      ($K\rightarrow\pm\sqrt{3 \Lambda}$).
      For the second subfigure $\Lambda=1$, $M=3/10$.}}
\end{figure}

Suppose that $r_{\min} = r_{\mathrm{b}}$ for some $(K,C)\in \overline{\mathscr{KC}_0}$; it then follows
from $D(r_{\mathrm{b}})=0$ that $C = (r_{\mathrm{b}}^3/3) K$.
Conversely, consider $(K, C)$ with $C=(r_{\mathrm{b}}^3/3) K$; from
\begin{equation}
D(r) =
1 - \frac{2 M}{r} -\frac{\Lambda r^2}{3} + \left( \frac{K r}{3} - \frac{K r_{\mathrm{b}}^3}{3 r^2}\right)^2
\end{equation}
we obtain $D(r_{\mathrm{b}})=0$ and
$dD/d r|_{r_{\mathrm{b}}} > 0$, therefore $r_{\min} = r_{\mathrm{b}}$.
Hence
\begin{equation}\label{lineofmaxrmin}
r_{\min}(K,C) = r_{\mathrm{b}} \qquad\Longleftrightarrow \qquad C = \frac{r_{\mathrm{b}}^3}{3} \,K \:.
\end{equation}
We conclude that there exists a straight line in $\overline{\mathscr{KC}_0}$, given by $C=
(r_{\mathrm{b}}^3/3) K$, along which $r_{\min}$ attains the maximal possible value $r_{\mathrm{b}}$. This straight line
intersects the boundary $\partial(\mathscr{KC}_0)$ in the $K=\pm \sqrt{3\Lambda}$ vertical lines
(and has no intersection with $C_{b,t}$). To establish this result we verify that $(r_{\mathrm{b}}^3/3)
\sqrt{3 \Lambda} < C_t(\sqrt{3\Lambda})$ and we note that $r_{-} < r_{\mathrm{b}}$ everywhere, so that the
slope of $C_t(K)$, cf.~(\ref{Ctmon}), is always less than the slope of the straight line. Hence,
this ``line of maximal $r_{\min}$'' divides $\overline{\mathscr{KC}_0}$ into two regions, an upper
(left) half and a lower (right) half. In each of the two halves $[ K r_{\min}/3 - C/r_{\min}^2 ]
\neq  0$ holds, since $[ K r_{\min}/3 - C/r_{\min}^2 ] = \pm \sqrt{-V(r_{\min})}$; it follows from
the connectedness of the regions and the continuity of the function that $[ K r_{\min}/3 -
C/r_{\min}^2 ] < 0$ in the upper left half and $[ K r_{\min}/3 - C/r_{\min}^2 ] > 0$ in the lower
right half, see Fig.~\ref{divrmin}.

Analogously, we find a straight line of minimal $r_{\max}$ in $\overline{\mathscr{KC}_0}$,
\begin{equation}\label{lineofminrmax}
r_{\max}(K,C) = r_{\mathrm{c}} \qquad\Longleftrightarrow \qquad C = \frac{r_{\mathrm{c}}^3}{3} \,K \:.
\end{equation}
The straight line of minimal $r_{\max}$, $C=(r_{\mathrm{c}}^3/3) K$, intersects $C_{b,t}(K)$ and so
defines an (upper) left half of $\overline{\mathscr{KC}_0}$, where $[ K r_{\max}/3 - C/r_{\max}^2 ] < 0$
and a (lower) right half,
where $[ K r_{\max}/3 - C/r_{\max}^2 ] > 0$ holds, see Fig.~\ref{divrmax}.

Consider a curve $(K,C)(\nu)$ and regard $r_{\min}$
and $r_{\max}$ as functions of $\nu$. By differentiating the equation $D(r_{\min})=0$ w.r.t.\
$\nu$ we obtain
\begin{equation}\label{drminds}
\dot{r}_{\min} =
- 2 \left(\frac{dD}{d r}\big|_{r_{\min}}\right)^{-1}
\left[ \frac{K r_{\min}}{3} -\frac{C}{r_{\min}^2}\right]\:
\left[ \frac{\dot{K} r_{\min}}{3} -\frac{\dot{C}}{r_{\min}^2}\right]\:,
\end{equation}
and the analogous result for $r_{\max}$; the overdot denotes differentiation w.r.t.\ $\nu$.

Consider the boundaries $C_{b,t}$ as parametrized curves $\big(K(\nu), C_{b,t}(K(\nu))\big)$
and assume $\dot{K} > 0$.
By definition, $r_{\min} = r_{-}$ on $C_{b,t}$,
whereby $dD/d r = 0$ at $r_{\min}$, so that~(\ref{drminds}) is not
applicable. However, by differentiating the equation $dD/d r|_{r_{\min}} = 0$
we are able to express $\dot{r}_{\min}$ in terms
of regular expressions: using $\dot{C} = (r_{\min}^3/3) \dot{K}$,
see~(\ref{Ctmon}), we obtain
\begin{equation}\label{drmindsalongCt}
\dot{r}_{\min} =
-2 \left(\frac{d^2D}{d r^2}\big|_{r_{\min}}\right)^{-1} \,
\dot{K}\:
\left[ \frac{K r_{\min}}{3} -\frac{C}{r_{\min}^2}\right]\:\qquad\text{along \,$C_{b,t}$.}
\end{equation}
Since $[K r_{\min}/3 -C/r_{\min}^2] < 0$ on $C_t$ and $d^2D/dr^2 |_{r_{\min}} >0$
we observe that $\dot{r}_{\min} > 0$ along $C_t$;
by virtue of the reflection symmetry, $\dot{r}_{\min} < 0$ along $C_b$.

From the monotonicity properties of $r_{\min}$ along $C_{b,t}$ and
the lines $K=\pm \sqrt{3 \Lambda}$,
we conclude that $r_{\min}$ assumes its global minimum at the point
$(K,C) = (\sqrt{3 \Lambda}, C_b(\sqrt{3 \Lambda}))$
and at the reflected point,
see Fig.~\ref{divrmin}.

Using~(\ref{Ctmon}) in the $r_{\max}$-analogue of~(\ref{drminds}) we get
\begin{equation}
\dot{r}_{\max} =
- 2 \left(\frac{dD}{d r}\big|_{r_{\max}}\right)^{-1}
\left[ \frac{K r_{\max}}{3} -\frac{C}{r_{\max}^2}\right]\:
\frac{\dot{K} r_{\max}}{3} \:
\left[ 1 - \frac{r_{-}^3}{r_{\max}^3}\right]\qquad\text{along \,$C_{b,t}$.}
\end{equation}
We infer the properties
of $r_{\max}$ along $C_{b,t}$ as depicted in Fig.~\ref{divrmax}.
In contrast to $r_{\min}$, $r_{\max}$ is unbounded
as $K \rightarrow \sqrt{3 \Lambda}$.

For small $\barL$ we can calculate $r_{\min}$ and $r_{\max}$
along $C_{b,t}$ up to any desired order of $\barL$:
\begin{equation}\label{rminalongCt}
r_{\min} = \pm\frac{\sqrt{3}}{K} \left(\pm 1 \mp \sqrt{1 \mp \sqrt{3} K M}\right)
\left(\pm 1 +
\frac{3 \sqrt{3} M}{4} \frac{\barL}{K}
\frac{(\pm 1 \mp  \sqrt{1\mp\sqrt{3} K M})}{\sqrt{1\mp\sqrt{3} K M}}\right)
+ O(\barL^2)
\end{equation}
along $C_{b,t}$ for $\barL\rightarrow 0$, where the upper sign applies to $C_{t}$, the lower sign
to $C_b$. Setting $\barL=0$, Eq.~(\ref{rminalongCt}) becomes
\begin{equation}\label{rminextrem}
r_{\min} =  \frac{1}{\sqrt{\Lambda}} \left(\pm 1 \mp \sqrt{1 \mp 3 \sqrt{\Lambda} M}\right)
\qquad
\text{for $K= \sqrt{3 \Lambda}$, $C= C_{b,t}(\sqrt{3\Lambda})$}.
\end{equation}
Finally, along $C_{b,t}$, $r_{\max}$ is given by
\begin{equation}\label{rmaxforfiniteC}
r_{\max} = \frac{2}{\sqrt{\barL}} \cos\left(\frac{\bar{\xi}-\pi}{3}\right) + O(\barL^{3/2})
\qquad\text{where}\quad
\cos \bar{\xi} := 3 \barM \sqrt{\barL}\:,
\end{equation}
which follows when $C^2/r^4$ is neglected against $\barM/r$ in $D(r)$.

\begin{figure}[htp]
    \psfrag{C}[cc][cc]{{\scriptsize $C$}}
    \psfrag{K}[cc][cc]{{\scriptsize $K$}}
    \centering
    \subfigure{\label{linesofrmin}
      \includegraphics[width=0.35\textwidth]{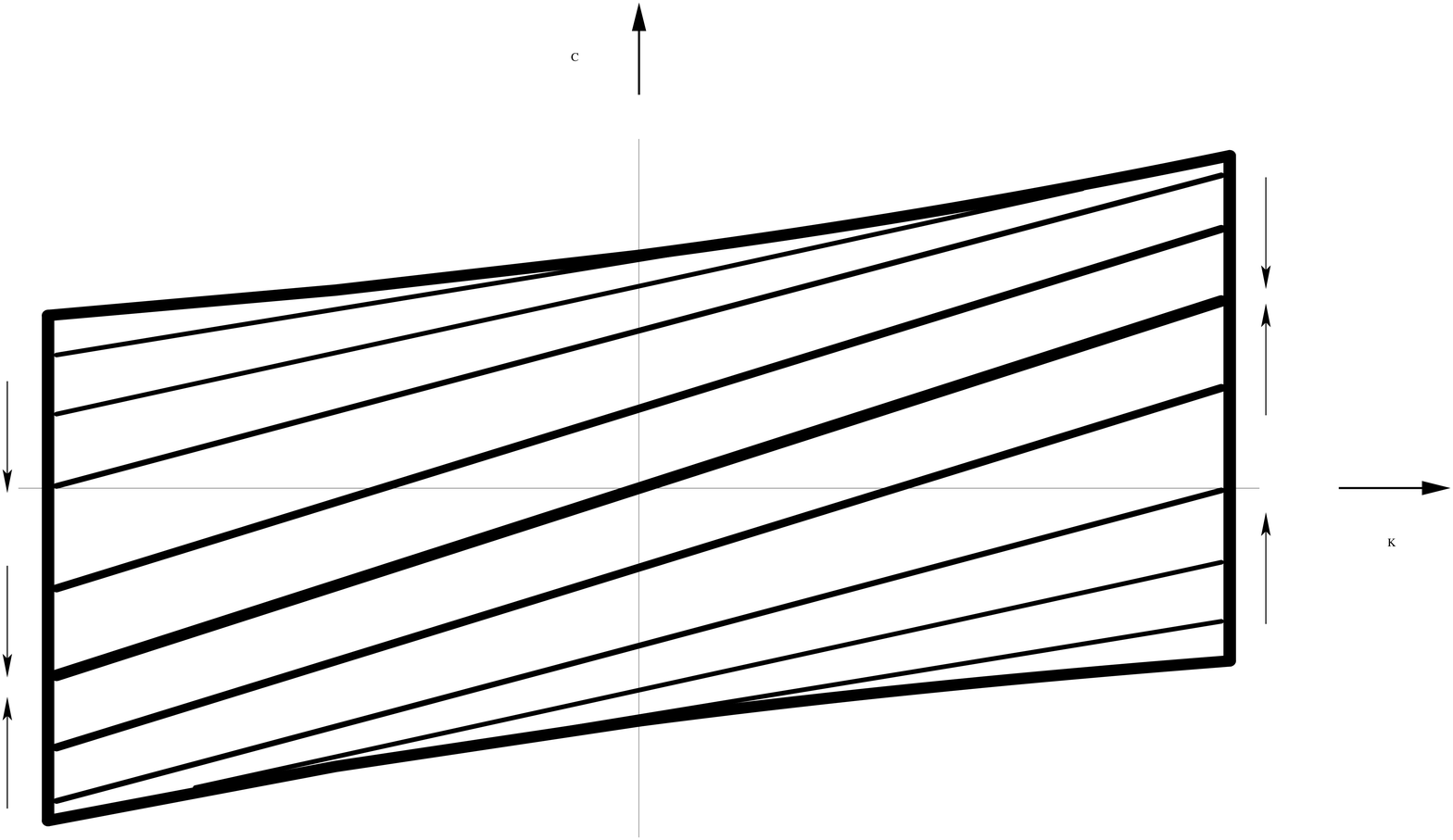}}\qquad
    \subfigure{\label{linesofrmax}
      \includegraphics[width=0.35\textwidth]{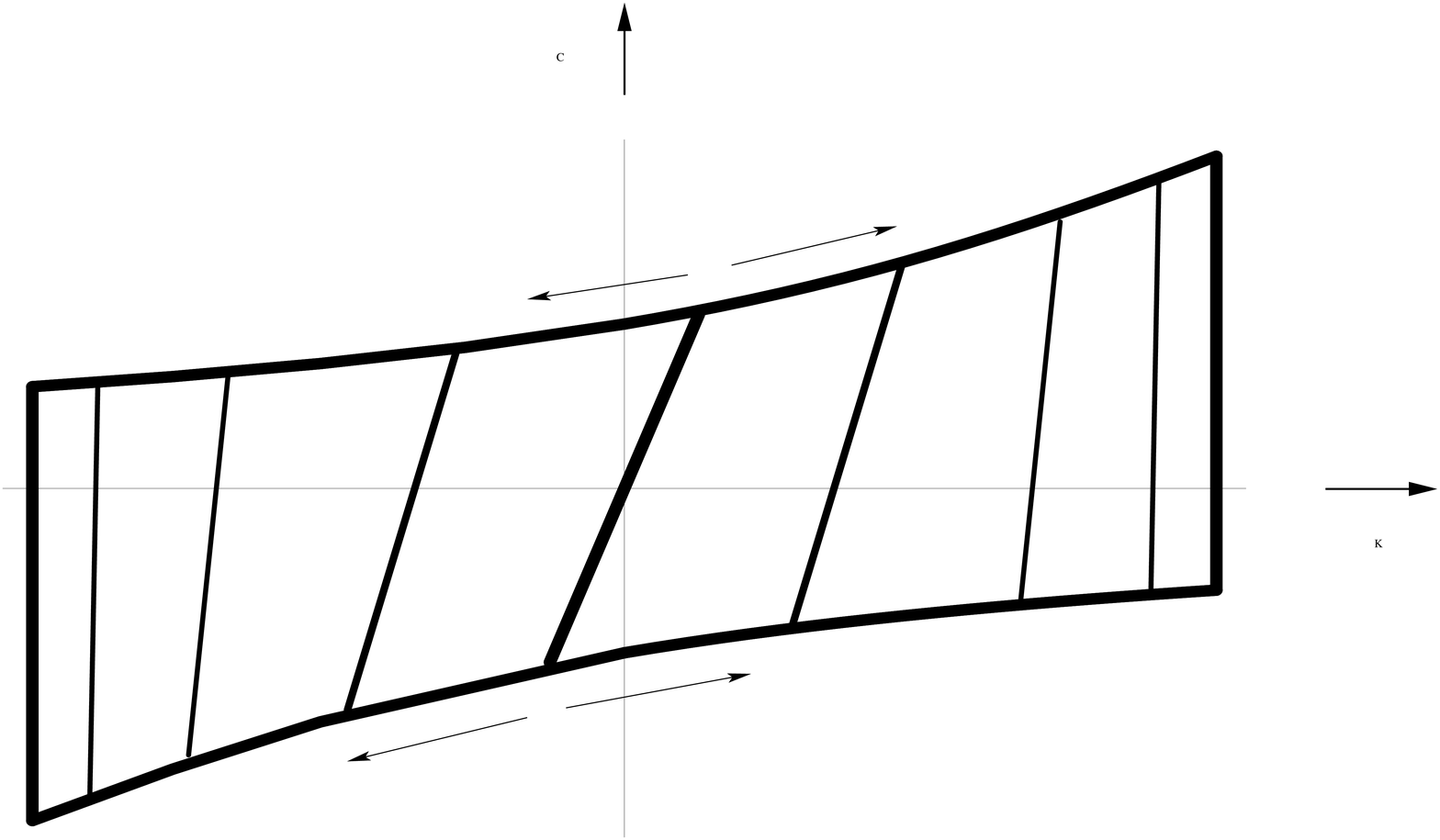}}
    \caption{{\small $r_{\min}$ and $r_{\max}$ are constant along straight lines in $\mathscr{KC}$.
    We chose $\Lambda=1$ and $M=1/5$, $M=3/10$.}}
        \label{linesofrminmax}
\end{figure}

For a given $r_0 \in (\sqrt{\Lambda}^{-1} (1 - \sqrt{1 - 3 \sqrt{\Lambda} M}), r_{\mathrm{b}})$,
the set of all $(K,C)$ such
that $r_{\min}(K,C) = r_0$ is (a segment of) a straight line
in $\overline{\mathscr{KC}_0}$:
$D(r_0) =0$ holds iff $(K,C)$ is chosen according to
\begin{equation}\label{r0straightline}
C = \frac{r_0^3}{3} \, K \mp r_0^2 \sqrt{-V(r_0)}\:;
\end{equation}
moreover, $d D/d r|_{r_0} >0$ holds for (a segment of) the described straight line;
hence $r_{\min} = r_0$ there. The lines of constant $r_{\min}$ are
depicted in Fig.~\ref{linesofrmin}.
Analogously, $r_{\max}$ is constant along straight lines~(\ref{r0straightline}),
where $r_0 \in (r_{\mathrm{c}},\infty)$; see Fig.~\ref{linesofrmax}.

\subsection{Hypersurfaces $r=\mathrm{const}$}

An alternative way of investigating the space $\mathscr{KC}$
is based on an analysis of the CMC-data associated with $r=\mathrm{const}$ hypersurfaces.
In KSSdS the hypersurfaces $r=\mathrm{const}$ are
spacelike hypersurfaces with unit normal $\sqrt{-V(r)} \,\partial/\partial r$
when $r \in (0, r_{\mathrm{b}}) \cup (r_{\mathrm{c}},\infty)$.
A hypersurface $r=\mathrm{const}$ possesses constant mean curvature, the induced CMC-data is represented by the constants $K$, $C$,
\begin{equation}\label{KCrconst}
K = \mp \frac{1}{r \sqrt{-V(r)}}\,\left( -2 +\frac{3 M}{r} + \Lambda r^2\right)\:,
\qquad
C = \pm \frac{r^2}{\sqrt{3}}\, \sqrt{1 - \left(\Lambda - \frac{K^2}{3}\right)\, r^2}\:,
\end{equation}
The induced metric is $d l^2 +r^2 d\Omega^2$ with $l=t \sqrt{-V(r)} \in (-\infty,\infty)$;
identifying
$l=l_0$ with $l=l_1 > l_0$ leads to compact CMC-data, however, no
canonical choice of $l_0$, $l_1$ exists.

The $r=\mathrm{const}$ initial data sets play the role
of borderline cases: $r=\mathrm{const}$ data arises
when the function $D(r)$ possesses a double zero.
We infer that the pair $(K,C)$ of~(\ref{KCrconst}) lies
on the boundary of $\mathscr{KC}_0$:
\begin{equation}\label{rconstint}
(K,C) \in C_{b,t} \qquad\text{when}\quad
\frac{1}{\sqrt{\Lambda}} \left( 1 - \sqrt{1 - 3 \sqrt{\Lambda} M}\right) < r <
\frac{1}{\sqrt{\Lambda}} \left( -1 +  \sqrt{1 + 3 \sqrt{\Lambda} M}\right) \:.
\end{equation}
Eqs.~(\ref{KCrconst}) thus constitute a parametric representation of the boundaries $C_{b,t}$ of
$\mathscr{KC}_0$.

\section{The lapse equation}
\label{thelapseequation}

We consider the equation
\begin{equation}\label{homeq}
\Delta \,\alpha + a\,\alpha \:=\: 0 \:,
\qquad\text{i.e.}\qquad
\alpha^{\prime\prime} + \frac{2}{r}\:r^\prime \alpha^\prime + a \,\alpha = 0 \:,
\end{equation}
where $a(l) = \Lambda - K^2/3 - 6 C^2/r^6(l)$;
note that there do not exist pairs $(K , C)\in \mathscr{KC}_0$
such that $a(l)$ is non-positive, which follows from algebraic computations based on the
results of~Appendix~\ref{KCspaceapp}.
We seek the general solution on the domain $\mathbb{R}$.
Equation~(\ref{homeq}) is homogeneous with
periodic, even and odd coefficient functions.
The general solution of~(\ref{homeq}) is a superposition of two principal solutions
with symmetry properties:
\begin{itemize}
\item Let $\alpha_0(l)$ denote the function that solves~(\ref{homeq})
with initial conditions $\alpha_0(0) = 1$ and $\alpha_0^\prime(0) = 0$; it is even.
\item The function $\alpha_\xi(l) = r^\prime(l)$ is odd; it solves~(\ref{homeq}) with initial conditions
$\alpha_\xi(0)=0$, $\alpha_\xi^{\prime}(0) = r^{\prime\prime}_{\min}$.
\end{itemize}
The function $\alpha_\xi$ is positive
in $(0,L)$ but negative in $(-L,0)$; $\alpha_\xi(0) = \alpha_\xi(L) =0$;
$\alpha_\xi$ is a periodic function.
Geometrically speaking, $\alpha_\xi(l) = r^\prime(l)$
is the lapse function associated with the Killing vector $\xi$,
cf.~(\ref{KID}).

To investigate the general solution of~(\ref{homeq}) we use
Floquet's theorem.
Let us write~(\ref{homeq}) as a system of first order,
\begin{equation}\label{eqfirstorder}
\begin{pmatrix}
\alpha \\
\alpha^\prime
\end{pmatrix}^\prime
=
\begin{pmatrix}
0 & 1 \\
-a & -2 r^\prime/r
\end{pmatrix}
\:
\begin{pmatrix}
\alpha \\
\alpha^\prime
\end{pmatrix}
=
A(l)
\begin{pmatrix}
\alpha \\
\alpha^\prime
\end{pmatrix}
\ .
\end{equation}
The principal solution matrix is given by
\begin{equation}\label{solmat}
\Pi(l,0) =
\left(\begin{array}{cc}
\alpha_0(l) & r^\prime(l)/r^{\prime\prime}_{\min} \\
\alpha_0^\prime(l) &  r^{\prime\prime}(l)/r^{\prime\prime}_{\min}
\end{array}\right)\,,
\qquad
\Pi^\prime(l,0) = A(l) \Pi(l,0)\,,
\qquad
\Pi(0,0) =
\left(\begin{array}{cc}
1 & 0\\
0 & 1
\end{array}\right)\ .
\end{equation}
The Wronskian $W(l,0) = \det \Pi(l,0)$ satisfies the equation
$W^\prime(l,0) = (\mathrm{tr} A(l))\, W(l,0)$, thus
\begin{equation}
W(l,0) = \det \Pi(l,0) = \exp\int_0^l \mathrm{tr} A(\tilde{l}) \: d\tilde{l} = r_{\min}^2/r(l)^2\:.
\end{equation}
From $W(2 L, 0) =1$ we see that the so-called monodromy matrix $\Pi(2 L,0)$ fulfills
\begin{equation}\label{Pi2L0}
\Pi(2 L,0) =
\left(\begin{array}{cc}
1 & 0\\
\alpha^\prime_0(2 L) & 1
\end{array}\right)\:.
\end{equation}
Floquet's theorem states that
$\Pi(l,0) = P(l,0) \exp ( l Q(0) )$, where $P(l,0)$ is a periodic matrix
and $Q(0)$ is such that $\exp(2 L Q(0) ) = \Pi(2 L,0)$.
Thus,
\begin{equation}
\Pi(l, 0) = P(l, 0) \exp \left[l\, \left(\begin{array}{cc}
0 & 0\\
\alpha^\prime_0(2 L)/(2 L) & 0
\end{array}\right)\right] =
P(l,0)
\left(\begin{array}{cc}
1 & 0\,\\
(l \alpha^\prime_0(2 L))/(2 L) & 1
\end{array}\right)
\:.
\end{equation}
It follows that
\begin{equation}\label{alpha0peri}
\alpha_0(2 n L + l) = \alpha_0(l) + n \,[\alpha_0^\prime(2 L)/r^{\prime\prime}_{\min}]\, \alpha_\xi(l)
\qquad \forall l\:,\quad \forall n\in\mathbb{Z}\:,
\end{equation}
in particular, $\alpha_0(2 n L) =\alpha_0(0)=1$.
When we employ the symmetry properties of the coefficients in
the differential equation~(\ref{eqfirstorder}) we are able to also
establish a direct relation between the initial data at $l=0$ and
the solution at the half-period $l=L$:
\begin{equation}\label{PiL0}
\Pi(L, 0) = \begin{pmatrix}
\frac{r^{\prime\prime}_{\min}}{r^{\prime\prime}_{\max}}\: \frac{r_{\min}^2}{r_{\max}^2} & 0\\
\alpha^\prime_0(L) &  \frac{r^{\prime\prime}_{\max}}{r^{\prime\prime}_{\min}}
\end{pmatrix} \:, \qquad \text{where}
\quad \alpha^\prime_0(2 L) = 2 \:\alpha^\prime_0(L) \,\frac{r^{\prime\prime}_{\min}}{r^{\prime\prime}_{\max}}\:.
\end{equation}
We infer that $\alpha_0(l)$ (and thus the general solution of~(\ref{homeq})) is periodic
iff $\alpha_0^\prime(L) = 0$.
We prove $\alpha^\prime_0(L)\neq 0$, which implies that the even solutions of~(\ref{homeq})
are not periodic, so that Lemma~\ref{isomlemma} is established.

We analyze an explicit representation of $\alpha_0$, which we obtain via
an ansatz $\alpha_0(l) = \beta(l) r^\prime(l)$; we get
\begin{equation}\label{alpha0stamm}
\alpha_0(l) = -\dot{C} r^\prime \int\limits^l \frac{1}{r^2 r^{\prime\,2}} d l \,+\, k r^\prime \:,
\end{equation}
where $\dot{C}$ abbreviates $\dot{C}= r_{\min}^2 r^{\prime\prime}_{\min}$,
and $k = \mathrm{const}$ is such that the function is even.
The integral diverges as $l\rightarrow 0$ and $l\rightarrow \pm L$; however, by applying
de l'Hospital's rule it is confirmed that $\alpha_0(0) =1$ and
$\alpha_0(\pm L) = \pm (r^{\prime\prime}_{\min}/r^{\prime\prime}_{\max})\:(r_{\min}^2/r_{\max}^2)$,
cf.~(\ref{PiL0}).

When we view (the first half-period of) the function
$\alpha_0$ as a function of $r$ we can write
\begin{equation}\label{alpha0inr}
\alpha_0(r) = \dot{C} \sqrt{D(r)}
\left(-\int\limits^r \frac{1}{\hat{r}^2} \frac{1}{D(\hat{r})^{3/2}} d \hat{r} + k\right)\:.
\end{equation}
We introduce $\delta(r)$ by defining
$\delta(r) = r^4 D(r) (r-r_{\min})^{-1} (r_{\max}-r)^{-1}$; the function $\delta(r)$ is
positive in $[r_{\min},r_{\max}]$.
Define
\begin{equation}
\phi(r) = \frac{4 r - 2 (r_{\max}+ r_{\min})}{(r_{\max}-r_{\min})^2 \sqrt{(r-r_{\min})(r_{\max}-r)}}
= \int \frac{1}{(r-r_{\min})^{3/2} (r_{\max}-r)^{3/2}} d r\:,
\end{equation}
then $\alpha_0(r)$ becomes
\begin{equation}\label{alpharstamm}
\alpha_0(r) = -\dot{C} \sqrt{D(r)} \phi(r) r^4 \delta(r)^{-3/2} +
\dot{C} \sqrt{D(r)} \int_{r_{\min}}^r \phi(r) \left( r^4 \delta(r)^{-3/2}\right)^\prime\, d r\:,
\end{equation}
where the prime denotes differentiation w.r.t.\ $r$ in the present context.
Note that $\sqrt{D(r)} \phi(r)$ is a regular bounded function so that
the first term in~(\ref{alpharstamm}) is regular and bounded.
Using that $d\alpha_0/d l = \sqrt{D}\, d\alpha_0/d r$ we obtain by differentiating~(\ref{alpharstamm})
\begin{equation}
\alpha_0^\prime(L) =
\frac{\dot{C}}{2} D^\prime(r_{\min})
\int\limits_{r_{\min}}^{r_{\max}} \phi(r) \left( r^4 \delta(r)^{-3/2}\right)^\prime\, d r
=
-\frac{\dot{C}}{2} D^\prime(r_{\min})
\int\limits_{r_{\min}}^{r_{\max}} \Phi(r) \left( r^4 \delta(r)^{-3/2}\right)^{\prime\prime} \,d r\:,
\end{equation}
where $\Phi(r) = - 4 \sqrt{(r-r_{\min})(r_{\max}-r)} (r_{\max}-r_{\min})^{-2}$
is the integral of $\phi(r)$.
Therefore, since $\Phi(r) < 0$ in $(r_{\min},r_{\max})$,
to show that $\alpha_0^\prime(L)\neq 0$ it is sufficient to show
that the function $r^4 \delta(r)^{-3/2}$ is convex.

The special case $C=0$ is easy to treat.
In this case, $\delta(r) = (\barL/3) r^3 (r-r_{\rm{neg}})$, where $r_{\rm{neg}} < 0$,
hence
\begin{equation}
\left( r^4 \delta(r)^{-3/2}\right)^{\prime\prime} \propto
\left( \frac{1}{\sqrt{r \,(r-r_{\rm{neg}})^3\,}} \right)^{\prime\prime}
=
\frac{3 (r-r_{\rm{neg}})^4 ( 8 r^2 - 4 r_{\rm{neg}} r + r_{\rm{neg}}^2)}%
{4 [r (r-r_{\rm{neg}})^3]^{5/2}} > 0 \:,
\end{equation}
i.e.\ $r^4 \delta(r)^{-3/2}$ is convex.

In the general case we show that $r^3 \delta(r)^{-1}$ is convex.
Then, $r^{8/3} \delta^{-1} = r^{-1/3} r^3 \delta(r)^{-1}$ is convex,
since $r^3 \delta(r)^{-1}$ is decreasing, and consequently
$r^4 \delta(r)^{-3/2} = (r^{8/3} \delta^{-1})^{3/2}$ is convex.
An in-depth analysis of the properties of the
zeros of $\delta(r)$ is
essential to establish the claim.
Asymptotically, for $C\rightarrow C_{b,t}$,
the zeros $r_{\min}$ and $r_{\max}$ of $D(r)$
coincide to form the double zero $r_-$, which is know explicitly, see~(\ref{rcupcap}).
Hence $\delta(r)$ is known explicitly for $C\rightarrow C_{b,t}$,
\begin{subequations}
\begin{align}
\delta(r) & = -\frac{r_0}{3}\, \frac{(r_0-3 M)^2}{3 r_0^2 V(r_0)} \:\pi\big(\frac{r}{r_0}\big) \:, \\
\text{where} \quad\pi(x) & = 1 + 2 x + 3 x^2 + 2 F x^3 + F x^4
\quad \text{with}\quad
F = 1 +\frac{3 r_0^2 V(r_0)}{(r_0-3 M)^2}\:,
\end{align}
\end{subequations}
and convexity of $r^3 \delta(r)^{-1}$ can be established;
here, $r_0$ is such that $(K(r_0),C(r_0))$ describes a point on $C_{b,t}$
via~(\ref{KCrconst}).
Since $\delta(r)$ is a fourth-order
polynomial, also its zeros are known explicitly
for $C\rightarrow C_{b,t}$.
Combining this with an analysis of the variation of the zeros
of $\delta$ as $C$ varies, the claim can be established;
the details are omitted here.

In order to differentiate~(\ref{alpha0stamm}), i.e.\ to be able
to write down $\alpha_0^\prime(2 L)$ in terms of quadratures,
the integral must be regularized appropriately.
The integral representation~(\ref{alpha0stamm}) of $\alpha_0$
can be ``regularized'' in several ways, e.g.\
\begin{subequations}\label{alpha0reg}
\begin{align}
\label{alpha0reg1}
\alpha_0 & = 1 - r^\prime \int\limits_0^l \frac{1}{r^{\prime\,2}}
\,\left( r^{\prime\prime}_{\min}  \frac{r_{\min}^2}{r^2} -r^{\prime\prime}\right)\,d l \:, \\
\label{alpha0reg2}
\alpha_0 & = \frac{r_{\min}^2}{r^2} + r_{\min}^2 r^\prime
\int\limits_0^l \frac{2}{r^3} d l -
r_{\min}^2 r^\prime
\int\limits_0^l \frac{1}{r^2 r^{\prime \,2}}
\left(r^{\prime\prime}_{\min} -r^{\prime\prime}\right)d l\:,
\end{align}
\end{subequations}
where the integrands are now regular at $l=0$.
When we differentiate the expression~(\ref{alpha0reg2}), and
manipulate the arising terms so that divergencies cancel, we obtain
\begin{equation}
\begin{split}
\frac{\alpha_0^\prime r^{\prime\prime}_{\max}}{r_{\min}^2} =
-\frac{1}{r^2 r^\prime}
(r^{\prime\prime}_{\max} - r^{\prime\prime})(r^{\prime\prime}_{\min} - r^{\prime\prime})
+
r^{\prime\prime} \int\limits_0^l \frac{2}{r^3}
\left( r^{\prime\prime}_{\min} + r^{\prime\prime}_{\max} -r^{\prime\prime} \right) d l\:
+ \quad \\
+ \,
r^{\prime\prime}
\int\limits_0^l \frac{r^{\prime\prime\prime}}{r^2 r^\prime} d l
-
r^{\prime\prime}
\int\limits_0^l \frac{1}{r^2 r^{\prime\, 2}} (r^{\prime\prime}_{\max}-r^{\prime\prime})
(r^{\prime\prime}_{\min}-r^{\prime\prime}) d l\:,
\end{split}
\end{equation}
which can be evaluated at $L$ to obtain $\alpha_0^\prime(L)$ and thus $\alpha_0^\prime(2 L)$;
the equation is mainly useful for numerical purposes.

We consider now the lapse equation
\begin{equation}\label{inhomeq}
\Delta \,\alpha + a\,\alpha \:=\: \dot{K }\:,
\qquad\text{i.e.}\qquad
\alpha^{\prime\prime} + \frac{2}{r}\:r^\prime \alpha^\prime + a \,\alpha = \dot{K} =\mathrm{const}\:,
\end{equation}
where $a(l) = \Lambda - K^2/3 - 6 C^2/r^6(l)$, cf.~(\ref{lapseeq}).
In first order from
the system corresponds to~(\ref{eqfirstorder}) with an additional inhomogeneity $(0,\dot{K})$.
Using the principal solution matrix~(\ref{solmat}) of the homogeneous system,
we obtain
\begin{equation}\label{solinhom}
\begin{pmatrix}
\alpha \\
\alpha^\prime
\end{pmatrix}
=
\Pi(l,0)
\begin{pmatrix}
\alpha \\
\alpha^\prime
\end{pmatrix}(0)
+
\int\limits_0^l \Pi(l,s)
\begin{pmatrix}
0 \\ \dot{K}
\end{pmatrix}
\,d s
\end{equation}
by the method of variation of constants.

Corollary~\ref{lapsecorr} states that there exists a unique even periodic solution of~(\ref{inhomeq});
we give an alternative argument here.
Eq.~(\ref{solinhom}) describes a periodic function if and only if
$(\alpha,\alpha^\prime)(2 L) = (\alpha,\alpha^\prime)(0)$, i.e.\ iff
\begin{equation}\label{periocondi}
\big(\mathrm{id} - \Pi(2 L,0)\big)\,
\begin{pmatrix}
\alpha \\ \alpha^\prime
\end{pmatrix}(0)
=
\int\limits_0^{2 L} \Pi(2 L, s)
\begin{pmatrix}
0 \\ \dot{K}
\end{pmatrix}
\,d s \:,
\end{equation}
where $\Pi(2 L,0)$ is given by~(\ref{Pi2L0});
$\Pi(2 L,s)$ can be computed easily,
\begin{equation}
\Pi(2 L, s) = \Pi(2 L,0) \Pi(s, 0)^{-1}
=
\frac{r(s)^2}{r_{\min}^2}\,
\begin{pmatrix}
1 & 0 \\
\alpha_0^\prime(2 L) & 1
\end{pmatrix}\,
\begin{pmatrix}
\frac{r^{\prime\prime}(s)}{r^{\prime\prime}_{\min}} & -\frac{r^\prime(s)}{r^{\prime\prime}_{\min}} \\
- \alpha_0^\prime(s) & \alpha_0(s)
\end{pmatrix}\:.
\end{equation}
Integration of the first component of $\Pi(2 L, s) \,(0,1)$ yields
\begin{equation}
\int\limits_0^{2 L} \frac{r^2}{r_{\min}^2} \left(- \frac{r^\prime}{r^{\prime\prime}_{\min}}\right)\:d s =
-\frac{1}{r_{\min}^2 r^{\prime\prime}_{\min}} \:\frac{r^3}{3}\: \Big|_0^{2 L} = 0\:,
\end{equation}
i.e.\ the first component of Eq.~(\ref{periocondi}) is satisfied identically.
The second condition is a condition for $\alpha(0)$; it is fulfilled iff
\begin{equation}\label{al0}
\alpha(0) =
-\frac{\dot{K}}{\alpha^\prime_0(2 L)} \frac{1}{r_{\min}^2} \int\limits_0^{2 L} r(s)^2 \alpha_0(s) d s =: \alpha_{\min} \:.
\end{equation}
Eq.~(\ref{periocondi}) does not impose a condition on $\alpha^\prime(0)$;
however, when $\alpha$ is required to be even, $\alpha^\prime(0) =0$ is necessary.
Hence we have reproduced the result that
there exists a unique even periodic solution of~(\ref{inhomeq}).

The general solution of~(\ref{inhomeq}) is the linear combination
$\hat{\alpha} = \alpha + k_0 \alpha_0 + k_\xi \alpha_\xi$, where
$k_0$ and $k_\xi$ are constants;
\begin{equation}\label{alphagen}
\hat{\alpha}(2 n L + l) = \alpha(l) + k_0 \alpha_0(l) +
[k_\xi + k_0 n \alpha_0^\prime(2 L)/r^{\prime\prime}_{\min}]\, \alpha_\xi(l) \qquad
\forall l, \forall n \in \mathbb{Z}
\end{equation}
follows from~(\ref{alpha0peri}).
When $k_0 = 0$, $k_\xi \neq 0$, the solution is periodic but not even;
when $k_0 \neq 0$, $k_\xi = 0$, the solution is even but not periodic.

To express $\alpha_{\min}$, cf.~(\ref{al0}), in terms
of quadratures we may use the regularization~(\ref{alpha0reg2}).
We obtain
\begin{equation}
\nonumber
\frac{\alpha_{\min}}{\dot{K}} = -\frac{2}{3 \alpha_0^{\prime}(2 L)}
\left[ L + 2 r_{\max}^3 \int\limits_0^L\frac{d l}{r^3} -
\int\limits_0^L\frac{(r_{\max}^3 -r^3)(r_{\min}^{\prime\prime}-r^{\prime\prime})}{r^2 r^{\prime\:2}} d l\right]
-\frac{1}{3}\frac{r_{\min}^3-r_{\max}^3}{r_{\min}^{\prime\prime}}\:.
\end{equation}
By using~(\ref{alpha0reg1}) and slightly different conventions we derive
\begin{equation}\label{alphaminnulliff}
\alpha_{\min} = 0 \quad\Leftrightarrow\quad
\int\limits_{r_{\min}}^{r_{\max}} \frac{r^2 d r}{\sqrt{D(r)}} -
\frac{1}{6} \int\limits_{r_{\min}}^{r_{\max}} \frac{r_{\min}^3 - r^3}{D^{3/2}(r)}
\left(D^\prime_{\max} \frac{r_{\max}^2}{r^2} - D^\prime(r)\right)d r = 0 \:,
\end{equation}
where $D^\prime(r) = d D(r)/d r$ and $D^\prime_{\max} = D^\prime(r_{\max})$.

We conclude this section by proving the claim made in the proof of Theorem~\ref{uniqueslicing},
i.e.\ that the system
\begin{equation}\label{KdotCdotsys}
\Delta \beta + a \beta = \dot{K} \:,\qquad
\Delta \beta -  3  \frac{r^\prime}{r} \beta^\prime - \frac{3 b}{r^3} \beta = \frac{3 \dot{C}}{r^3} \:,
\end{equation}
where $a(l) = \Lambda - K^2/3 - 6 C^2/r^6(l)$, and $b(l) = M + K C/3 -4 C^2/r^3(l)$,
has a unique even solution $\beta(l)$ (on the domain $\mathbb{R}$)
for given even functions $\dot{K}$ and $\dot{C}$. It is straightforward to see
that the system~(\ref{KdotCdotsys}) is equivalent to the equation
\begin{equation}
r^\prime \beta^\prime - r^{\prime\prime} \beta =
\frac{\dot{K} r}{3} -\frac{\dot{C}}{r^2}\:,
\end{equation}
which we have encountered already in~(\ref{Cdoteq}).
Since the coefficient $r^\prime$ is odd, $r^{\prime\prime}$ even,
there exists a unique solution $\beta$ that is even;
the general solution $\hat{\beta}$ is a linear
combination $\hat{\beta} = \beta + \mathrm{const}\: r^\prime$.

\section{The space $\mathscr{KC}_{0+}$ and foliations $\mathcal{S}_\tau$}
\label{foliapp}

In this section we discuss analytical and numerical results concerning the question of when a
compact CMC-slicing $\mathcal{S}_\tau$ in the spacetime KSSdS[T] is a foliation. These results
strengthen the statement of Theorem~\ref{foliationthm}.

Let $\alpha(K,C;l)$ denote the unique even solution of the lapse equation~(\ref{lapseeq})
associated with $(K,C)\in\mathscr{KC}_0$ and a given constant $\dot{K}>0$, 
cf.~Corollary~\ref{lapsecorr}. We make the following
\begin{definition}
$\mathscr{KC}_{0+}$ is defined as
the set of all $(K,C)\in\mathscr{KC}_0$ such that
the associated lapse function $\alpha(K,C;l)$ is positive.
\end{definition}

\begin{proposition}
There exists a neighborhood $W$ of the line $C=0$ in $\mathscr{KC}_0$ such that
\begin{equation}
W \subseteq \mathscr{KC}_{0+}\:.
\end{equation}
\end{proposition}

\proof
The proof is similar to the proof of Theorem~\ref{foliationthm}:
$\alpha_{\min}(K,C)$ and thus $\alpha(K,C;l)$ depends continuously
on $(K,C)\in\mathscr{KC}_0$,
when $\dot{K}$ is a given constant (or a continuous function on $\mathscr{KC}_0$),
see Appendix~\ref{thelapseequation}.
Hence, since $\alpha(K,C;l) =\mathrm{const} > 0$ on the line $C=0$, cf.~(\ref{alphaCeq0}),
there exists a neighborhood $W$
of $C=0$ which is contained in $\mathscr{KC}_{0+}$.
\proofend

\begin{proposition}\label{KC0+Prop}
There exists a neighborhood $V(C_b)$ in $\overline{\mathscr{KC}_0}$ of the curve
$\big\{(K,C_b(K))\:|$ \linebreak $|\:K\in(-\sqrt{3\Lambda},\sqrt{3\Lambda})\big\}$
and a neighborhood $V(C_t)$
of $\{(K,C_t(K))\:|\: K\in(-\sqrt{3\Lambda},\sqrt{3\Lambda})\}$ such that
\begin{equation}
V(C_b) \cap \mathscr{KC}_{0+} = \emptyset \:,\qquad
V(C_t) \cap \mathscr{KC}_{0+} = \emptyset \:.
\end{equation}
There exists a neighborhood $U(\sqrt{3\Lambda})$ of the line
$\{(\sqrt{3\Lambda},C)\:|\: C\in(C_b(\sqrt{3\Lambda}),C_t(\sqrt{3\Lambda}))\}$
in $\overline{\mathscr{KC}_0}$ and an analogous neighborhood $U(-\sqrt{3\Lambda})$
of 
$\{(-\sqrt{3\Lambda},C)\:|\: C\in(C_b(-\sqrt{3\Lambda}),C_t(-\sqrt{3\Lambda}))\}$
such that
\begin{equation}
U(\sqrt{3\Lambda}) \cap \mathscr{KC}_0 \subseteq \mathscr{KC}_{0+}\:, \qquad
U(-\sqrt{3\Lambda}) \cap \mathscr{KC}_0 \subseteq \mathscr{KC}_{0+} \:.
\end{equation}
\end{proposition}

\proof
In KSSdS[T] consider the foliation of the black hole region by $r=\mathrm{const}$
hypersurfaces. Recall that $r=r_0= \mathrm{const}$ is a CMC-hypersurface with metric $d l^2 + r_0^2
d\Omega^2$ and $K= K_0$, $C=C_0$ given by~(\ref{KCrconst}). The lapse function $\alpha_r$ of the
$r=\mathrm{const}$ foliation at $r=r_0$ is given by
\begin{equation}
\alpha_r(l) \equiv \frac{\dot{K}}{a(r_0)} \:,\qquad\quad
\text{where}\quad  a(r_0) = \Lambda - \frac{K_0^2}{3} -\frac{6 C_0^2}{r_0^6} =\mathrm{const}\:.
\end{equation}
It is a solution of the lapse equation~(\ref{lapseeq}), where $r^\prime=0$, i.e. $\alpha_r$ solves
$\alpha_r^{\prime\prime} + a(r_0) \alpha_r = \dot{K}$. We consider a hypersurface $r=r_0$ such that
$r_0$ lies in the interval given in~(\ref{rconstint}), so that we have $K_0\in (-\sqrt{3
\Lambda},\sqrt{3 \Lambda})$ and (without loss of generality) $C_0 = C_t(K_0)$.

Consider a neighborhood of $(K_0, C_0)$ in $\overline{\mathscr{KC}_0}$.
Choose a pair $(K,C)\in\mathscr{KC}_0$ of that neighborhood and consider the 
associated CMC-hypersurface which is determined by the embedding $t(r)$, see~(\ref{tofl}).
For all small $\epsilon > 0$, for all large $E>0$,
\begin{equation}\label{tEeps12}
|t(r_{\min}(K,C)+\epsilon)| =
\left|\int\limits_{r_{\min}(K,C)}^{r_{\min}(K,C)+\epsilon}
\frac{V^{-1}}{\sqrt{D(K,C;r)}}\:\left(\frac{K r}{3} -\frac{C}{r^2}\right) d r \right| > E\:
\end{equation}
and $|r_{\min}(K,C) - r_0| < \epsilon$, provided that $(K,C)$ is sufficiently close to $(K_0,C_0)$.
This follows from~(\ref{tofl}) using the same techniques
as in the proof of Lemma~\ref{Tlemma} and Proposition~\ref{slicingends}.
In the subset $t\in [-E,E]$ of the black hole region,
the (monotonic) function $r(t)$ thus satisfies 
$|r(t) - r_0| + | \partial r/\partial t| \leq 3 \epsilon$,
if $(K,C)$ is sufficiently close to $(K_0,C_0)$,
i.e., $r(t)$ is approximately constant for $t\in [-E,E]$.

Let, now, $\mathcal{S}$ be a
compact CMC-hypersurface whose values $(K_{\mathcal{S}},C_{\mathcal{S}})$ are sufficiently close to
$(K_0, C_0)$, i.e.\
$\big|(K_{\mathcal{S}},C_{\mathcal{S}})- (K_0, C_0)\big| < \delta$;
$\mathcal{S}$ can be approximated by the hypersurface
$r = r_0$ in a region $t\in[-E,E]$ of the black hole.
Consider the compact CMC-slicing $\mathcal{S}_\tau$ that contains the CMC-hypersurface
$\mathcal{S}$ associated with $(K_{\mathcal{S}},C_{\mathcal{S}})$.
Since the oriented direction field on $\mathscr{KC}_0$ is tangential to
$\partial(\mathscr{KC}_0)$ in the limit $(K,C)\rightarrow (K_0,C_0)$,
the integral curve $(K,C)(\tau)$ that represents $\mathcal{S}_\tau$
can be approximated (in at least the $\mathcal{C}^1$-sense) 
by the curve $\big(K(\tau),C_t(K(\tau))\big)$
for all $\tau$ of some (small) $\tau$-interval (independent of $\delta$).

It follows that the above statement carries over to the slicings $\mathcal{S}_\tau$:
in some region $t\in[-E,E]$ of the black hole, for all $\tau$ in some small interval,
the slicing $\mathcal{S}_\tau$ can be approximated by the slicing of $r=\mathrm{const}$ hypersurfaces
through $r=r_0$.
We conclude that
\begin{equation}\label{alphaminneg}
\alpha_{\min}(K,C) \:\rightarrow\: \alpha_r = \frac{\dot{K}}{a(r_0)} \qquad \text{for }
(K,C)\rightarrow (K_0,C_0)\:.
\end{equation}
Standard algebraic manipulations reveal that $a(r_0) < 0$ for all $(K_0,C_0) \in C_t$. Therefore,
$\alpha_{\min}(K,C) < 0$ for all $(K,C)$ sufficiently close to $C_t$, i.e.\ there exists a
neighborhood $V(C_t)$ of the line $\{(K,C_t(K))\:|\: K\in(-\sqrt{3\Lambda},\sqrt{3\Lambda})\}$ in
$\overline{\mathscr{KC}_0}$ such that $V(C_t) \cap \mathscr{KC}_{0+} = \emptyset$. The statement
for $C_b$ follows via the symmetry property of $\mathscr{KC}_0$, hence the first claim is
established.

The proof of the second claim of the proposition follows the same principle: we consider a
CMC-hyper\-surface $H_0$ in KSSdS that is associated with 
mean curvature $K_0 = \sqrt{3 \Lambda}$ and $C_0 \in
(C_b(\sqrt{3\Lambda}), C_t(\sqrt{3\Lambda}))$ and we exploit the fact that the family of
CMC-hypersurfaces $H_\tau$ generated by $(\sqrt{3 \Lambda},C(\tau))$, where $C(\tau)$ is running in
$(C_b(\sqrt{3\Lambda}), C_t(\sqrt{3\Lambda}))$, forms a foliation of (a part of) KSSdS; in
particular, $\alpha(r) >0$ for $H_0$. The slicing $H_\tau$ has been investigated
in~\cite{Nakao/Maeda/Nakamura/Oohara:1991}. However, a proof of the positivity of the associated
lapse function is missing. This gap can be closed via considerations similar to those of
Appendix~\ref{thelapseequation}, for a detailed discussion see~\cite{NonCompPaper}.

When $\mathcal{S}$ is a compact CMC-hypersurface whose values
$(K_{\mathcal{S}}, C_{\mathcal{S}})$ are sufficiently close to $(K_0=\sqrt{3\Lambda}, C_0)$,
it can be suitably approximated by the hypersurface $H_0$,
because $r_{\min}(K,C) \rightarrow r_{\min}(\sqrt{3 \Lambda}, C_0)$ and
\begin{equation}
t(K,C;r) \rightarrow t(\sqrt{3 \Lambda}, C_0; r)
\qquad \text{when}\quad(K,C) \rightarrow (\sqrt{3 \Lambda},C_0)
\end{equation}
at least in $\mathcal{C}^1$, uniformly on every compact $r$-interval.
Since the oriented direction field on $\mathscr{KC}_0$
is tangential to $\partial(\mathscr{KC}_0)$ in the limit
$(K,C)\rightarrow (\sqrt{3\Lambda},C_0)$
we infer
\begin{equation}
\alpha(K,C;r) \:\rightarrow\: \alpha(\sqrt{3 \Lambda}, C_0;r)
\qquad \text{for }
(K,C)\rightarrow (\sqrt{3 \Lambda},C_0)\:,
\end{equation}
hence, if $(K,C)$ is in a sufficiently small neighborhood
of $(\sqrt{3 \Lambda},C_0)$, then $\alpha(K,C;r)$ is positive
since $\alpha(\sqrt{3 \Lambda}, C_0;r)$ is positive.
Accordingly, there exists a neighborhood $U(\sqrt{3\Lambda})$ of the line
$\{(\sqrt{3\Lambda},C)\:|\: C\in(C_b(\sqrt{3\Lambda}),C_t(\sqrt{3\Lambda}))\}$
in $\overline{\mathscr{KC}_0}$
such that $U(\sqrt{3\Lambda}) \cap \mathscr{KC}_0 \subseteq \mathscr{KC}_{0+}$.
\proofend

\begin{corollary}\label{folicor}
The compact CMC-slicing $\mathcal{S}_\tau$, $\tau\in(\tau_-,\tau_+)$, in the spacetime
KSSdS[T], $|\text{T}|$ sufficiently large, cannot be a foliation, i.e.\
$(\bar{\tau}_-,\bar{\tau}_+) \neq(\tau_-,\tau_+)$ in Theorem~\ref{foliationthm}.
\end{corollary}

\proof
The integral curve $(K,C)(\tau)$ associated
with the slicing $\mathcal{S}_\tau$
is characterized by $\mathcal{T}(\tau) \equiv \text{T}$;
when $|\text{T}|$ is sufficiently large,
it must pass through a given neighborhood of $C_{b}$ or $C_t$
(where $\mathcal{T} = \pm \infty$, cf.~Fig.~\ref{direconKC})
and thus through $V(C_b)$ or $V(C_t)$;
in $V(C_b)$ and $V(C_t)$ the lapse function is not positive.
\proofend

\begin{remark}
Let $\mathcal{S}_\tau^\prime$ be an arbitrary slicing in KSSdS[T] that
is not reflection symmetric and let $\hat{\alpha}$ denote its lapse function.
$\mathcal{S}_\tau^\prime$ arises from $\mathcal{S}_\tau$ by combining the flow
of $\mathcal{S}_\tau$ with an appropriate admixture of the Killing flow, see Theorem~\ref{uniqueslicing},
therefore $\hat{\alpha}(l) = \alpha(l) + k_\xi \alpha_\xi(l) = \alpha(l) + k_\xi r^\prime(l)$
for a constant $k_\xi$ (depending on $\tau$), see Appendix~\ref{thelapseequation}.
Since $r^\prime(l)$ is odd, $\hat{\alpha} \not> 0$ whenever $\alpha \not> 0$, i.e.\
if $\mathcal{S}_\tau$ is not a foliation then there exists no other slicing
$\mathcal{S}_\tau^\prime$ in KSSdS[T] that is.
\end{remark}

\begin{figure}[htp]
    \psfrag{Ct}[cc][cc][0.9][7]{$\leftarrow\, C_t(K)\, \rightarrow$}
    \psfrag{Cb}[cc][cc][0.9][7]{$\leftarrow\, C_b(K)\,\rightarrow$}
    \psfrag{Pt}[cc][cc][0.75][10]{$P_t(K)$}
    \psfrag{Pb}[cc][cc][0.75][10]{$P_b(K)$}
    \psfrag{E}[cc][cc][0.9][-90]{$K=\sqrt{3 \Lambda}$}
    \psfrag{F}[cc][cc][0.9][90]{$K=-\sqrt{3\Lambda}$}
    \psfrag{KC}[cc][cc][1.5][0]{$\mathscr{KC}_{0+}$}
    \centering
    \includegraphics[width=0.6\textwidth]{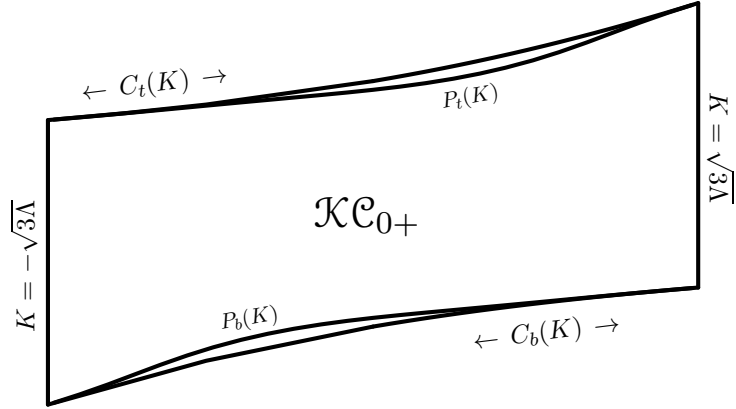}
    \caption{{\small $\mathscr{KC}_{0+}$ and its boundaries $P_t$, $P_b$ in the case $\Lambda=1$, $M=1/4$.}}
        \label{KC0+}
\end{figure}

\begin{figure}[htp]
    \psfrag{E}[cc][cc][0.7][0]{$\sqrt{3 \Lambda}$}
    \psfrag{F}[cc][cc][0.7][0]{$-\sqrt{3\Lambda}$}
    \psfrag{A}[lc][cc][0.8][0]{$1$}
    \psfrag{B}[lc][cc][0.8][0]{$0.95$}
    \centering
    \includegraphics[width=0.6\textwidth]{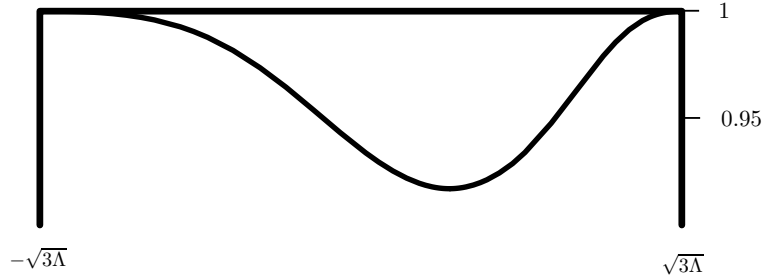}
    \caption{{\small The depicted function represents $P_t(K)/C_t(K)$.}}
        \label{PtFig}
\end{figure}

Numerical investigations suggest that $\mathscr{KC}_{0+}$ is a connected
set; the boundary $\partial(\mathscr{KC}_{0+})$ consists of two
smooth curves:
one curve, $P_t$, that connects the points $(-\sqrt{3 \Lambda}, C_t(-\sqrt{3 \Lambda}))$
and $(\sqrt{3 \Lambda}, C_t(\sqrt{3 \Lambda}))$, and
a second curve, $P_b$, that connects the points $(-\sqrt{3 \Lambda}, C_b(-\sqrt{3 \Lambda}))$
and $(\sqrt{3 \Lambda}, C_b(\sqrt{3 \Lambda}))$.
The curve $P_t$ is given by
\begin{equation}
P_t = \{(K,C)\:|\: (\alpha_{\min}(K,C) = 0) \wedge (C > 0)\} \:,
\end{equation}
cf.~(\ref{alphaminnulliff}), $P_b$ is the reflected curve, see Figs.~\ref{KC0+} and~\ref{PtFig}.
As with $C_{b,t}$ we use the same symbols when we
write the curves in parametric form, i.e.\ we also write $C=P_b(K)$ and $C=P_t(K)$.

\renewcommand{\theenumi}{\roman{enumi}}
A slicing $\mathcal{S}_\tau$, $\tau\in(\tau_-,\tau_+)$, in KSSdS[T] is a foliation as long as the
associated integral curve $(K,C)(\tau)$ of the oriented direction field lies in
$\mathscr{KC}_{0+}$; if the integral curve intersects $P_t$ or $P_b$, $\mathcal{S}_\tau$ cannot be
a foliation for all $\tau$. Since the integral curve is characterized by $\mathcal{T}(\tau) =
\mathcal{T}(K(\tau),C(\tau)) \equiv \mathrm{T}$, at a possible intersection point $(K,P_t(K))$,
$\mathcal{T}(K,P_t(K)) = \mathrm{T}$ must hold, and analogously $\mathcal{T}(K,P_b(K)) =
\mathrm{T}$ for $P_b$. Therefore, the problem of whether or under what conditions a slicing is a
foliation can be investigated by analyzing the function $\mathcal{T}(K,P_t(K))$ and the equation
$\mathcal{T}(K,P_t(K)) = \mathrm{T}$. Let $\mathrm{T}_{\mathrm{f}} := \inf_K
\mathcal{T}(K,P_t(K))$;
\begin{enumerate}
\item if $\mathrm{T}_{\mathrm{f}} \not> 0$, then, for each T,
the slicing $\mathcal{S}_\tau$ in KSSdS[T] is not a foliation.
\item If $\mathrm{T}_{\mathrm{f}}> 0$,
then slicings $\mathcal{S}_\tau$ contained in spacetimes KSSdS[T] with
$|\mathrm{T}| < \mathrm{T}_{\mathrm{f}}$
are foliations; slicings in KSSdS[T] with $|\text{T}| \geq \mathrm{T}_{\mathrm{f}}$ are not.
\end{enumerate}
To see (i) and (ii), we note that
the equation $\mathcal{T}(K,P_t(K)) = \text{T}$ has a solution $K$
iff $\text{T} \geq \mathrm{T}_{\mathrm{f}}$.
By virtue of the reflection symmetry in $\mathscr{KC}_0$,
$\mathcal{T}(K,P_b(K)) = -\mathcal{T}(-K,P_t(-K))$, thus
the equation $\mathcal{T}(K,P_b(K)) = \text{T}$
is equivalent to the equation $\mathcal{T}(-K,P_t(-K)) = -\text{T}$;
it has a solution iff $\text{T} \leq -\mathrm{T}_{\mathrm{f}}$.
If $\mathrm{T}_{\mathrm{f}} \leq 0$, then, for all $\text{T}$,
at least one of the equations has a solution.
If $\mathrm{T}_{\mathrm{f}} > 0$, then, for all $|\text{T}| \geq \mathrm{T}_{\mathrm{f}}$,
at least one of the equations has a solution; however,
for all $|\text{T}| < \mathrm{T}_{\mathrm{f}}$,
neither of the equations has a solution.
This implies that the integral curve $\mathcal{T}(\tau) = \text{T}$ (with $|\text{T}| < \mathrm{T}_{\mathrm{f}}$)
in $\mathscr{KC}_0$
does neither intersect $P_t$ nor $P_b$, but is entirely contained in $\mathscr{KC}_{0+}$;
thus the associated slicing $\mathcal{S}_\tau$, $\tau \in (\tau_-,\tau_+)$, is a foliation.

\begin{figure}[htp]
    \psfrag{K}[lc][rc][0.8][0]{$K$}
    \psfrag{T}[cb][cc][0.8][0]{$\mathcal{T}$}
    \psfrag{1}[cc][cc][0.5][0]{$1$}
    \psfrag{-1}[rc][rc][0.5][0]{$-1$}
    \psfrag{-3}[rc][rc][0.5][0]{$-3$}
    \psfrag{-5}[rc][rc][0.5][0]{$-5$}
    \psfrag{0.5}[ct][ct][0.5][0]{$0.5$}
    \centering
    \subfigure[$M=0.100$]{\label{tofrmax100}\psfrag{A}[ct][ct][0.5][0]{$\sqrt{3 \Lambda}$}
      \includegraphics[width=0.35\textwidth]{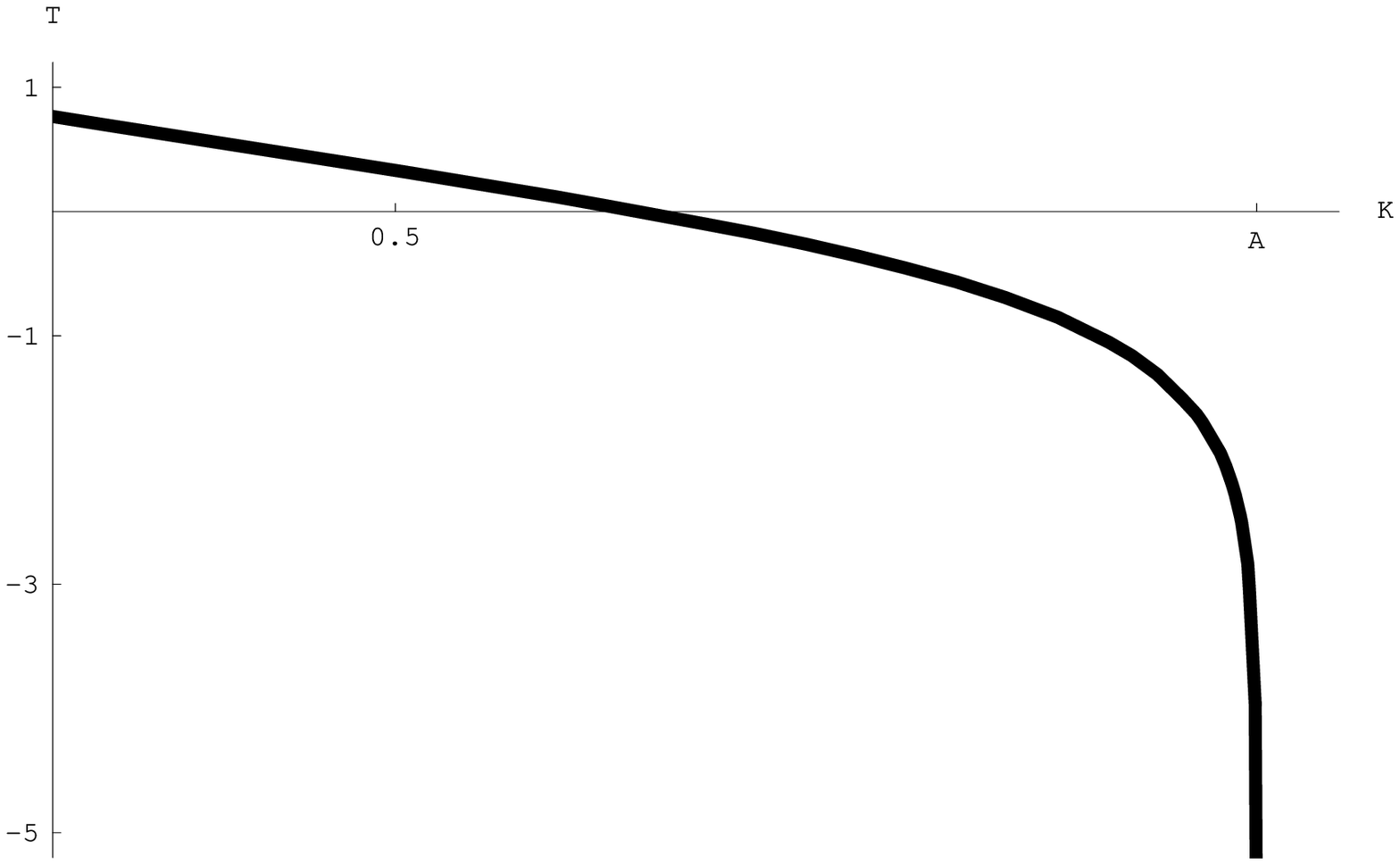}}\qquad\qquad
    \subfigure[$M=0.200$]{\label{tofrmax200}\psfrag{A}[cc][cc][0.5][0]{$\sqrt{3 \Lambda}$}
      \includegraphics[width=0.35\textwidth]{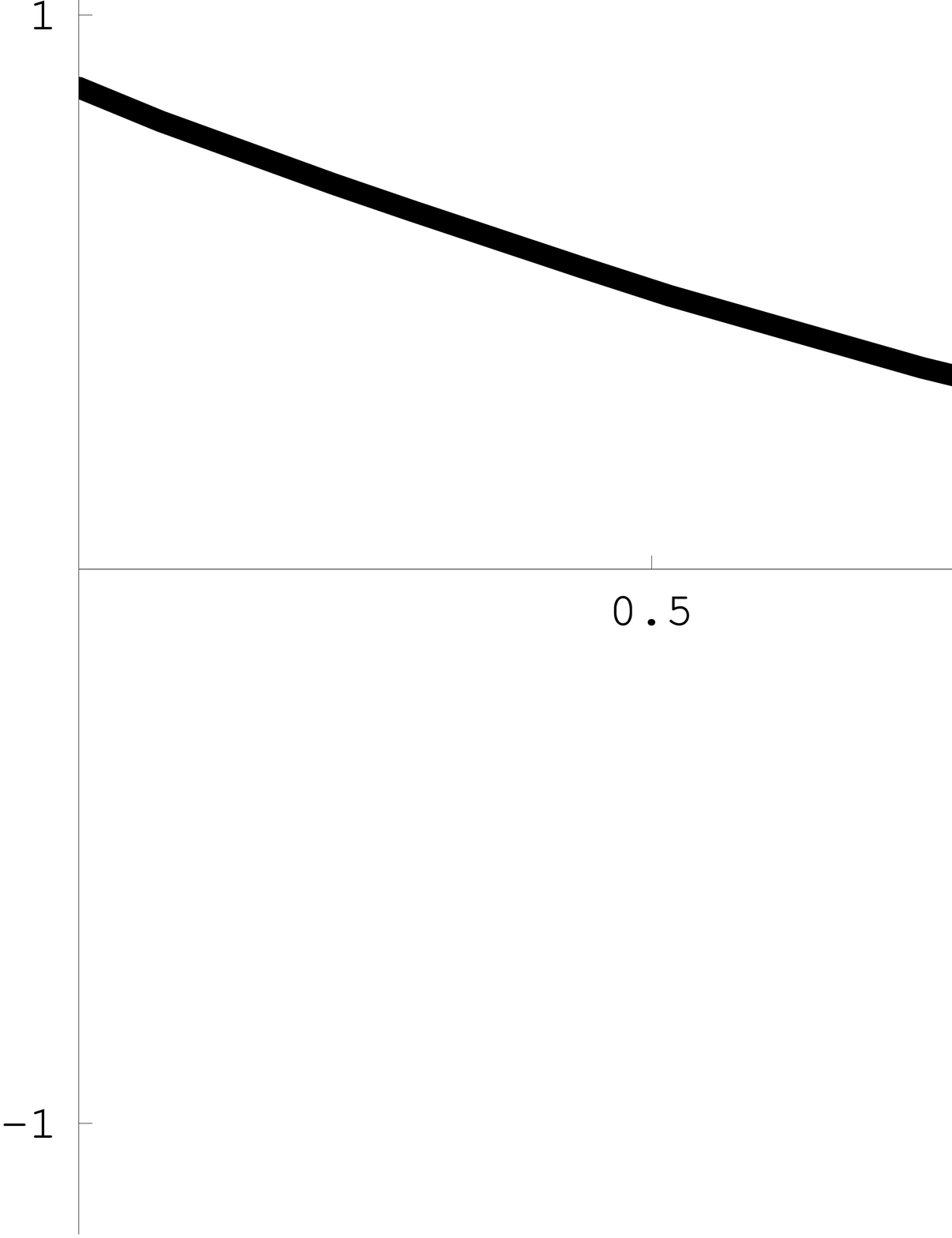}}
    \subfigure[$M=0.215$]{\label{tofrmax215}
      \includegraphics[width=0.35\textwidth]{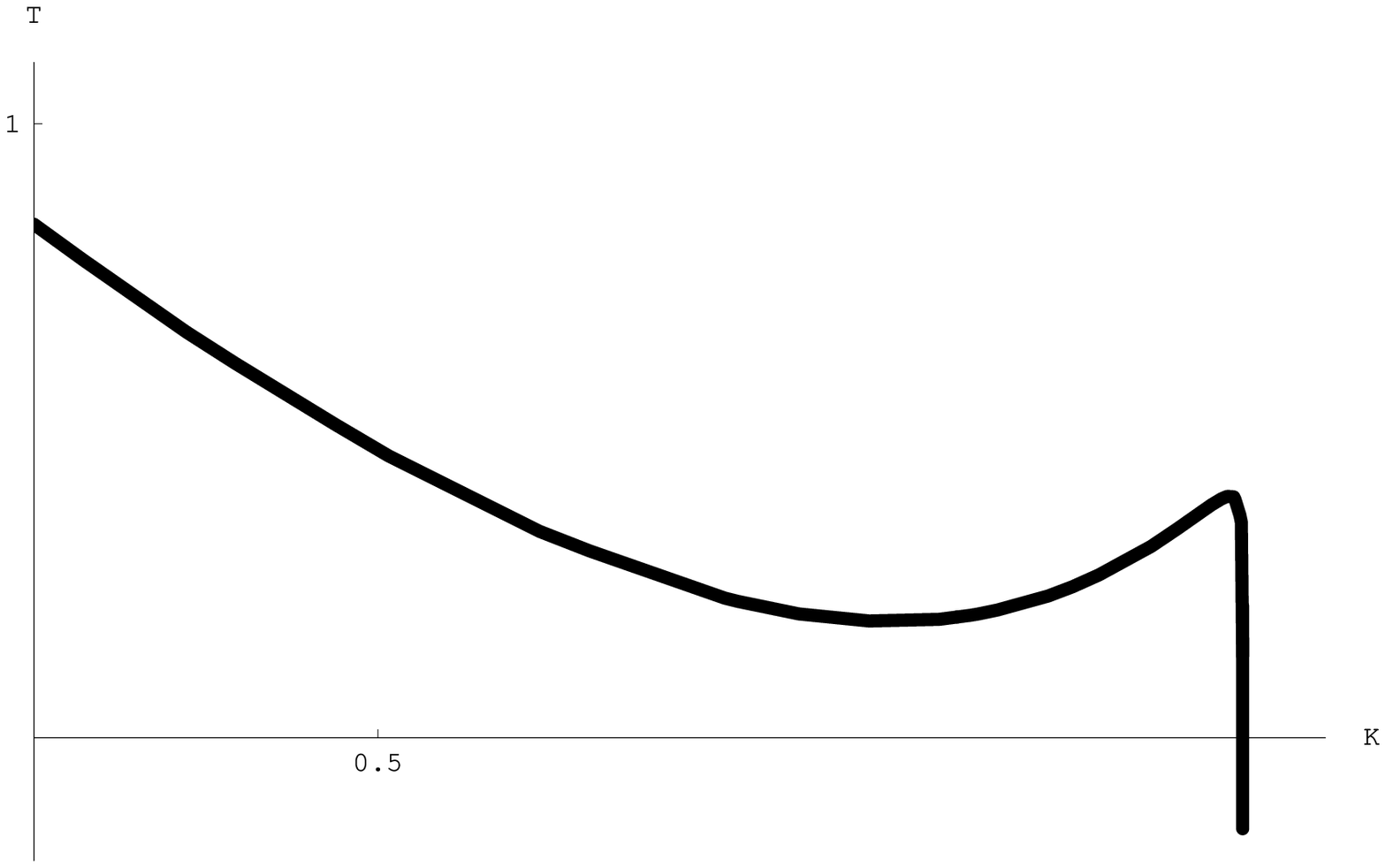}}\qquad\qquad
    \subfigure[$M=0.220$]{\label{tofrmax220}\psfrag{A}[ct][ct][0.5][0]{$\sqrt{3 \Lambda}$}
      \includegraphics[width=0.35\textwidth]{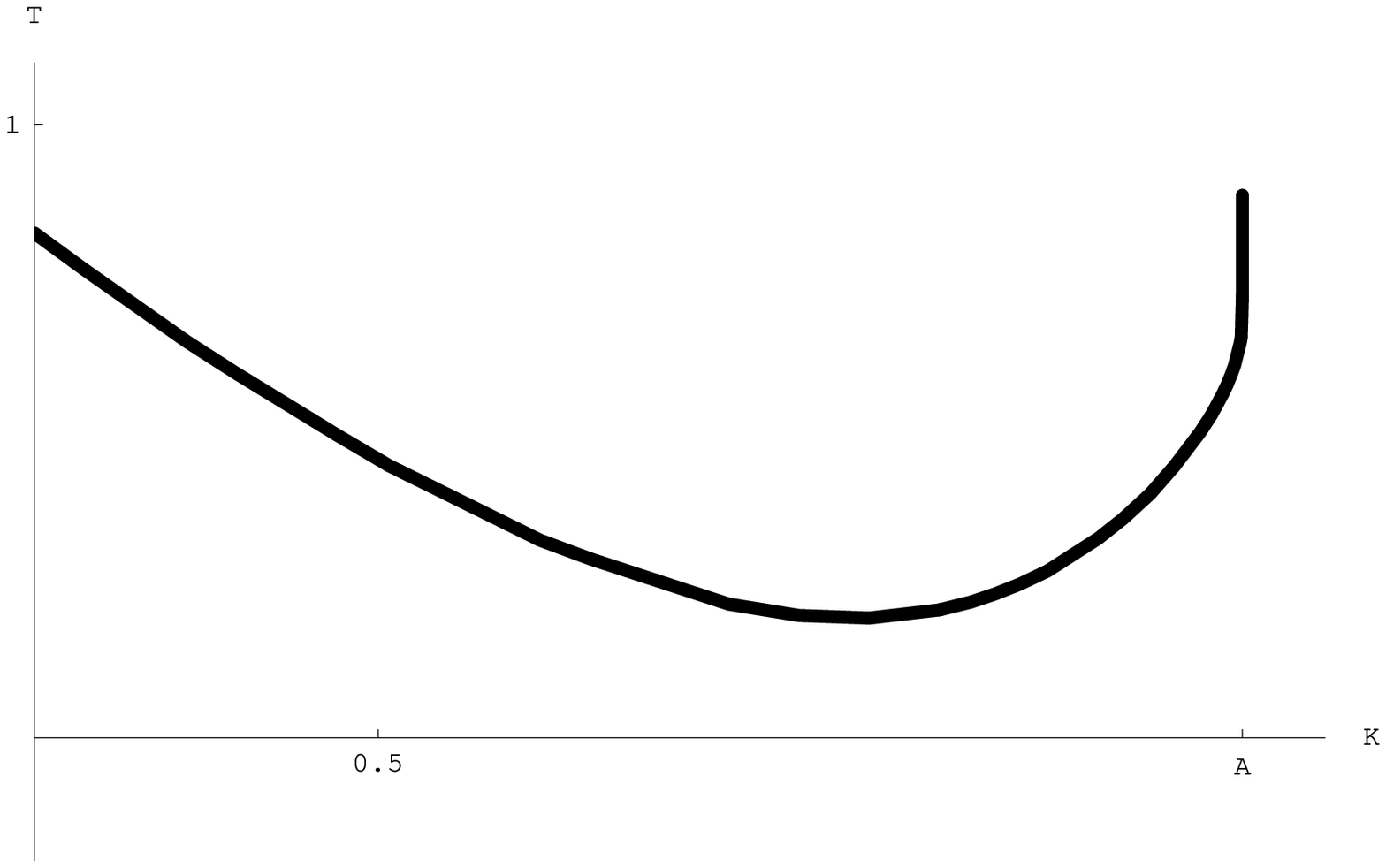}}
    \subfigure[$M=0.250$]{\label{tofrmax250}\psfrag{A}[ct][ct][0.5][0]{$\sqrt{3 \Lambda}$}
      \includegraphics[width=0.35\textwidth]{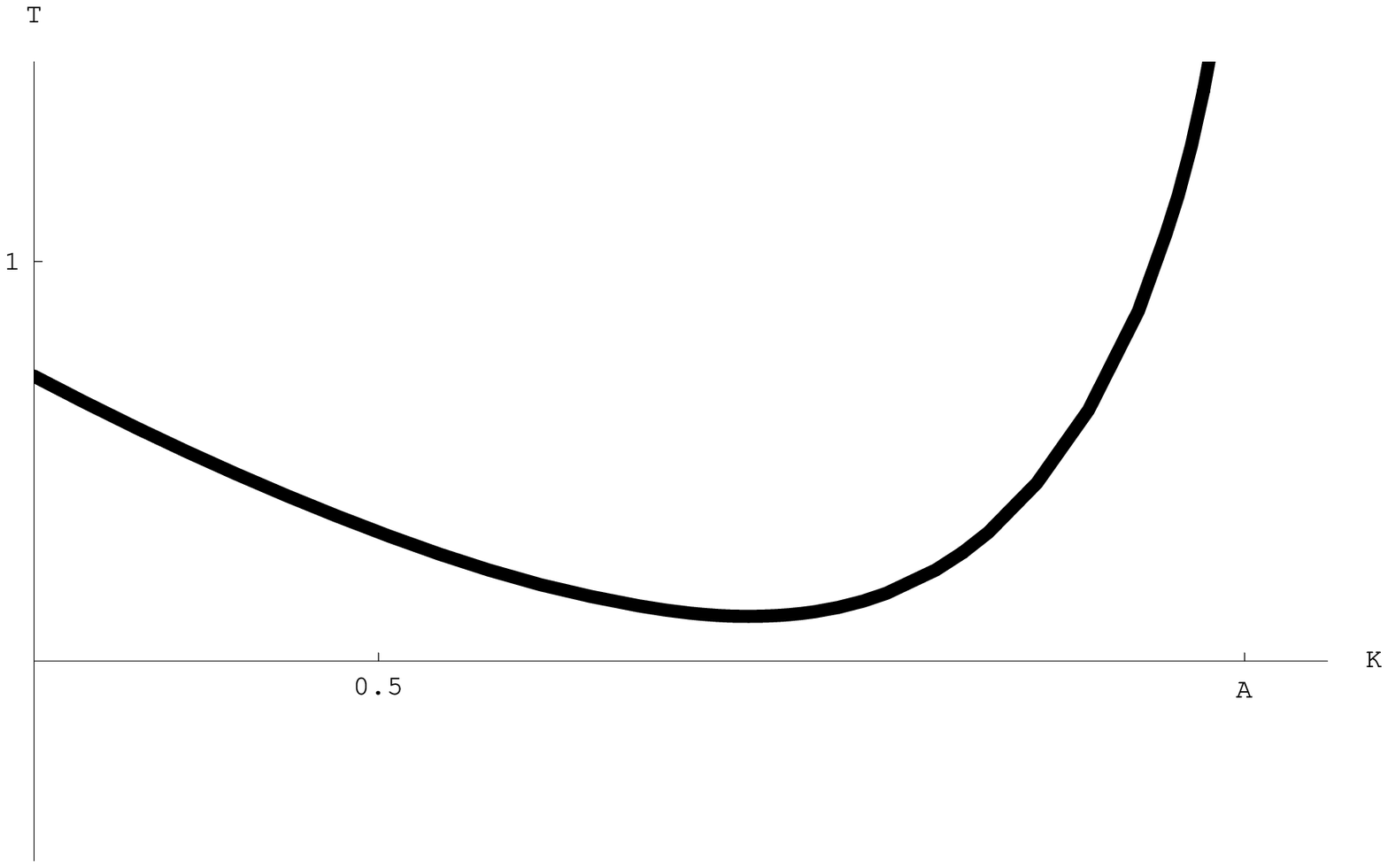}}\qquad\qquad
    \subfigure[$M=0.330$]{\label{tofrmax330}\psfrag{A}[ct][ct][0.5][0]{$\sqrt{3 \Lambda}$}
      \includegraphics[width=0.35\textwidth]{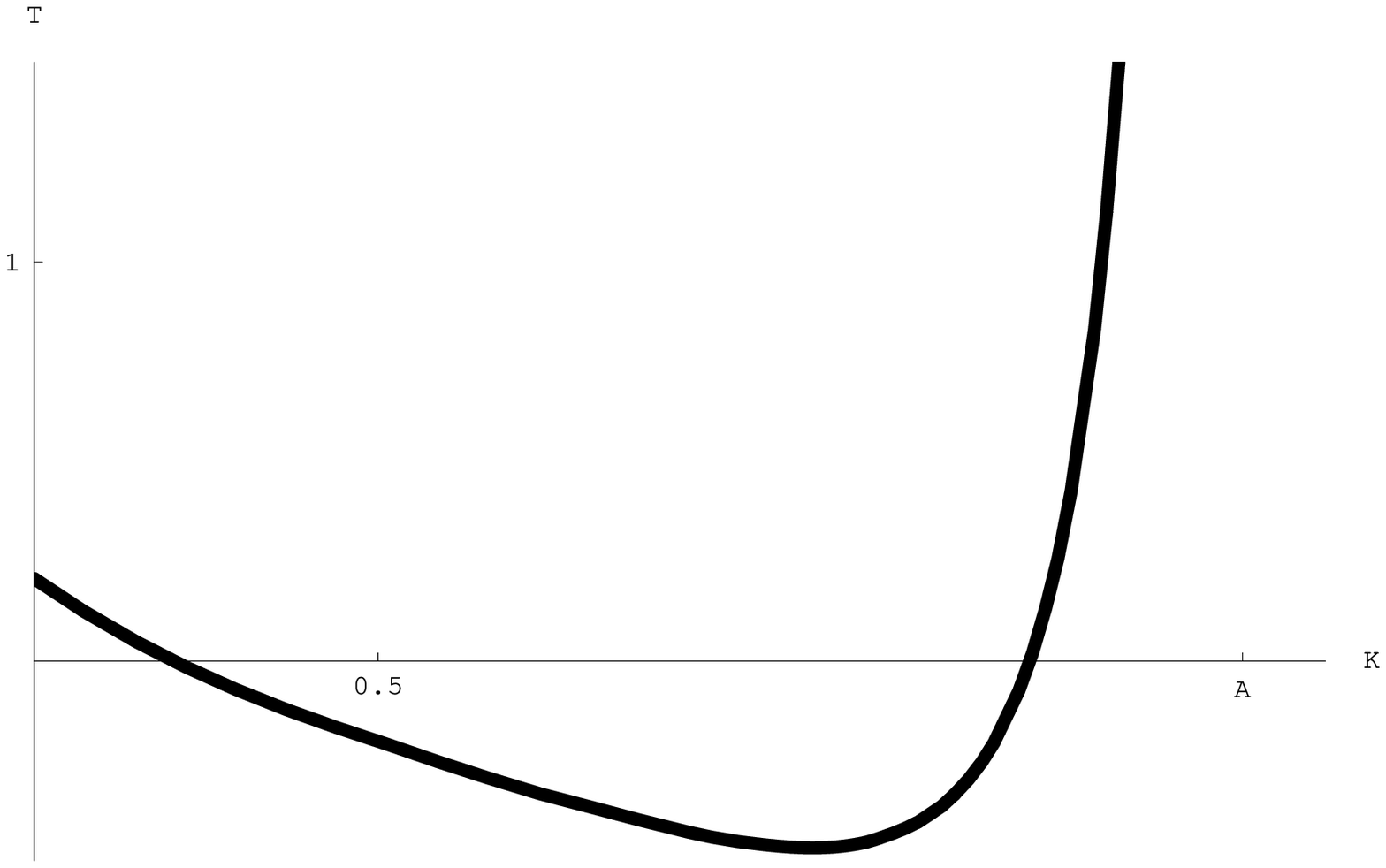}}
    \caption{{\small The figures show the function $\mathcal{T}(K,P_t(K))$ for $K>0$
        for several values of $M$; $\Lambda=1$ in all cases.
        $\mathcal{T}(K,P_t(K))\rightarrow -\infty$ ($K\rightarrow \sqrt{3\Lambda}$) for all $M < M_{\mathrm{c}}$,
        i.e.\ for (a)--(c);
        $\mathcal{T}(K,P_t(K))\rightarrow +\infty$ ($K\rightarrow \sqrt{3\Lambda}$) for all $M > M_{\mathrm{c}}$,
        i.e.\ for (d)--(f);
        in the case $M = M_{\mathrm{c}}$ the function converges to a constant;
        $M_{\mathrm{c}} \approx 0.218945$. $\mathcal{T}(K,P_t(K))\geq\mathrm{const}>0$ for all $K$
        when $M\in [M_{\mathrm{c}}, M_0)$, i.e.\ for (d) and (e); $M_0 \approx 0.268516$.}}
        \label{tofrmaxalongPt}
\end{figure}

Lemma~\ref{Tlemma} entails that $\mathcal{T}(K,P_t(K)) \rightarrow \infty$ for $K\rightarrow -\sqrt{3 \Lambda}$;
moreover, the numerical results suggest that
the function $\mathcal{T}(K,P_t(K))$ is always strictly monotonically decreasing for
$K<0$. However, for $K>0$, the properties of $\mathcal{T}(K,P_t(K))$ depend
on the chosen family of KSSdS[T]-spacetimes, i.e.\ on $(\Lambda, M)$!
In Fig.~\ref{tofrmaxalongPt} we show the function $\mathcal{T}(K,P_t(K))$
for several cases of $M$, where $\Lambda =1$.

Numerical results suggest that
the asymptotic behavior of $\mathcal{T}(K,P_t(K))$ as $K\rightarrow \sqrt{3\Lambda}$
can be approximated by
\begin{equation}
\mathcal{T}(K,P_t(K)) = c_1 + c_2 \log (\sqrt{3\Lambda} - K) \:,
\end{equation}
where the constants $c_1$, $c_2$ depend on $(\Lambda, M)$. This is consistent with the
asymptotics of $\mathcal{T}$ obtained in the proof of Lemma~\ref{Tlemma}.
The constant $c_2$ is positive (so that $\mathcal{T}(K,P_t(K)) \rightarrow -\infty$
as $K\rightarrow \sqrt{3\Lambda}$)
for all $M$ satisfying $M < M_{\mathrm{c}}$; $c_2 = 0$ for $M = M_{\mathrm{c}}$,
and $c_2 < 0$ in the case $M_{\mathrm{c}}< M < (3\sqrt{\Lambda})^{-1}$, cf.~Fig.~\ref{tofrmaxalongPt}.

In the case $M \geq M_{\mathrm{c}}$, positivity of $\mathcal{T}(K,P_t(K))$ is possible;
indeed, there exists $M_0$ such that $\mathrm{T}_{\mathrm{f}}>0$
for all $M_{\mathrm{c}}\leq M < M_0$, see Figs.~\ref{tofrmax220} and~\ref{tofrmax250}.
Hence, for $M_{\mathrm{c}}\leq M < M_0$, the slicings $\mathcal{S}_\tau$ in
the spacetimes KSSdS[T] whose $|\mathrm{T}|$ is small enough,
are foliations, cf.~(ii). Thus there is strong numerical evidence that
Conjecture~\ref{foliconj} is true.

From Fig.~\ref{tofrmax215} we see that the equation $\mathcal{T}(K,P_t(K)) = \text{T}$
can have up to three solutions. Since in that case $\mathcal{T}(K,P_b(K)) = \text{T}$
has also one solution,
an integral curve of the
oriented direction field can switch between $\mathscr{KC}_{0+}$
and $\mathscr{KC}_{0}\backslash\mathscr{KC}_{0+}$ up to four times.
The different scenarios can be read off from Fig.~\ref{tofrmaxalongPt}.

We conclude this section by noting that the asymptotic behavior of
$P_t(K)$ is given by
\begin{equation}
P_t(K) = C_t(K) \left( 1 - [d_1 + d_2 (\sqrt{3\Lambda}-K)] (\sqrt{3\Lambda}- K)^2 \right)\:,
\end{equation}
where $d_1$ and $d_2$ are constants, $d_1 > 0$, that depend on $(M,\Lambda)$.

\end{appendix}

\bibliographystyle{plain}

\end{document}